\documentclass[preprint,nonatbib]{elsarticle}

\usepackage{booktabs}
\usepackage{todonotes}
\usepackage{graphics}
\usepackage{booktabs}
\usepackage{tabularx}
\usepackage{caption}
\usepackage{tcolorbox}
\usepackage{fbox}
\usepackage{footnote}
\usepackage[flushleft]{threeparttable}
\usepackage{dblfloatfix} 
\usepackage{changepage}
\usepackage{float}
\usepackage{longtable} % for 'longtable' environment
\usepackage{pdflscape} % for 'landscape' environment

\usepackage{url}

\usepackage{breakurl}

\usepackage{hyperref}
\usepackage[labeled,resetlabels]{multibib}

\newcites{S}{References for Academic Literature}
\newcites{G}{References for Grey Literature}

\newtcolorbox{qoutebox}[3][]
{
  colframe=black!30!white,
  colback  = #2!10,
  #1,
}

\journal{Journal of Systems and Software}

\begin{document}

\title{Test Flakiness'  Causes, Detection, Impact and Responses: A Multivocal Review}
%\titlerunning{Understanding flaky tests}

\author[massey]{Shawn Rasheed\footnotemark[1]}
\author[massey]{Amjed Tahir\thanks{some info}}
\author[jens]{Jens Dietrich}
\author[massey]{Negar Hashemi}
\author[zhang]{Lu Zhang}

\address[massey]{School of Mathematical and Computational Sciences, Massey University, New Zealand}
\address[jens]{School of Engineering and Computer Science, Victoria University of Wellington, New Zealand}
\address[zhang]{Key Laboratory of High Confidence Software Technologies, MoE and the Department of Computer Science and Technology, Peking University, China}

\begin{keyword}
flaky tests; non-deterministic tests; test bugs; software testing; multivocal review
\end{keyword}%}

%\UseRawInputEncoding

% \IEEEtitleabstractindextext{%
\begin{abstract}
Flaky tests (tests with non-deterministic outcomes) pose a major challenge for software testing. They are known to cause significant issues such as reducing the effectiveness and efficiency of testing and delaying software releases. In recent years, there has been an increased interest in flaky tests, with research focusing on different aspects of flakiness, such as identifying causes, detection methods and mitigation strategies. 
Test flakiness has also become a key discussion point for practitioners (in blog posts, technical magazines, etc.) as the impact of flaky tests is felt across the industry. 
This paper presents a \textit{multivocal review} that investigates how flaky tests, as a topic, have been addressed in both research and practice. We cover a total of 651 articles (560 academic articles and 91 grey literature articles/posts), and structure the body of relevant research and knowledge using four different dimensions: \textit{causes}, \textit{detection}, \textit{impact} and \textit{responses}. For each of those dimensions we provide a categorisation, and classify existing research, discussions, methods and tools. 
With this, we provide a comprehensive and current snapshot of existing thinking on test flakiness, covering both academic views and industrial practices, and identify limitations and opportunities for future research. 

\end{abstract}

\maketitle

\footnotetext[1]{Joint first authors.}

\section{Introduction}
 \label{sec:introduction}

Software testing is a standard method used to uncover defects. Developers use tests early during development to uncover software defects when corrective actions are relatively inexpensive. A test can only provide useful feedback if it has the same outcome (either pass or fail) for every execution with the same version of code. Tests with non-deterministic outcomes (known as \textit{flaky tests}) are tests that may pass in some runs and fail on others. Such flaky behaviour is problematic as it leads to uncertainty in choosing corrective measures \cite{harman2018start}. They also incur heavy costs in developers' time and other resources, particularly when the test suites are large and development follows an agile methodology, requiring frequent regression testing on code changes to safeguard releases. 

Test flakiness has been attracting more attention in recent years. In particular, there are several studies on the causes and impact of flaky tests in both open-source and proprietary software. In a study of open source projects, it was observed that 13\% of failed builds are due to flaky tests \cite{Labuschagne2017}. At Google, it was reported that around 16\%  of their tests were flaky, and 1 in 7 of the tests written by their engineers occasionally fail in a way that is not caused by changes to the code or tests \cite{googleFlaky2016}. GitHub  also reported that, in 2020, one in eleven commits (9\%) had at least one red build caused by a flaky test \cite{github2020reducing}.
Other industrial reports have shown that flaky tests present a real problem in practice that have a wider impact on product quality and delivery \cite{fowler2011eradicating,SandhuTesting2015,palmer2019}. 
Studies of test flakiness have also been covered in the context of several programming languages including Java \cite{luo2014empirical}, Python \cite{gruber2021empirical} and, more recently, JavaScript \cite{Hashemi2022flakyJS}.

Awareness that more research on test flakiness is needed has increased in recent years \cite{harman2018start}. Currently, studies on test flakiness and its causes largely focus on specific sources of test flakiness, such as order-dependency \cite{gambi2018practical}, concurrency \cite{dong2021flaky} or UI-specific flakiness \cite{memon2013automated,romano2021empirical}. 
Given that test flakiness is an issue in both research and practice, we deem it is important to integrate knowledge about flaky tests from both academic literature and grey literature in order to provide insights into the state-of-the-practice. 

In order to address this, 
we performed a multivocal literature review on flaky tests. 
A multivocal review is a form of a \textit{systematic literature review} \cite{kitchenham2007guidelines} which includes sources from both academic (formal) and grey literature \cite{Garousi2019Guidelines}. Such reviews in computer science and software engineering have become popular over the past few years  \cite{Tom2013TD,garousi2018smell,Islam2019Security,Butijn2020Blockchains}  as it is acknowledged that the majority of developers and practitioners do not publish their work or thoughts through peer-reviewed academic channels \cite{Garousi2016Multivocalreviews,Glass2006Creativity}, but rather in blogs, discussion boards and Q\&A sites \cite{Williams2019Grey}.

This research summarizes existing work and current thinking on test flakiness from both academic and grey literature. 
We hope that this can help a reader to develop an in-depth understanding of common causes of test flakiness, methods used to detect flaky tests, strategies used to avoid and eliminate them, and the impact flaky tests have.  We identify current challenges and suggest possible future directions for research in this area.

The remaining part of the paper is structured as follows: Section \ref{sec:background} presents recent reviews and surveys on the topic. Our review methodology is explained in Section \ref{sec:design}. We present our results answering all four research questions in Section \ref{sec:results}, followed by a discussion of the results in Section \ref{sec:discussion}. Threats to validity are presented in Section \ref{sec:threats}, and finally we present our conclusion in Section \ref{sec:conclusion}. 
\section{Related Work}
\label{sec:background}
There has been a growing interest in flaky tests in recent years, especially after the publication of Martin Fowler's article on the potential issues with non-deterministic tests \cite{fowler2011eradicating}, and Luo et al.'s \cite{luo2014empirical} seminal study. 

To the best of our knowledge, there have been three reviews of studies on test flakiness: two systematic literature reviews, one by Zolfaghari et al. and the other by Zheng et al. \cite{zolfaghari2020root,zheng2021research} and a survey by Parry et al. \cite{parry2021survey}.

There have also been some developers' surveys that aimed to understand how developers perceive and deal with flaky tests in practice. A developer survey conducted by Eck et al. \cite{eck2019understanding} with 21 Mozilla developers studied the nature and the origin of 200 flaky tests that had been fixed by the same developers. The survey looked into how those tests were introduced and fixed, and found that there are 11 main causes for those 200 flaky tests (including concurrency, async wait and test order dependency). It was also pointed out that flaky tests can be the result of issues in the production code (code under test) rather than in the test. The authors also surveyed another 121 developers about their experience with flaky tests. It was found that flakiness is
perceived as a significant issue by the vast majority of developers they surveyed. 
The study reported that developers found flaky tests to have a wider impact on the reliability of the test suites. 
As part of their survey with developers, the authors also conducted a mini-multivocal review study to collect evidence from the literature on the challenges to deal with flaky tests. However, this was a small, targeted review to address only the challenges to deal with flaky tests. The study included a review of only a few (19) articles.
A recent developers' survey \cite{habchi2022qualitative} echoed the results found in Eck et al., noting that flakiness can result from interactions between the system components, the testing infrastructure, and other external factors. 

Ahmed et al. \cite{ahmad2021empirical} conducted a similar survey with developers aiming to understand developers' perception of test flakiness (e.g., how developers define flaky tests, and what factors are known to impact the presence of flaky tests).
The study identified several key factors that are believed to be impacted by the presence of test flakiness, such as software product quality and the quality of the test suite.

The systematic review by Zolfaghari et al. \cite{zolfaghari2020root} identified what has been done so far on test flakiness in general, and presented points for future research directions. The authors identified main methods behind approaches for detecting flaky tests, methods for fixing flaky tests, empirical studies on test flakiness, root causes of flakiness and listed tools for detecting flaky tests. The study suggested investigation into building a taxonomy of flaky tests that covers all dimensions (causes, impact, detection), formal modelling of flaky tests, setting standards for flakiness-free testing and investigating the application of AI-based approaches to the problem, and automated flaky test repair.

Zheng et al. \cite{zheng2021research} also discussed current trends and research progress in flaky tests. The study analysed similar questions to the research reported in this paper on causes and detection techniques of flaky tests in 31 primary studies on flaky tests. Hence, this review  was limited, and it did not discuss in detail the mechanism of current detection approaches or the wider impact of flaky tests on other techniques. 
There was a short scoping  grey literature review by Barboni et al. \cite{barboni2021we} that  focused on investigating the definition of flaky tests in grey literature by analysing flaky-test related blogs posted on \emph{Medium}. The study is limited to understanding the definition of flaky tests (highlighting the problem of inconsistent terminology used in the surveyed articles), and covered a small subset of the posts (analysing only 12 articles in total).

Parry et al. \cite{parry2021survey} conducted a more recent comprehensive survey of academic literature on the topic of test flakiness. The study addressed similar research questions to our review, and to those in the previous reviews, by studying causes, detection and mitigation of flaky tests. The study reviewed 76 articles that focused on flaky tests.

The review presented in this paper covers a longer period of time than those previous reviews \cite{parry2021survey,zolfaghari2020root,zheng2021research}, which includes work dating back further (on ``non-deterministic tests''), before the term ``flaky tests'' became popular.  The review contains a discussion of publications through the end of April 2022, where the most recent review of Parry et al \cite{parry2021survey} covers publications through April 2021. We found a significant number of academic articles published in the period between the two reviews (229 articles published between 2021-2022). In general, our study complements previous work in that, 1) we gather more detailed evidence about causes of flaky tests, and investigate the relationships between different causes, 2) we investigate both the impact of and responses to flaky tests in both research and practice, 3) we list the \textit{indirect} impact of flaky tests on other analysis methods and techniques (e.g., software debugging and maintenance).

All previous reviews focused on academic literature. The review by Zolfaghari et al. \cite{zolfaghari2020root} covered a total of 43 articles, Parry et al. \cite{parry2021survey} covered 76 articles, and Zheng et al. \cite{zheng2021research} covered 31 articles.  Our study complements these reviews by providing a much wider coverage and in-depth perspective on the topic of flaky tests. We include a total of 602 academic articles, as well as reviewing 91 grey literature entries (details in Section \ref{sec:data-extraction}). We cover not only studies that directly report on flaky tests, but also those that reference or discuss the issue of test flakiness while it is not the focus of the study. We also discussed a wide range of flaky tests related tools used in research and practice (including industrial and open-source tools), and discuss the impact of flaky tests from different perspectives.

A comparative summary of this review with the previous three reviews is shown in Table \ref{tab:reviews-summary}.

\begin{table*}[]
\caption{Summary of prior reviews on test flakiness compared with our review}
\label{tab:reviews-summary}
\centering
\resizebox{\linewidth}{!}{
\begin{tabular}{@{}lllll@{}}
\toprule
Paper & Period covered & \begin{tabular}[c]{@{}l@{}}\# of reviewed\\  articles\end{tabular} & \begin{tabular}[c]{@{}l@{}}Grey \\ literature\end{tabular} & Focus \\ \midrule
\cite{zolfaghari2020root} Zolfaghari et al.  & 2013 - 2020 & 43 & -- & causes and detection techniques   \\
\cite{zheng2021research} Zheng et al.  &  2014 - 2020 & 31 & -- & causes, impact, detection and fixing approaches  \\
\cite{parry2021survey} Parry et al.  & 2009 - 4/2021 & 76 & -- &
\begin{tabular}[c]{@{}l@{}}causes, costs and consequences, detection and approaches for \\ mitigation and repair \end{tabular} \\
This review & 1994 - 5/2022 & 560 & 91 & \begin{tabular}[c]{@{}l@{}}taxonomy of causes, detection and responses techniques, and \\impact on developers, process and product in research and practice\end{tabular}   \\ \bottomrule
\end{tabular}}
\end{table*}

\section{Study Design}
\label{sec:design}

\begin{figure*}[h]
\centering  
% \resizebox{0.5\linewidth}{!}{
\includegraphics[width=0.70\linewidth]
    	{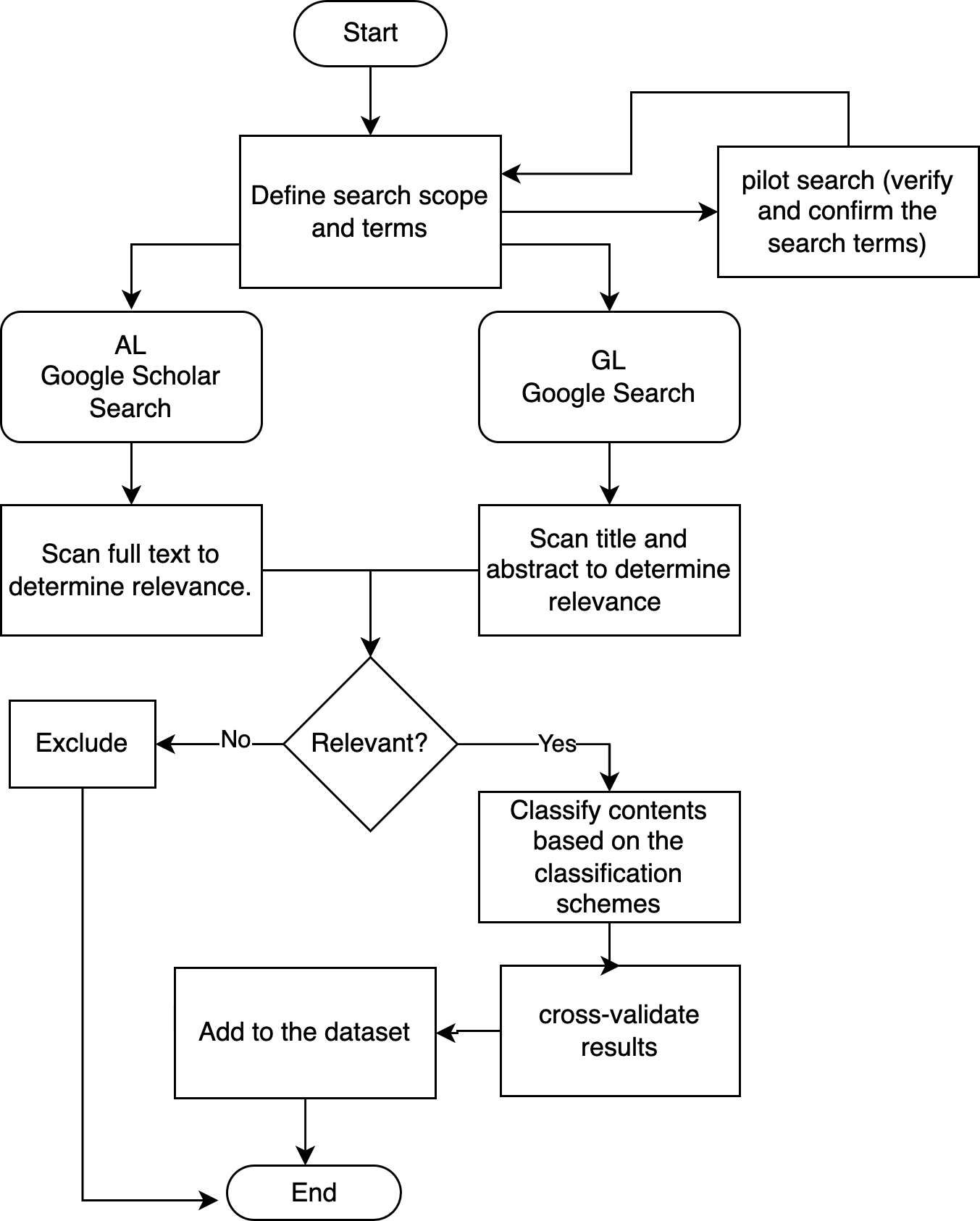}%}
        	        	\caption{An overview of our review process}
        	\label{fig:protocol}
\end{figure*}

We designed this review following the Systematic Literature Review (SLR) guidelines by Kitchenham and Charters \cite{kitchenham2007guidelines}, and the guidelines of Garousi et al. \cite{Garousi2019Guidelines} on multivocal review studies in software engineering. The review process is summarized in Fig. \ref{fig:protocol}.

\subsection{Research Questions}
\label{sec:questions}
This review addresses a set of research questions that we categorized along four main dimensions: \textit{causes}, \textit{detection}, \textit{impact} and \textit{responses}. We list our research questions below, with the rationale for each.

\subsubsection{Common Causes of Flaky Tests:}
 
\noindent\textbf{RQ1. What are the common causes of flaky tests?
}

The rationale behind this question is to list the common causes of flaky tests and then group similar categories of causes together. 
We also investigate the cause-effect relationships between different flakiness causes as reported in the reviewed studies, as we believe that some causes are interrelated. For example, flakiness related to the User Interface (UI)  could be attributed to the underlying environment (e.g., the Operating System (OS) used). Understanding the causes and their relationships is key to dealing with flaky tests (i.e., detection, quarantine or elimination).
 
\subsubsection{Detection of Flaky Tests} %(defn: detection is finding tests that are certainly flaky or potentially flaky): \\
 
 \noindent\textbf{RQ2. How are flaky tests being detected?}\\

 \noindent To better understand flaky tests detection, we divide this research question into the following two sub-questions.\\
 
\noindent\textbf{RQ2.1. What \textit{methods} have been used to detect flaky tests?\\
}
 
\noindent\textbf{RQ2.2. What \textit{tools} have been developed to detect flaky tests?
} \\

In RQs 2.1 and 2.2, we gather evidence regarding methods/tools that have been proposed/used to detect flaky tests. We seek to understand how these methods work. We later discuss the potential advantages and limitations of current approaches.
 
\subsubsection{Impact of Flaky Tests}
 
\noindent\textbf{RQ3. What is the impact of flaky tests?}

As reported in previous studies, flaky tests are generally perceived to have a negative impact on software product and process \cite{fowler2011eradicating,googleFlaky2016,harman2018start,eck2019understanding}. However, it is  important to understand the extent of this impact, and what exactly is affected (e.g., process, product, personnel).
 
\subsubsection{Responses to Flaky Tests}% (mitigation, elimination and fixing):\\
 
\noindent\textbf{RQ4. How do developers/organizations respond to flaky tests when detected? }

Here we are looking at the responses and mitigation strategies employed by developers, development teams and organisations. We note that there are both technical (i.e., how to fix the test or the underlying code that causes the flakiness) and management (i.e., what are the processes followed to manage test suites that are flaky) responses.

\subsection{Review Process}
\label{sec:review-process}
Since this is a multivocal review where we search for academic and grey literature in different forums, the search process for each of the two parts (academic and grey literature) is different and requires different steps.
The systematic literature review search targets peer-reviewed publications that have been published in relevant international journals, conference proceedings and theses in the areas of software engineering, software testing and software maintenance. The search also covers preprint and postprint articles available in open access repositories such as \textit{arXiv}. For the grey literature review, we searched for top-ranked online posts on flaky tests. This includes blog posts, technical reports, white papers, and official documentation for tools.
 
We used \textit{Google Scholar} to search for academic literature and \textit{Google} search engine to search for grey literature. 
Both \textit{Google Scholar} and \textit{Google search} have been used in similar multivocal studies in software engineering \cite{garousi2018smell,myrbakken2017devsecops,garousi2016and} and other areas of computer science  \cite{Islam2019Security,pereira2021security}.
Google Scholar indexes most major publication databases relevant to computer science and software engineering \cite{neuhaus2006depth}, in particular the ACM digital library, IEEE Xplore, ScienceDirect and SpringerLink, thus providing a much wider coverage compared to those individual libraries. A recent study on the use of Google Scholar in software engineering reviews found it to be very effective, as it was able to retrieve $\sim$96\% of primary studies in other review studies \cite{yasin2020using}. 
Similarly, it has been suggested that a regular \textit{Google Search} is sufficient to search for grey literature material online \cite{mahood2014searching,adams2016searching}.

\subsubsection{Searching Academic Publications}
We closely followed Kitchenham and Charters's guidelines \cite{kitchenham2007guidelines} to conduct a full systematic literature review. 
The goal here is to identify and analyse primary studies relating to test flakiness. We defined a search strategy and search string that covers the terminology associated with flaky tests. The search string was tested and refined multiple times during our pilot runs to ensure coverage. We then defined a set of inclusion and exclusion criteria.  
We included a quality assessment of the selection process to ensure that we include all relevant primary studies. We explain those steps in detail below.

 We define the following criteria for our search:
 
\begin{enumerate}
\item [] \textbf{Search engine:} Google Scholar.
% \item []\textbf{Search terms:} ``flaky test'', ``test flakiness'', ``nondeterministic test'', ``non deterministic test''.
\item []\textbf{Search String:} ``flaky test'' OR ``test flakiness'' OR ``flaky tests'' OR ``nondeterministic tests'' OR ``non deterministic tests'' OR ``nondeterministic test'' OR ``non deterministic test''
\item [] \textbf{Search scope:} all academic articles published until 30 April 2022.
\end{enumerate}
 
In case that an article appears in multiple venues (e.g., a conference paper that was also published on arXiv under a different title, or material from a thesis that was subsequently published in a journal or conference proceedings), we only include the published articles/papers over the other available versions. This was to ensure that we include as much peer-reviewed material as possible.
We conducted this search over two iterations. The first iteration covers the period until 31 December 2020, where the second iteration covers the period between 1 January 2021 and 30 April 2022. Results from the two searches were then combined.

\noindent Our review for academic articles follow the following steps:
\begin{enumerate}
	\item Search and retrieve relevant articles using the defined search terms using Google Scholar.
	\item Read the title, abstract and the full text (if needed) to determine relevance by one of the co-authors and apply inclusion and exclusion criteria.
% 	\item Cross-check relevance using a randomly selected set of articles by another co-author.
	\item Cross-validate a randomly selected set of articles by another co-author.
	\item Apply inclusion and exclusion criteria.
% 	\item Snowball references articles that pass the inclusion criteria.
	\item Classify all included articles based on the classification we have for each question (details of those classifications are provided for each research question in Section \ref{sec:results}).
\end{enumerate}
 
\subsubsection{Searching for Grey Literature}
Here we followed the recommendations made in previous multivocal review studies \cite{garousi2018smell,Garousi2019Guidelines} and reported the results obtained only from the first 10 pages (10 items each) of the Google search. It was reported that relevant results usually only appear in the first few pages of the search \cite{Garousi2019Guidelines}. We observed that the results in the last five pages were less relevant compared to those that appeared in the first five. 

For the grey literature search, we define the following criteria:
\begin{enumerate}
\item [] \textbf{Search engine:} Google Search.
% \item []\textbf{Search terms:} ``flaky test", ``test flakiness'', ``non-deterministic test'', ``test bugs''.
\item []\textbf{Search String:} \textit{``flaky test'' OR ``test flakiness'' OR ``non-deterministic test''}.
\item [] \textbf{Search scope:} pages that appear in the first 10 pages of the search. 
Note that this search was conducted over two iterations. We searched for material published up until 31 December 2020, then in the second iteration we searched for material published up to 30 April 2022 (we removed duplication found between the two searches).%We did not restrict the search by a specific timeline.
% TODO: report iterations?
\end{enumerate}
 
\noindent The grey literature review consists of the following steps:
 
\begin{enumerate}
	\item Search and retrieve relevant results using the search terms in Google Search.
	\item Read the title and full article (if needed) to determine relevance.
	\item Cross-validate a randomly selected set of articles by another co-author.
	\item Check external links and other external resources that have been referred to in the articles/posts. Add new results to the dataset.
	\item Classify all posts based on the classification scheme.
\end{enumerate}
 
\subsubsection{Selection Criteria}
 
\noindent We selected articles based on the three following inclusion criteria:
 
\begin{itemize}
	\item Studies discussed test flakiness as the main topic of the article.
	\item Studies discussed test flakiness as an impact of using a specific technique, tool or in an experiment.
	\item Studies discussed test flakiness as a limitation of a technique, tool or an experiment.
\end{itemize}
 
\noindent We apply the following exclusion criteria:
 
\begin{enumerate}
    \item [--] Articles only mentioning flakiness, but without substantial discussion on the subject. 
    \item [--] Articles that are not accessible electronically, or the full text is not available for downloading\footnote{In case the article is not available either through a known digital library such as  ACM Digital Library, IEEE Xplore, ScienceDirect and SpringerLink; or not publicly available through other open-access repositories such as arXiv or ResearchGate}.
	\item [--] Studies on nondeterminism in hardware and embedded systems.
	\item [--] Studies on nondeterminism in algorithms testing (e.g., when nondeterminism is intentionally introduced).
  	\item [--] Duplicate studies (e.g., reports of the same study published in different places or on different dates, or studies that appeared in both academic and grey literature searches).
  	\item [--] Secondary studies on test flakiness (previous review articles).
    \item [--] Editorials, prefaces, books, news, tutorials and summaries of workshops and symposia. 
   \item [--] Multimedia material (videos, podcasts, etc.) and patents.
	\item [--] Studies written in a language other than English.
\end{enumerate}
 
\noindent For the grey literature study, we also exclude the following:
 
\begin{enumerate}
  	\item [--]  Academic articles, as those are covered by our academic literature search using Google Scholar.
  	\item [--]  Tools description pages (such as GitHub pages) with little or no discussion about the causes of flakiness or its impact.
  	\item [--]  Web pages that mention flaky tests with no substantial discussion (e.g., just provided a definition of flakiness).
\end{enumerate}

% \textcolor{red}{\subsubsection{Snowballing and Reference Checking}

\subsubsection{Pilot Run}
\label{sec:pilot}
Before we finalized our search keywords and search strings, and defined our inclusion and exclusion criteria, we conducted a pilot run using a simplified search string to validate the study selection criteria, refine/confirm the search strategy and refine the classification scheme before conducting the full-scale review. Our pilot run was conducted using a short, intentionally inclusive string (\textit{``flaky test'' OR ``test flakiness'' OR ``non-deterministic test''}) using both Google and Google Scholar. These keywords were drawn from two key influential articles and blog posts (either used in the title or as keywords) that the authors are aware of - i.e.,  the highly cited work on the topic by Luo et al. \cite{luo2014empirical}, and the well-known blog post by Martin Fowler \cite{fowler2011eradicating}). This was done for the period until 31 December 2020. 

In the first iteration, we retrieved 10 academic articles and 10 grey literature results, and then in the second iteration we obtained another 10 academic articles (next 10 results) and 10 grey literature results.
We validated the results of this pilot run based on articles' relevance as well as our familiarity with the field. We validated the retrieved articles and classified all retrieved results using the defined classification scheme in order to answer the four research questions. We used this pilot run to improve and update our research questions and our classification scheme. We classified a total of 20 articles in each group (academic and grey literature).
With respect to the former, we were able to identify 14 of the 20 articles found by the search as being familiar to us, lending a degree of confidence that our search would at minimum find papers relevant to our research questions.

\subsection{Data Extraction and Classification}
 \label{sec:data-extraction}

\begin{figure*}[h]
	\centering
 \resizebox{0.6\linewidth}{!}{
	\includegraphics[width=\linewidth]{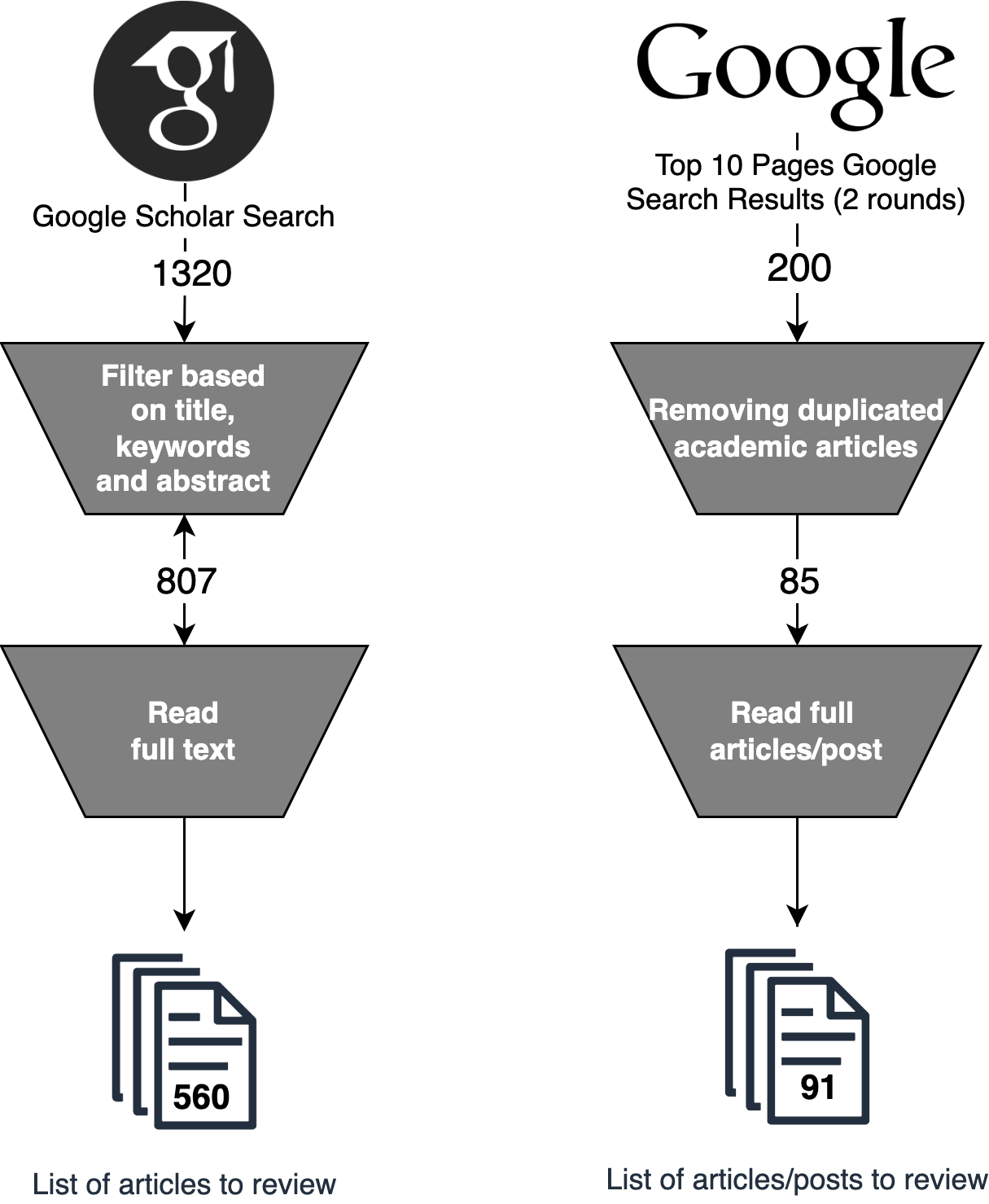}}
	\caption{Results of the review process}
	\label{fig:results_stats}
\end{figure*}

We extracted relevant data from all reviewed articles in order to answer the different research questions. Our search results (following the different steps explained above) are shown in Fig. \ref{fig:results_stats}. 

As explained in Section \ref{sec:review-process}, we conducted our search over two iterations, covering two periods. The first period covers articles published up until 31 December 2020, while the second cover the period from 1 January 2021 until 30 April 2022. 
In the first iteration we retrieved a total of 1092 results, with 992 articles obtained from the Google Scholar search and 100 grey literature articles obtained from Google Search (i.e., the first 10 pages). After filtering the relevant articles and applying the inclusion and exclusion criteria, %and reference snowballing,
we ended up with a total of 408 articles to analysis (354 academic articles and 54 grey literature articles). 
In the second iteration (from January 2021 until April 2022), we retrieved 330 academic articles from Google Scholar and 100 articles from Google Search (results from the first 10 pages). We removed the duplicates (e.g., results that might appear twice such as the same publication appeared in multiple publication venues, or grey literature article that appeared in the top 10 pages over the two iterations). For this iteration, after filtering the relevant articles, we ended up with 243 results (206 academic articles and 37 grey literature posts). Collectively, we identified a total of 560 academic and 91 grey literature articles for our analysis.

%Each of those articles were read in full and analysed to answer our research questions.
The analysis of articles was done by three coders (co-authors). We split the articles between coders, where each coder read the articles and obtained data to answer our research questions. For each article, we first tried to understand the context of flakiness. We then looked for the following: 1) the discussed causes of flakiness (RQ1), 2) how flakiness is detected (methods, tools, etc.) (RQ2), 3) the noted impact of flakiness (R3), and 4) the approach used to respond to or deal with flaky tests (RQ4). %Given their different nature and structure, we analysed the academic and grey literature articles separately.

\subsection{Reliability Assessment}
We conducted a reliability check of our filtration and classification. We cross-validated a randomly selected sample of 88/992 academic articles and 49/100 grey literature articles, as obtained from the first iterations (obtaining a confidence level = 95\% and confidence interval = 10). Two of the authors cross-validated those articles, with each classifying half of these articles (44 of academic literature and 25 of the grey literature articles). In addition, a third co-author (who was not involved in the initial classification) cross-validated another 25 randomly selected academic/grey articles. 
On those cross-validation, we reached an agreement level of $\geq$ 90\%.\\

We provide the full list of articles that we reviewed (both academic and grey) online\url{https://docs.google.com/spreadsheets/d/1eQhEAUSMXzeiMatw-Sc8dqvftzLp8crC3-B9v5qHEuE}.

\section{Results}
\label{sec:results}

\subsection{Overview of the publications}

\begin{figure}[h]
\centering
	\includegraphics[width=\linewidth]
        {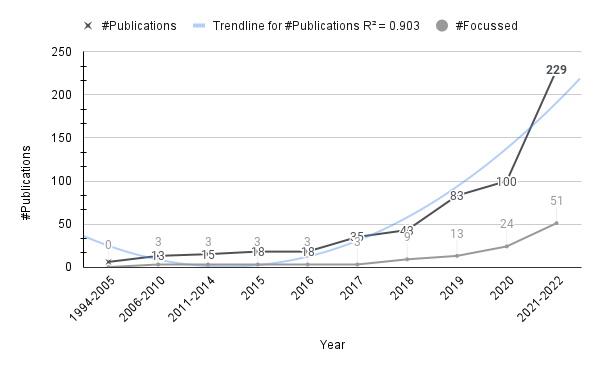}
    \caption*{\scriptsize{* The 2021-2022 numbers include article published between Jan 2021 and April 2022.}}
	\caption{Timeline of articles published on flaky tests, including the focused articles}
 	\label{fig:timeline}
\end{figure}

We first provide an overview of the timeline of publications on flaky tests in order to understand how the topic has been viewed and developed over the years. The timeline of publications is shown in Fig. \ref{fig:timeline}. Based on our search for academic articles, we found that there have been articles that discuss the issue of \emph{nondeterminism} in test outcomes dating back to 1994, with 34 articles found between 1994 and 2014. However, the number of articles has significantly increased since 2014. 
There has been an exponential growth in publications addressing flaky tests in the past 6 years (between 2016 and 2022). 
We attributed this increase to the rising attention to flaky tests by the research community since the publication of the first empirical study on the causes of flaky tests in Apache projects by Luo et al. in 2014 \citeS{S1}, which was the first study that directly addressed the issue of flaky tests in great detail (in terms of common causes and fixes). Over 93\% of the articles were published after the publication of this study, with 41\% of those articles (229) published between January 2021 and April 2022 only, indicating an increased popularity over the years.\\ 

\begin{figure}[h]
\centering
	\includegraphics[width=\linewidth]
        {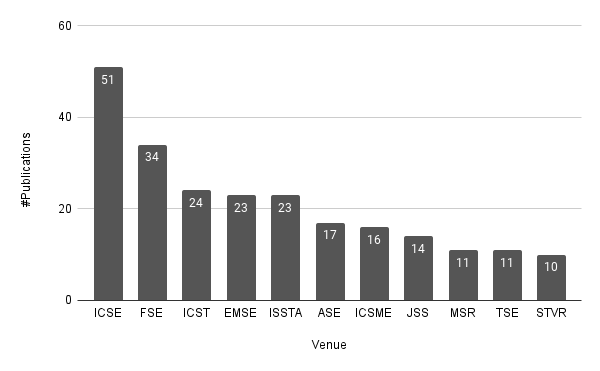}
	\caption{Distribution of publications based on publication venues}
 	\label{fig:venues}
\end{figure}

In terms of publication types and venues, more than 40\% of these publications have been published in reputable and highly rated software engineering venues. Top publications venues include the premier software engineering conferences: International Conference on Software Engineering (ICSE), Joint European Software Engineering Conference and Symposium on the Foundations of Software Engineering (ESEC/FSE) and the International Conference on Automated Software Engineering (ASE). Publications on flaky tests has also appeared in the main software testing conferences International Symposium on Software Testing and Analysis (ISSTA) and International Conference on Software Testing, Verification and Validation (ICST).  They also appear in software maintenance conferences, International Conference on Software Maintenance and Evolution (ICSME) and International Conference on Software Analysis, Evolution and Reengineering (SANER). A few articles ($\sim$10\%) were published in premier software engineering journals, including Empirical Software Engineering (EMSE), Journal of Systems and Software (JSS), IEEE Transaction in Software Engineering (TSE) and Software Testing, Verification and Reliability (STVR) journal. The distribution of publications in key software engineering venues is shown in Fig. \ref{fig:venues}).

To expand on the methodology for including and excluding articles, we did not perform a quality assessment of articles based on venues or citation statistics. We focused on primary studies (excluding the three reviews and secondary studies). Furthermore, looking deeper into focused studies provided insights into their quality.

In terms of programming languages, the vast majority of the studies have focused on Java (49\% of those studies), with only a few other studies that discuss flakiness in other languages such as Python, JavaScript and .NET, with 49 (14\%) studies used multiple languages (results listed in Table \ref{tab:languages}). 

\begin{table}[h]
\centering
\caption{Flaky tests in terms of languages studied}
% \resizebox{0.8\linewidth}{!}{
\begin{tabular}{lc}
\toprule
\textbf{Language}   & \textbf{\#  Articles} \\ \midrule
Java       & 212                                     \\
Python     & 25                                      \\

JavaScript & 11                                      \\
.NET languages      & 6                                      \\

Other languages  & 27                                      \\
Multiple languages   & 58                                      \\
Not specified/unknown  & 221                                      \\
\bottomrule
\end{tabular}%}
\label{tab:languages}
\end{table}

We classified all articles into three main categories: 1) studies focusing on test flakiness, where flakiness is the focal point  (e.g., an empirical investigation into test flakiness due to a particular cause such as concurrency or order dependency), 2) studies that explain how test flakiness impacts tools/techniques  (e.g., the impact of test flakiness on program repair or fault localization) or 3) studies that just mention or reference test flakiness (but it is not a focus of the study). A breakdown of the focus of the articles and the nature of the discussion on test flakiness in academic literature is shown in Table \ref{tab:focus}.

We observed that only 109 articles ($\sim$20\%) from all the collected studies focused on test flakiness as the main subject of the study. The majority of the reviewed articles (297, representing $\sim$53\%) just mentioned test flakiness as a related issue or as a threat to validity. The remaining 154 articles($\sim$27\%) discussed flakiness in terms of their impact on a proposed technique or tool that is the subject of the study.
 
\begin{table}[!h]
	\caption{Focus of academic articles}
\resizebox{\linewidth}{!}{
\begin{tabular}{llc}
\toprule
\textbf{Type} & \textbf{Description}                                                          & \multicolumn{1}{l}{\textbf{\# of articles}} \\ \toprule
Focused & \begin{tabular}[c]{@{}l@{}}Studies focusing on test flakiness\end{tabular}                              & 109                                \\ %\midrule
Impact & \begin{tabular}[c]{@{}l@{}}Studies that explain how test flakiness impacts tools/techniques \end{tabular} & 154                               \\ %\midrule
Referenced & \begin{tabular}[c]{@{}l@{}}Studies that just mention or reference test flakiness\end{tabular}             & 297                               \\ \bottomrule
\end{tabular}}
  	\label{tab:focus}
\end{table}

%\textcolor{green}{
As for the grey literature results, all articles we found were published following the publication of Martin Fowler's influential blog post on test nondeterminism in early 2011 \cite{fowler2011eradicating}.

Similar to the recent increased attention on those academic literature articles, we found that almost 50\% of the grey literature articles (26) have been published between 2019 and 2020, indicating a growing interest in flaky tests, and shedding light on the importance of the topic in practice.

In the following subsections, we present the results of the study by answering each of our four research questions in detail.

\subsection{Causes of Test Flakiness (RQ1)}
\label{sec:causes}

We analysed causes of flaky tests, as noted in both academic and grey literature articles. We looked for the quoted reasons for why a test is flaky, and in most cases, we note multiple (sometimes connected) causes  being the reason for flakiness. We then grouped those causes into categories based on their overall nature. \\
The most widely discussed causes in literature are those that are already in the empirical study  on flaky tests by Luo et al. \citeS{S1}. The study provides a classification of causes as the result of analysing 201 commits that fix flaky tests across Apache projects, which are diverse across languages and maturity. Their methodology is centred around examining commits that \textit{fix} flaky tests. In addition to classifying root causes of test flakiness into categories, they present approaches to manifest flakiness and strategies used to fix flaky tests. The classification consists of 10 categories that are the \textit{root causes} of flaky tests in the commits, which are  \emph{async-wait}, \emph{concurrency}, and \emph{test order dependency}, \emph{resource leak}, \emph{network}, \emph{time}, \emph{IO}, \emph{randomness}, \emph{floating point operations} and  \emph{unordered collections}. Thorve et al. \citeS{S7} listed additional causes identified from a study of Android commits fixing flaky tests: \emph{dependency}, \emph{program logic}, and \emph{UI}. Dutta et al. \citeS{S14} noted subcategories of \emph{randomness}, and Moritz et al. \cite{S8} identified three additional causes: \emph{timeout}, \emph{platform dependency} and \emph{too restrictive range} from a survey of developers.

We mapped all causes found in all surveyed publications, and categorized them into the following major categories (based on their nature): \emph{concurrency}, \emph{test order dependency}, \emph{network}, \emph{randomness}, \emph{platform dependency}, \emph{external state/behaviour dependency}, \emph{hardware}, \emph{time} and \emph{other}. A summary of the causes we classified is provided in Table \ref{tab:causes} and discussed below.

    \noindent\textbf{Concurrency.} This categorizes flakiness due to concurrency issues resulting from concurrency-related bugs. These bugs can be race conditions, data races, atomicity violations or deadlocks. 
   \textbf{\emph{Async-wait}} is investigated as one of the major causes of flakiness under concurrency. This occurs when an application or test makes an asynchronous call and does not correctly wait for the result before proceeding.
   This category accounts for nearly half of the studied flaky test fixing commits \citeS{S1}. Thorve et al.\citeS{S7} and Luo et al \citeS{S1} classified async-wait related flakiness under concurrency. Lam et al. \citeS{S10,S6} reported async-wait as the main cause of flakiness in Microsoft projects. %Lam empirical studies that async/wait include \citeS{S6}. 
   Other articles cited async-wait in relation to root-cause identification \citeS{S6}, detection \citeS{S15} and analysis \citeS{S72}.
    Luo et al. \citeS{S1} identified an additional subcategory \textit{``bug in condition''} for concurrency-related flakiness, where the guard for code that determines which thread can call it is either too restrictive or permissive. Concurrency is also identified as a cause in articles on detection \citeS{S6}, \citeS{S15}, \citeS{S14} and \citeS{S7}. Another subcategory identified from browser applications is event races~\citeS{S1008}.  
    
    \noindent\textbf{Test order dependency.} The test independence assumption implies that tests can be executed in any order and produce expected outcomes. This is not the case in practice \cite{S40} as tests may exhibit different behaviour when executed in different orders. This is due to a shared state where it can either be in memory or external (e.g. file system, database). Tests can expect a particular state before they can exhibit the expected outcome, which can be different if the state is not setup correctly or reset.  There can be multiple sources of test order dependency. Instances of shared state can be in the form of explicit or implicit data dependencies in tests, or even bugs such as resource leaks or failure to clean up resources between tests. Luo et al. \citeS{S1} listed these as separate root causes: resource leaks and I/O.   \emph{\noindent\textbf{Resource leak}} can be a source of test order dependency when the code under test (CUT) or test code is not properly managing shared resources (e.g. obtaining a resource and not releasing it). Empirical studies that discuss this resource leak related flakiness include \citeS{S1} and \citeS{S10}, as well other studies on root cause analysis, such as \citeS{S6} and \citeS{S210}, that find instances of flakiness in test code due to improper management of resources. Luo et al. \citeS{S1} identified I/O as a potential cause of flakiness. An example is a code that opens and reads from a file and does not close it, leaving it to garbage collection to manage it. If a test attempts to open the same file, it would only succeed if the garbage collector had processed the previous instance. In Luo et al. \citeS{S1}, 12\% of test flakiness in their study is due to order dependency. Articles that cite order dependency include those that propose detection methods \citeS{S147,S4}, and one that is an experimental study on flakiness in generated test suites \citeS{S110}. A shared state can also arise due to \emph{incorrect/flaky API usage} in tests. Tests may intermittently fail if programmers use such APIs in tests without accounting for such behaviour. Dutta et al. \citeS{S14} discussed this in the study of machine learning applications and cite an example where the underlying cause is the shared state between two tests that use the same API and one of the tests not resetting the fixture before executing the second.
    
    \noindent\textbf{Network.} Another common cause for flaky tests relates to network issues (connections, availability, and bandwidth). This has two subcategories:  local and remote issues. Local issues pertain to managing resources such as sockets (e.g. contention with other programs for ports that are hard-coded in tests) and remote issues concern failures in connecting to remote resources. Mor{\'a}n et al. \citeS{S99} studied network bandwidth in localizing flakiness causes. In a study consisting of Android projects \citeS{S7}, network is identified as a cause of flakiness of 8\% of the studied flaky tests.

    \noindent\textbf{Randomness.}  Tests or the code under test may depend on randomness, which can result in flakiness if the test does not consider all possible random values  that can be generated. This is listed as a main cause by Luo et al. \citeS{S1}. Dutta et al. \citeS{S14} identified subcategories of randomness in their investigation of flaky tests in probabilistic and machine learning applications. Such applications rely on machine learning frameworks that provide operations for inference and training, which are largely nondeterministic in nature. Writing tests can be challenging for such applications, which use these frameworks. The applications are written in Python, and they study applications that use the main ML frameworks for the language. They analysed 75 bug/commits that are linked to flaky tests and obtained three cause subcategories: 1) \textit{algorithmic nondeterminism}, 2) \textit{incorrect/flaky API usage} and 3) \textit{hardware}. They state that these categories are subcategories of the \textit{randomness} category in \citeG{S1}. The most common cause identified was algorithmic nondeterminism. They also present a technique to detect flaky tests due to such assertions. They evaluate the technique on 20 projects and found 11 previously unknown flaky tests. The subcategories identified are \emph{Algorithmic non-determinism} and \emph{Unsynchronized seeds} in ML applications. In these applications, as test input, developers use small datasets and models, expecting the results to converge to values within an expected range. Assertions are added to check if the inferred values are close to the expected ones. As there is a chance that the computed value may fall outside the expected range, this may result in flaky outcomes. Tests in ML applications may also use multiple libraries that need sources of randomness, and flakiness can arise if different random number seeds are used across these modules or if the seeds are not set.  We include a related category here, \emph{too restrictive ranges}, identified in \citeS{S8}. This is due to output values falling outside ranges or values in assertions determined at design time. 
    
\begin{landscape}
\begin{table}
 \caption{Causes of flaky tests}
 \label{tab:causes}
%  \centering
\resizebox{0.70\linewidth}{!}{
\begin{tabular}{@{}llp{10cm}l@{}}
\toprule
\textbf{Main category} & \textbf{Sub-category} & \textbf{Description} & \textbf{Example Articles} \\ \midrule
Concurrency & Synchronization  & Asynchronous call in test (or CUT) without proper synchronization before proceeding & \citeS{S1}, \citeS{S7},  \citeS{S10}, \citeS{S6} \\
& & & \citeS{S15}, \citeS{S72} \\
 & Event races &  Event racing due to a single UI thread and async events triggering UI changes & \citeS{S16,S93} \\
 & Bugs &  Other concurrency bugs (deadlocks, atomicity violations, different threads interacting in a non-desirable manner.) &  \citeS{S1} \\
 & Bug in condition  & A condition that inaccurately guards what thread can execute the guarded code. &  \citeS{S1} \\
 \midrule\\
 Test order dependency & Shared state & Tests having the same data dependencies can affect test outcome. &  \citeS{S147}, \citeS{S4}, \citeS{S110}\\
  & I/O & Local files  &   \citeS{S1,S357} \\
   & Resource leaks   & When an application does not properly manage the resources it acquires  & \citeS{S1}, \citeS{S10}, \citeS{S6},  \citeS{S210} \\
    \midrule\\
   Network & Remote &  Connection failure to remote host (latency, unavailability) & \citeS{S1,S357}\\
& Local & Bandwidth, local resource management issues (e.g. port collisions)& \\
 \midrule\\
Randomness & Data   & Input data or output from the CUT & \citeS{S6,S545} \\
 & Randomness seed   & If the seed is not fixed in either the CUT or test it may cause flakiness. & \citeS{S14} \\
 & Stochastic algorithms  & Probabilistic algorithms where the result is not always the same. & \citeS{S217}\\
 & Too restrictive range  & Valid output from the CUT are outside the assertion range. & \citeS{S8} \\
  \midrule\\
 Platform dependency & Hardware &   Environment that the test executes in (Development/Test/CI or Production.)  & \citeS{S1,S14,S548,S210} \\
 & OS &   Varying operating system & \citeS{S42,S99} \\
  & Compiler &   Difference in compiled code & \citeS{S1019} \\
 & Runtime &  e.g., Languages with virtual runtimes (Java, C\# .. etc) & \citeS{S1} \\
  & CI infra flakiness &  Build failures due to infrastructure flakiness. &  \citeS{S1007}\\
 & Browser &  A browser may render objects differently affecting tests. & \citeS{S42} \\
  \midrule\\
External state/behaviour & Reliance on production service   & Tests rely on production data that can change. & \\
dependency & & &  \\
 & Reliance on external resources   & Databases, web, shared memory... etc & \citeS{S23,S39} \\
 & API changes  & Evolving REST APIs due to changing requirements &  \\
 & External resources  & Relying on data from external resources (e.g., REST APIs, databases) & \citeS{S23,S713} \\
  \midrule\\
  Environmental dependencies & & Memory and performance & \citeS{S42} \\
  \midrule\\
 Hardware & Screen resolution  & UI elements may render differently on different screen resolutions causing problems for UI tests & \\
 & Hardware faults & & \citeS{S210} \\
  \midrule\\
Time & Timeouts & Test case/test suite timeouts. &  \citeS{S8} \\
 & System date/time  & Relying on system time can result in non-deterministic failure (e.g. time precision and changing UTC time) & \citeS{S6}, \citeS{S39} \\
  \midrule\\
Other & Floating point operations  & Use of floating point operations can result in non-deterministic cases &  \citeS{S14}, \citeS{S1}\\
 & UI & Incorrectly coding UI interactions & \citeS{S7} \\
 & Program logic  &  Incorrect assumptions about APIs & \citeS{S7} \\
  & Tests with manual steps  &  & \citeS{S1025} \\
 & Code transformations & Random amplification/instrumentation can cause flaky tests & \citeS{S258} \\
\bottomrule
\end{tabular}}
 \end{table}
 \end{landscape}

    \noindent\textbf{Platform dependency.} This causes flakiness when a test is designed to pass on a specific platform  but when executed on another platform it unexpectedly fails. A platform could be the hardware and OS and also any component on the software stack that test execution/compilation depends on. Tests that rely on platform dependency may fail on alternative platforms due to missing preconditions or even performance issues across them. The cause was initially described in Luo et al. \citeS{S1}, though it was not in the list of 10 main causes as it was a small category. It is discussed in more detail in \citeS{S8}. In Thorve et al \citeS{S7}, it was reported that dependency flakiness for Android projects are due to hardware, OS version or third-party libraries. The study consisted of 29 Android projects containing 77 flakiness related commits. We also include  \emph{implementation dependency}, differences in compilation \citeS{S1019} and infrastructure flakiness \citeS{S1007} under this category. Infrastructure flakiness could also be due to issues in setting up the required infrastructure for test execution, which could include setting up Virtual Machines (VM)/containers and downloading dependencies, which can results in flakiness. Environmental dependency flakiness due to dynamic aspects (performance or resources) are also included in this category.

    \noindent\textbf{Dependencies on external state/behaviour.} We include flakiness due to changes in external dependencies like state (e.g. reliance on external data from databases or obtained via REST API's) or behaviour (changes or assumptions about the behaviour of third-party libraries) in this category. Thorve et al ~\citeS{S7} included this under platform dependency.

    \noindent\textbf{Hardware.} Some ML applications/libraries may use specialized hardware, as discussed in \citeS{S14}. If the hardware produces nondeterministic results, this can cause flakiness. An example is where an accelerator is used that performs floating-point computations in parallel. The ordering of the computations can produce nondeterministic values, leading to flakiness when tests are involved. Note that this is distinct from platform dependency, which can also be at the hardware level, for instance, different processors or Android hardware.

    \noindent\textbf{Time.} Variations in time are also a cause of test flakiness (e.g. midnight changes in the UTC time zone, daylight saving time, etc.), and due to differences in precision across different platforms. Time is listed as a cause in root cause identification by Lam et al. \citeS{S6}. New subcategories, \textit{timeouts}, are listed by developers in the survey done in \citeS{S8}. Time precision across OS, platforms and different time zones are listed under this category \citeS{S39}. Another cause related to time, is that test cases may time out nondeterministically e.g. failing to obtain prerequisites or execution not completing within the specified time due to flaky performance.   
    A similar cause is test suite timeouts where no specific test case is responsible for it. Both of these causes were identified in the developers survey reported in \citeS{S8}.

    \noindent\textbf{Other causes.} We include causes listed in articles, which may already have relationships with the major causal categories listed above. Thorve et al. \citeS{S7} listed \emph{program logic} as one of them. This category consists of cases where programmers have made incorrect assumptions about the code's behaviour, which results in cases where tests may exhibit flakiness. The authors cited an example where the Code Under Test (CUT) may nondeterministically raise an I/O exception and the exception handling throws a runtime exception, causing the test to crash in that scenario. \emph{UI flakiness} can be caused due to developers either not understanding UI behaviour or incorrectly coding UI interactions \citeS{S7}. They can also be caused by concurrency (e.g., event races or async-wait) or platform dependency (e.g., dependence on the availability of a display, dependence on a particular browser \citeS{S99}). \emph{Floating-point operations.}  floating-point operations can be a cause of flakiness as they can be non-deterministic due to non-associative addition, overflows and underflows as described in \citeS{S1}. It is also discussed in the context of machine learning applications \citeS{S14}. Concurrency, hardware and platform dependency can be a source of nondeterminism in floating-point operations. 
    Luo et al. \citeS{S1} identified \emph{unordered collections}, where there are variations in outcomes due to a test's incorrect assumptions about an API. An example of this is the sets which can have specifications that are underdetermined. Code may assume behaviour such as the order of the collection from a certain execution/implementation, which is not deterministic.

\subsubsection{Ontology of causes of flaky tests}
\label{sec:ontology}
\begin{figure*}[h]
\centering
\includegraphics[width=0.9\linewidth]{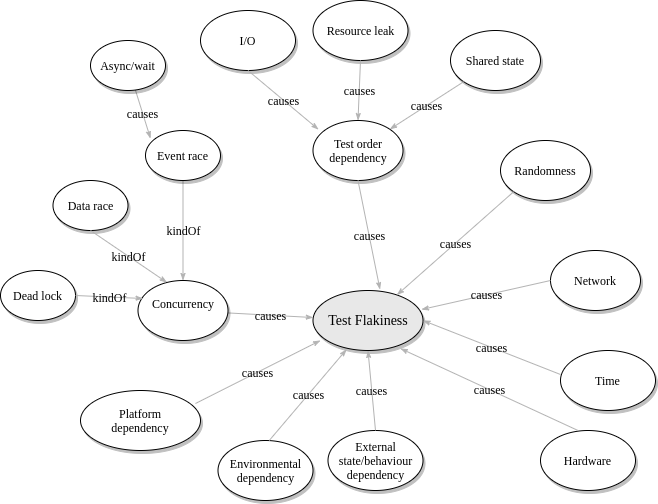}
\caption{Relationships between the different causes of flaky tests}
 \label{fig:ontologycause}
\end{figure*}

Fig.~\ref{fig:ontologycause} illustrates the different causes of flakiness. The figure uses Web Ontology Language (OWL) ~\cite{mcguinness2004owl} terminology  such as classes, subclasses and relations. We identify classes for causes of flakiness and flaky tests. Subclass relationships between classes are named `kindOf'  and `causes' is the relation for denoting causal relationships.  

Note that not all identified causes are shown in the diagram. For instance, causes listed under the other category may be due to sources already shown in the diagram. For instance, UI flakiness can be due to platform dependency or environmental dependency. An example that demonstrates the complex causal nature of flakiness is in \citeS{S14}, where the cause of flakiness is due to a hardware accelerator for deep learning, which performed fast parallel floating point computations. As different orderings of floating point operations can result in different outputs, this leads to test flakiness. In this case,  the causes are a combination of \textit{hardware}, \textit{concurrency}, and \textit{floating point operations}.
Network uncertainty can be attributed to multiple reasons, for instance, connection failure and bandwidth variance. Stochastic algorithms exhibit randomness, and concurrency related flakiness can be due to concurrency bugs such as races and deadlocks. Finally, order dependency is due to improper management of resources (e.g. leaks and not cleaning up after I/O operations) or hidden state sharing that may manifest in flakiness.

There are a number of factors that vary, which underlie those causes. For instance, \textit{random seed variability} can cause flakiness related to randomness and scheduling variability causes concurrency-related flakiness. \textit{Test execution order variability}, which causes order dependent test flakiness and types of \textit{platform variability} (e.g. hardware and browser that can, for instance, manifest in UI flakiness) are additional dimensions of variability.

%\todo[inline]{we know have the variability factors that we will also need to explain}

\begin{qoutebox}{white}{}
\textbf{RQ1 summary.}
Numerous causes of flakiness have been identified in literature, with factors related to concurrency, test order dependency, network availability and randomness are the most common causes for flaky test behaviour. Other factors related to specific types of systems such as \textit{algorithmic nondeterminism} and \textit{unsynchronised seeds} impacting testing in ML applications. There is also a casual relationship between some of these factors (i.e., they impact each other - for example, UI flakiness is mostly due to concurrency issues).
\end{qoutebox}

\subsection{Flaky Tests Detection (RQ2)}
\label{sec:results:detection}

One of the dimensions we studied is how flaky tests are identified and/or detected.  In this section, we present methods used to detect and identify or locate causes of flakiness. We make a distinction between these three goals in our listing of techniques found in the reviewed literature. %These methods detect flakiness either statically or dynamically. 
We look at methods identified in both academic and grey literature. RQ2 is divided into two sub-questions, we answer each subquestion separately, as shown below: 

\subsubsection{Methods Used to Detect Flaky Tests (RQ2.1)}

\begin{figure*}[htp]
\centering
\includegraphics[width=\linewidth]{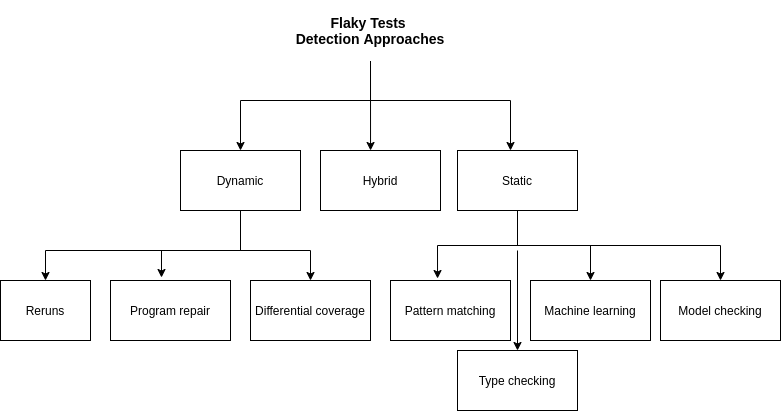}
\caption{Taxonomy of detection methods}
 \label{fig:taxonomydetection}
\end{figure*}

There are two distinctive approaches towards the detection of flaky tests, which are either by using dynamic techniques that involve the execution of tests or using static techniques that rely only on analysing the test code without actually executing tests. %There are two techniques that use a combination of hybrid and static approaches. 
Figure~\ref{fig:taxonomydetection} depicts a broad overview of these strategies. Dynamic methods are based mostly on multiple test runs whilst also using techniques to perturb specific variability factors (e.g. environment, test execution order, event schedules or random seeds) that quickly manifests flakiness. There is one study on using program repair \citeS{S28} to induce test flakiness and two studies on using differential coverage to detect flakiness without resorting to reruns \citeS{S2}. Under static approaches, studies have employed  machine learning (3 studies), model checking for implementation dependent tests and similarity patterns techniques for identifying flaky tests. There are only two studies that leverage hybrid approaches (one for order dependent tests and another for async-wait). \\

\noindent\textbf{Static methods}: Static approaches that do not execute tests are mostly classification-based that use machine learning techniques \citeS{S607,S118,S31}. Other static methods use pattern matching  \citeS{S357} and association rule learning \citeS{SB1}. Model checking using Java PathFinder \cite{visser2003model} has also been used for detecting flakiness due to implementation dependent tests \citeS{SB28}.\\
% shawn: todo: expand above
Ahmad et al. \citeS{S607} evaluated a number of machine learning methods for predicting flaky tests. They used projects from the iDFlakies dataset \citeS{S4}. There is also a suggestion that the evaluation also covered another language (Python) besides the data from the original dataset (which is in Java), though this is not made clear, and the set of Python programs or tests is not listed. The study built on the work of Pinto et al. \citeS{S11}, which is an evaluation of five machine learning classifiers (Naive Bayes, Random Forest, Decision Tree, Support Vector Machine and Nearest Neighbour) that predict flaky tests. In comparison to \citeS{S11}, the study of Ahmad et al. \cite{S607} answers two additional research questions: how classifiers perform with regard to another programming language, and the predictive power of the classifiers. Another static technique based on patterns in code have been used to predict flakiness \citeS{S357}.\\

% shawn: TODO it is multiple times (e.g. rerun, could be single if its differential coverage based)
\noindent\textbf{Dynamic methods:} Dynamic techniques to detect flakiness are built on  executions of tests (single or multiple runs). Those techniques are centred around making reruns less expensive by accelerating ways to manifest flakiness, i.e., fewer number of reruns or rerunning fewer tests. Methods to manifest flakiness include varying causal factors such as random number seeds \citeS{S14}, event order \citeS{S93}, environment (e.g. browser, display) \citeS{S99}, and test ordering \citeS{S147, S4}. Test code has also been varied using program repair \citeS{S28} to induce flakiness. Fewer tests are run by selecting them based on differential code coverage  or those with state dependencies.  \\

\noindent\textbf{Hybrid methods:} Dynamic and static techniques are known to make different trade-offs between desirable attributes such as recall, precision and scalability \cite{ernst2003static}. As in other applications of program analysis, hybrid techniques have been proposed to  combine the strength of different techniques, whilst avoiding their weaknesses. 
One of the tools, FLAST \citeS{S118}, proposes a hybrid approach where the tool uses a static technique but suggests that dynamic analysis can be used to detect cases missed by the tool. Malm et al. \citeS{S72} proposed a hybrid analysis approach to detect delays used in tests that cause flakiness. Zhang et al. \citeS{S591} proposed a tool for dependent test detection, and they use static analysis to determine side-effect free methods, whose field accesses are ignored when determining inter-test dependence in the dynamic analysis. Some tools, stated earlier under static methods (e.g., \citeS{S607}), may need access to historic execution data for analysis or training.

\subsubsection{Tools to Detect Flaky Tests (RQ2.2)}
Table~\ref{tab:detection} lists the tools that detect test flakiness, which are described in the literature. Most of the tools detect flakiness manifested in test outcomes.
% correct type of flakiness column. e.g. dynamic dataflow analysis checks for shared state. @Shawn
The majority of the tools found in academic articles work on Java programs, with only three for Python and a single tool for JavaScript.
These tools can be grouped by the source of flakiness they target: UI, test order, concurrency and platform dependency (implementation dependency to a particular runtime). Some of these tools identify the cause of flakiness as well (which may already be a part of the tool's output if the  source of flakiness they detect is closely associated with a cause: e.g., test execution order dependency arising from a shared state can be detected by executing tests under different orders).\\

\begin{landscape}
\begin{table*}[!b]
\caption{Flaky tests detection tools as reported in academic studies}
\centering
\label{tab:detection}
\resizebox{0.8\linewidth}{!}{
\begin{tabular}{lllllll}
\toprule
\textbf{Detection type} & \textbf{Category} & \textbf{Language} & \textbf{Method type} & \textbf{Method} & \textbf{Tool name}  & \textbf{Article} \\
\midrule
Outcomes & Order & Java & Dynamic & Rerun (Vary orders) & - & \citeS{S1014} \\
Outcomes & Android & Java & Dynamic & Rerun (Vary event schedules) & FlakeScanner & \citeS{S1008} \\
Outcomes & General & Java & Dynamic & Rerun (twice) & - & \citeS{S1011} \\
Cause & Web & Java & Dynamic & Rerun (Vary environment) & FlakyLoc & \citeS{S99} \\
Location & General & - & Dynamic & Log analysis & RootFinder & \citeS{S6} \\
Outcomes & UI & Java & Dynamic & Rerun (Vary event schedules) & FlakeShovel & \citeS{S16} \\
Outcomes & General & Java & Hybrid & Machine learning & FlakeFlagger & \citeS{S1027} \\
Outcome & General & Mixed & Dynamic & Rerun (Environment) & - & \citeS{S27} \\
Outcomes & General & Java & Static & Machine learning & - & \citeS{S607} \\
Outcomes & ML & Python & Dynamic & Rerun (Vary random number seeds) & FLASH & \citeS{S14} \\
Outcomes & Concurrency & JavaScript & Dynamic & Rerun (Vary event order) & NodeRacer & \citeS{S93} \\
Outcomes & General & Java & Static & Machine learning (test code similarity) & FLAST & \citeS{S118} \\
Outcomes & General & Python & Dynamic & Rerun (Vary test code) & FITTER & \citeS{S28} \\
Outcomes & Concurrency & Java & Dynamic & Rerun (Add noise to environment) & Shaker & \citeS{S24} \\
Outcomes & General & Python & Dynamic & Test execution history & GreedyFlake & \citeS{S78} \\
Outcomes & General & Java & Dynamic & Rerun & iDFlakies & \citeS{S4} \\
Location & Assumptions & Java & Dynamic & Rerun (vary API implementation ) & NonDex & \citeS{S152} \\
Cause & Order & Java & Dynamic & Rerun and delta debugging & iFixFlakies & \citeS{S135} \\
Outcomes & General & Java & Dynamic &  Differential coverage   & DeFlaker & \citeS{S2} \\
 & & & & and test execution history & &  \\
Cause, location & General & C++ / Java & Dynamic & Rerun & Flakiness  & \citeS{S25} \\
 & & & & & Debugger & \\
 & UI & JavaScript & Dynamic & Machine learning (Bayesian network) & - & \citeS{S31}\\
 Cause & Order & Java & Dynamic & - & PolDet &  \citeS{S460}\\
 Outcome & General & - & Static & Machine learning & Flakify &  \citeS{S1022}\\
 Cause,Outcome & IO/Concurrency/Network & - & Dynamic & Rerun in varied containers & - &  \citeS{S19}\\
 Cause,Outcome & - & - & Dynamic & Rerun in varied containers & TEDD &  \citeS{S274}\\
 Cause,Outcome & - & C & Static & Dependency analysis & - &  \citeS{S780}\\
 Outcomes & Order & Java & Dynamic &  Rerun (Dynamic dataflow analysis) & PRADET & \citeS{S147} \\
  Outcomes & Order & Java & Dynamic &  Rerun (Vary order) & DTDetector & \citeS{S591} \\
 Outcomes & Order & Java & Dynamic &  Rerun (Dynamic dataflow analysis) & ElectricTest & \citeS{S391} \\
 Outcome & Order and Async/Wait & Java & Static & Pattern matching & - &  \citeS{S357}\\ % implementation using PMD
 Outcome & Order & Python & Dynamic & Rerun (varying orders) & iPFlakies &  \citeS{S1031}\\ 
 Outcome & - & Multilanguage & Dynamic & Machine learning & Fair &  \citeS{S1050}\\ 
 Outcome & Order & Java & Static & Model checking & PolDet (JPF) &  \citeS{S1051}\\ 
 Outcome & Nondeterminism & Java & Static & Type checking & Determinism Checker &  \citeS{S1054}\\ 
\bottomrule

\end{tabular}}
\end{table*}
\end{landscape}

% Tools that identify Causes
% important to discuss how combinations are determined.. pair-wise. t-wise
FlakyLoc \citeS{S99} does not detect flaky tests, but identifies causes for a given flaky test. The tool executes the known flaky test in different environment configurations. These configurations are composed of  environment factors (i.e., memory sizes, CPU cores, browsers and screen resolutions) that are varied in each execution. The results are analysed using a spectrum-based localization technique \cite{wong2016survey}, which analyses the factors that cause flakiness and assigns a ranking and a suspiciousness value to determine the most likely factors. The tool was evaluated on a single flaky test from a Java web application (with several end-to-end flaky tests). The results for this particular test indicate that the technique is successfully able to rank the cause of flakiness (e.g., low screen resolution) for the test.

RootFinder \citeS{S6} identifies causes as well as the location in code that cause flakiness. It can identify nine causes (network, time, I/O, randomness, floating-point operations, test order dependency, unordered collections, concurrency). The tool adds instrumentation at  API calls during test execution, which can log interesting values (time, context, return value) as well as add additional behaviour (e.g., add a delay to identify causes involving concurrency and async wait). Post-execution, the logs are analysed by evaluating predicates (e.g., if the return value was the same at this point compared to previous times) at each point where it was logged. Predicates that evaluate to consistent values in passing and failing runs are likely to be useful in identifying the causes, as they can explain what was different during passing and failing runs.

DeFlaker \citeS{S2} detects flaky tests using differential coverage to avoid reruns (as rerun can be expensive). If a test has a different outcome compared to a previous run and the code covered by the test has not changed, then it can be determined to be flaky. The study also examines if a particular rerun strategy has an impact on flakiness detection. With Java projects, there can be many such strategies (e.g., five reruns in the same JVM, forking with each run in a new JVM, rebooting the machine and cleaning files generated by builds between runs).
%\todo[inline]{perhaps expand this text on deflaker a bit, as this is one of the most known papers on using reruns} DONE

% Tools that look for specific kinds of:
% for concurrency flakiness
NodeRacer \citeS{S93}, Shaker \citeS{S24} and FlakeShovel \citeS{S16} specifically detect concurrency related flakiness. NodeRacer analyses JavaScript programs and accelerates manifestation of event races that can cause test flakiness. It uses instrumentation and builds a model consisting of a happens-after relation for callbacks. During the guided execution phase, this relation is used to explore  postponing of events  such that callback interleaving is realistic with regard to  actual executions. In Shaker, it is suggested that the tool exposes flakiness faster than rerun by adding noise to the environment in the form of tasks that also stress the CPU and memory whilst the test suite is executed. FlakeShovel targets the same type of cause as NodeRacer by similarly exploring different yet feasible event execution orders, but only for GUI tests in Android applications.\\

% test order
A number of detection tools are built to detect order dependent tests. In the case of iDFlakies \citeS{S4}, which uses rerun by randomizing the order of their execution, it classifies flaky tests into two types: order-dependent and non-order dependent.  In this category there are four more studies: DTDetector \citeS{S591}, ElectricTest \citeS{S391}, PolDet  \citeS{S460}, and PRADET  \citeS{S147}. DTDetector presents four algorithms to check for dependent tests, which are manifested in test outcomes: reversal of test execution order, random test execution order, the exhaustive bounded algorithm (which executes bounded subsequences of the test suite instead of trying out all permutations), and the dependence-aware bounded algorithm that only tests subsequences that have data dependencies. ElectricTest checks for data dependencies between tests using a more sophisticated check for data dependencies. While DTDetector checks for writes/reads to/from static fields, ElectricTest checks for changes to any memory reachable from static fields. PRADET uses a similar technique to check for data dependencies, but it also refines the output by checking for manifest dependencies, i.e., data dependence that also influences flakiness in test outcomes.  Wei et al. \citeS{S1014} used a systematic and probabilistic approach to explore the most effective orders for manifesting order dependent flaky tests. Whereas tools such as PRADET and DTDetector explore randomized test orders, Wei et al. analyse the probability of randomized orders detecting flaky tests, and they propose an algorithm that explores consecutive tests to find all order-dependent tests that depend on one test.

Anjiang et al. \citeS{S1011} discussed a class of flakiness due to shared state, non-idempotent-outcome (NOP) tests, which are detected by executing the same test twice in the same VM.

NonDex \citeS{S152} is the only tool we found that detects flakiness caused by implementation dependency. The class of such dependencies it detects is limited to dependencies due to assumptions developers make about underdetermined APIs in the Java standard libraries, for instance the iteration order of data structures using hashing in the internal representation, such as Java's \texttt{HashMap}.

A number of studies discussed  machine learning approaches for flakiness prediction. Pontillo et al. \citeS{S1004} studied the use of test and production code factors that can be used to predict test flakiness using classifiers. Their evaluation uses a logistic regression model. Haben et al. \citeS{S1005} reproduced a Java study \citeS{S11} with a set of Python programs to confirm the effectiveness of code vocabularies for predicting test flakiness. Camara et al. \citeS{S1012} is another replication of the same study that extends it with additional classifiers and datasets.  Parry et al. \citeS{S1034} presented an evaluation of static and dynamic features that are more effective as predictors of flakiness in comparison to previous feature sets. Camara et al. \citeS{S1017} evaluated using test smells to predict flakiness.\\

\begin{qoutebox}{white}{}
\textbf{RQ2 summary.} 
A number of methods have been proposed to detect flaky tests, which include static, dynamic and hybrid methods. Most static approaches use machine learning. Rerun (in different forms) is the most common dynamic approach for detecting flaky tests. Approaches that use rerun focus on  making flaky tests detection less expensive by accelerating ways to manifest flakiness or running fewer tests. 
\end{qoutebox}

\begin{table*}[!b]
\centering
\caption{Detection tools as reported in grey literature}
\label{tab:detection-industry}
\resizebox{\linewidth}{!}{
\begin{tabular}{p{3cm}p{12cm}l}
\toprule
\textbf{Tool} & \textbf{Features} & \textbf{Articles}\\
\midrule
% Odeneye & Visualizes test suites and execution results & \citeG{G2} \\
Flakybot & Determines test(s) are flaky before merging commits. Flakybot can be invoked on a pull request, and tests will be exercised quickly and results reported & \citeG{G2} \\
Azure DevOps Services & Feature that enables the detection of flaky tests (based on changes and through reruns) & \citeG{G6} \\
Scope& Helps identify flaky tests, requiring a single execution based on the commit diff & \citeG{G8} \\
Cypress & Automatically rerun (retries) a failed test prior to marking it as fail & \citeG{G9} \\
Gradle Enterprise & Considers a test flaky if it fails and then succeeds within the same Gradle task & \citeG{G22} \\
pytest-flakefinder \& pytest-rerunfailures & Rerun failing tests multiple times without having to restart pytest (in Python) & \citeG{G31} \\
pytest-random-order \& pytest-randomly & Randomise test order so that it can detect flakiness due to order dependency and expose tests with state problems &  \citeG{G31} \\
BuildPluse &  Detect and categorise flaky tests in the build by checking changes in test outcomes between builds (cross-language) & \citeG{G92} \\
rspec-retry & Ruby scripts  that rerun flaky \texttt{RSpec} tests and obtain a success rate metric & \citeG{G35} \\
Quarantine & A tool that provides a run-time solution to diagnosing and disabling flaky tests and automates the workflow around test suite maintenance & \citeG{G36} \\
protractor-flake & Rerun failed tests to detect changes in test outcomes & \citeG{G50} \\
Shield34 & Designed to address the Selenium flaky tests issues & \citeG{G57} \\
Bazel & Build and auto testing tool, An option to mark tests as flaky, which will skip those marked tests & \citeG{G58,G71} \\
Flaky (pytest plugin) & Automatically rerunning failing tests & \citeG{G59,G67} \\
Capybara & Contains an option to prevent against race conditions & \citeG{G68} \\
Xunit.Skip-pableFact & Tests can be marked as SkippableFact allowing control over test execution & \citeG{G70} \\
timecop & ruby framework to test time-dependent tests & \citeG{G81,G96} \\
% RSpec::-OrderedCommandFormatter & & \citeG{G101} \\
Athena & Identifies commits that make a test nondeterministically fail, and notifying the author. Automatically quarantines flaky tests & \citeG{G108} \\
Datadog & Flaky test management through a visualisation of test outcomes, it shows which tests are flaky & \citeG{G116} \\
CircleCI dashboard &  The ``Test Insights'' dashboard provides information about all flaky tests, with an option to automate reruns of failed tests & \citeG{G122} \\
Flaky-test-extractor-maven-plugin & Maven plugin that filters out flaky tests from existing surefire reports. It generates additional XML files just for the flaky tests & \citeG{G140}  \\
TargetedAutoRetry & A tool to retry just the steps which are most likely to cause issues with flakiness (such as Apps launch, race conditions candidates etc..) & \citeG{G213} \\
Junit surefire plugin & an option to rerun failing tests in Junit surefire plugin (rerunFailingTestsCount) & \citeG{G192} \\
Test Failure Analytics & gradle's plugin that helps to identify flaky tests between different builds & \citeG{G142} \\
Test Analyzer Service & An internal tool at Uber used to manage the state of unit tests and to disable flaky tests & \citeG{G149} \\
TestRecall & Test analysis tool that provides insights about test suites, including tracking flaky tests &  \citeG{G202} \\
Katalon Studio & an option to retry all tests (or only failed tests) when the Test Suite finishes & \citeG{G203} \\
\bottomrule
\end{tabular}}
\end{table*}

\subsection{Flaky Tests Datasets (RQ1 and RQ2)}
\label{sec:datasets}

Datasets used in flakiness related studies can be divided into those used in empirical studies or for detection/causes analysis studies. Those are all obtained from academic literature studies. Table~\ref{tab:flakdatasets} lists these datasets with the type of flakiness in the programs, the programming language, the number of flaky tests identified and the total number of projects along with the names of the tools or if it's an empirical study.

As can be seen in Table \ref{tab:flakdatasets}, the dominant programming language is Java. There are a few studies in Python \citeS{S14}. Some of these datasets are used in multiple studies, for instance \citeS{S15} obtains its subjects from \citeS{S4}, and \citeS{S12} from \citeS{S2}.\footnote{A dataset on the relationship between test smells and flaky tests was largely used in multiple studies but recently was retracted \url{https://ieeexplore.ieee.org/document/8094404}.}

\begin{table*}[!b]

\centering
\caption{Datasets to Study Test Flakiness}
 \label{tab:flakdatasets}

\resizebox{\linewidth}{!}{
\begin{tabular}{llllll}
\toprule
\textbf{Study} & \textbf{Flakiness type} & \textbf{Language} & \textbf{\# flaky tests} & \textbf{\# projects} & \textbf{Article} \\
\midrule
 iDFlakies  & Order dep/Other & Java & 422 & 694 & \citeS{S4} \\
DeFlaker & General & Java & 87 & 96 & \citeS{S2} \\
NonDex & Wrong assumptions & Java & 21 & 8 & \citeS{S152} \\
iFixFlakies & Order dependent & Java & 184 & 10 & \citeS{S135} \\
FLASH & Machine learning & Python & 11 & 20 & \citeS{S14} \\
Shaker & Concurrency & Java/Kotlin (Android) & 75 & 11 & \citeS{S24} \\
FlakeShovel & Concurrency & Java (Android) & 19 & 28 & \citeS{S16} \\
NodeRacer & Concurrency & JavaScript & 2 & 8 & \citeS{S93} \\
GreedyFlake & Flaky coverage & Python & -- & 3 & \citeS{S78} \\
Travis-Listener & Flaky builds & Mixed & -- & 22,345 & \citeS{S136} \\
RootFinder & General & .Net & 44 & 22 & \citeS{S6} \\

\bottomrule

\end{tabular}}
 \end{table*}

\subsection{Impact of Flaky Tests (RQ3)}
Next, we explore the wider view of the impact flaky tests have on different aspects of software engineering. In addressing this research question, we look at the impact of flaky tests as discussed in the articles we reviewed, and then combine the evidence noted in academic and grey literature. We discuss this in detail in the following two subsections.

\subsubsection{Impact Noted in Academic Research}
For each article we included in our review, we look at the context of flaky tests in the study. We classify the impact of flaky tests as reported in academic literature into the following three categories:

% \todo[inline]{change the categories below to align it with the taxonomy}
\begin{enumerate}
\item \textbf{Testing (including testing techniques):} the impact on software testing process in general (i.e., impact on test coverage).
\item \textbf{Product quality:} impact on the software product itself, and its quality.
\item \textbf{Debugging and maintenance:} the impact on other software development and program analysis techniques.
\end{enumerate}

\begin{figure*}[!h]
\centering
\includegraphics[width=0.9\linewidth]{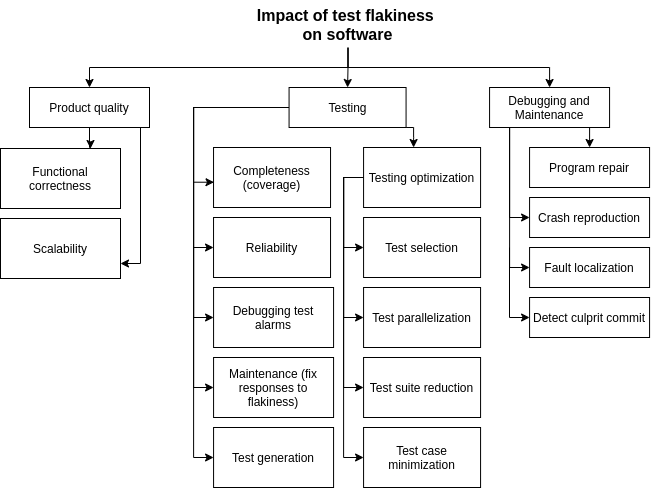}
\caption{Taxonomy for impact of test flakiness}
 \label{fig:taxonomyimpact}
\end{figure*}

Figure~\ref{fig:taxonomyimpact} illustrates these categories and provides a general taxonomy of impact points as noted in the reviewed studies. %For instance, order-dependent flakiness affects specific techniques related to testing, whilst regression testing is affected generally by test flakiness. 
Table \ref{tab:impact_academic} provides a summary of the impact of flaky tests as noted in academic literature. We discuss some examples for each of the three categories below.\\

\noindent \textbf{Impact on testing:}  Many aspects of testing are affected by test flakiness. This includes automatic test generation \citeS{S330}
, test quality characteristics \citeS{S57}, and techniques or tasks involved in test debugging and maintenance \citeS{S430}. 
% something about test quality
A number of testing techniques are based on the assumption that tests have deterministic outcomes, and when this assumption does not hold, they may not be reliable for their intended purposes. Test optimization techniques such as test suite reduction, test prioritization, test selection, and test parallelization rely on this assumption. For instance, flakiness can manifest in order dependent tests when test optimization is applied to test suites with such tests. Lam et al. \citeS{S40} studied the necessity of dependent-test-aware techniques to reduce flaky test failures, where they first  investigate the impact of flaky tests on three regression testing techniques: test prioritization, test selection and parallelization. Other testing techniques impacted are test amplification \citeS{S1156}, simulation testing \citeS{S1041} and manual testing \citeS{S1081}. \\
%\todo[inline]{add a couple more references here and expand the text a bit}

\noindent \textbf{Impact on product quality:} Several articles cite how test flakiness breaks builds \citeS{S336,S256}. Testing drives automated builds, which flakiness can break, resulting in delaying CI workflows. Zdun et al. \citeS{S429} highlighted how flaky tests can introduce noise into CI builds that can affect service deployment and operation (microservices and APIs in particular). B{\"o}hme \citeS{S57} discussed flakiness as one of the challenges for test assurance, i.e., executing tests as a means to increase confidence in the software. Product quality can be affected due to lack of test stability, which is cited as an issue by Hirsch et al. \citeS{S197}, in the context of a single Android application with many UI tests that are fragile. Several articles mention the issue of cost in detecting flaky tests, Pinto et al.~\citeS{S11} pointed out that it can be costly to run detectors after each change and hence organizations run them only on new or changed tests, which might not be the best approach as this would affect the recall. Vassallo et al. \citeS{S556} identified retrying failure to deal with flakiness as a CI smell, as it has a negative impact on development experience by slowing down progress and hiding bugs. Mascheroni et al. \citeS{S1043} proposed a model to improve continuous testing by presenting test reliability as a level in the improvement model, and flaky tests as a main cause for reliability issues with tests. They suggest good practices to achieve this.  \\
Multiple articles also discuss how test flakiness can affect developers, leading to a negative impact on product quality. This includes the developer's perception of tests, and the effort required to respond to events arising from test flakiness (build failures in CI, localizing causes, fixing faulty tests). Koivuniemi \citeS{S750} mentioned uncertainty and frustration due to developers attributing flaky failures to errors in the code where there are none. Eck et al. \citeS{S8} survey on developer's perception of flaky tests noted that flaky tests can have an impact on software projects, in particular on resource allocation and scheduling.\\

\noindent \textbf{Impact on debugging and maintenance:} Several techniques used in maintenance and debugging are known to be impacted by the presence of flaky tests. This includes all techniques that rely on tests such as test-based program repair, crash reproduction, test amplification and fault localization, which can all be negatively impacted by flakiness. Martinez et al. \citeS{S252} reported a flaky test in a commonly used bugs' dataset, Defects4J, and how the repair system's effectiveness can be affected (if the flaky test fails after a repair, the system would conclude that the repair introduced a regression). Chen et al. \citeS{S962} explained that subpar quality tests can affect their use for detecting performance regressions, and in the case of flaky tests they may introduce noise and require multiple executions. Dorward et al. \citeS{S1049} proposed an approach that is more efficient to find culprit commits using flaky tests, as bisect fails in this situation.

\subsubsection{Impact noted in grey literature}
We also analysed the impact of flaky test as found in grey literature articles. We checked if there was any discussion of the impact of flaky tests on certain techniques, tools, products or processes. We classify the noted impact of flaky tests into the following three categories: 

\begin{enumerate}
    \item \textbf{Code-base and product:}  the impact of flaky tests on the quality or the performance of the production code and the CUT.
    \item \textbf{Process:} the impact on the development pipeline and the delivery of the final product. 
    \item \textbf{Developers:} the `social' impact of flaky test on the developers/testers.
\end{enumerate}

\begin{table*}[]
\caption{Summary of the impact of flaky tests noted in academic literature}
\label{tab:impact_academic}
\resizebox{\linewidth}{!}{
\begin{tabular}{lll}
\toprule
% \multicolumn{3}{c}{\textbf{Academic Literature}} \\ \midrule
\textbf{Impact Type} & \textbf{Impact} & \textbf{Reference} \\ \midrule
\textbf{Product quality} & Breaking builds & \citeS{S336,S256}  \\
 & Service deployment and operation & \citeS{S429} \\
 & Test reliability & \citeS{S1043} \\
 & Test assurance & \citeS{S57} \\
 & Product quality & \citeS{S197}\\ 
 & Costly to detect & \citeS{S15,S11,S143} \\
 & Delays CI workflow & \citeS{S556,S136}\\
 & Maintenance effort & \citeS{S430}  \\
 & Uncertainty and frustration & \citeS{S750} \\
 & Trust in tools and perception & \citeS{S395,S159}\\ \midrule
\textbf{Testing} & Regression testing techniques & \citeS{S40}  \\
 & Simulation testing & \citeS{S1041} \\
 & Test amplification & \citeS{S1156} \\
 & Test suite/case reduction & \citeS{S662,S444}  \\ 
 & Mutation testing & \citeS{S174,S234}  \\
 & Manual testing & \citeS{S1081} \\
 & Test minimization & \citeS{S526} \\
 & Test coverage (ignored tests) & \citeS{S211}  \\
 & Test selection & \citeS{S207,S584} \\
 & Patch quality & \citeS{S269} \\
 & Test performance & \citeS{S704}  \\
 & Test suite efficiency & \citeS{S565}  \\
 & Test prioritization & \citeS{S73,S207} \\
 & Regressions & \citeS{S110} \\
 & Test suite diversity & \citeS{S73}  \\
 & Test generation & \citeS{S330}  \\
  & Differential testing & \citeS{S348} \\
 & Test assurance & \citeS{S57} \\ \midrule

\textbf{Debugging and maintenance} & Program repair & \citeS{S520,S252,S1070}\\
 & Determining culprit commits & \citeS{S1049} \\
 & Performance analysis & \citeS{S962} \\ 
 & Bug reproduction & \citeS{S229} \\
 & Crash reproduction & \citeS{S762} \\
 & Fault localization & \citeS{S17,S692} \\
 \bottomrule
\end{tabular}}
\end{table*}

\begin{table*}[]
\caption{Summary of the impact of flaky tests as noted in grey literature}
\label{tab:impact_grey}
\resizebox{\linewidth}{!}{
\begin{tabular}{lll}
\toprule
\textbf{Impact Type} & \textbf{Impact} & \textbf{Reference} \\ \midrule
 \textbf{Product} & Hard to debug & \citeG{G11,G52,G95} \\
  & Hard to reproduce & \citeG{G11} \\
 & Reduces test reliability & \citeG{G27,G103} \\
 & Expensive to repair & \citeG{G114} \\
 & \begin{tabular}[c]{@{}l@{}}Increase cost of testing as flaky \\ behaviour can spread to other tests\end{tabular} & \citeG{G8,G210} \\ \midrule
 \textbf{Developers}& Losing trust in builds & \citeG{G74,G81,G114,G127,G203} \\
%  & \begin{tabular}[c]{@{}l@{}}Losing confidence of the \\ usefulness of the test suite\end{tabular} & \citeG{G2,G22,G105} \\

  & Loss of productivity & \citeG{G8,G152,G165,G210} \\
& Time-consuming / wastes time & \citeG{G22,G95,G107,G134,G142,G144,G147,G149}\\ 
  & Resource consuming & \citeG{G26,G30,G127} \\
 & Demotivate/mislead developers & \citeG{G22,G134} \\
%  & \begin{tabular}[c]{@{}l@{}}Loss of faith that the \\ tests will catch bugs\end{tabular} & \citeG{G30,G89} \\ 
\midrule
%  \textbf{Delivery} & Reduces the value of the regression suite & \citeG{G4} \\
\textbf{Delivery} & Affects the quality of shipped code & \citeG{G6,G129,G202} \\ 
 & Slows down deployment pipeline & \citeG{G22,G95,G114,G142,G154} \\
  & Slows down the development & \citeG{G45,G95,G98,G22,G110} \\
  & Loses faith in tests catching  bugs & \citeG{G30,G89}
  \\
 & Causes unstable deployment  pipelines & \citeG{G35} \\
 & Slows down development and testing processes & \citeG{G45,G110} \\
 & Delays project release & \citeG{G107,G108,G213} \\
%  & Loss of faith in the tests to catch bugs & \\
 \bottomrule
\end{tabular}}
\end{table*}

%\todo[inline]{ [G53] (or [G1] as currently appear in table 9 next to ''Cause unstable deployment pipelines`` is ref to conference paper. Should be removed}
 A summary of the impact noted in the grey literature is shown in Table \ref{tab:impact_grey}. We discussed each of those three categories below.\\

\noindent \textbf{Impact on the code-base and product:} 
Several grey literature articles have discussed the wider impact of flaky tests on the production code and on the final software product.  
Among several issues reported by different developers, testers and managers, it was noted that the presence of flaky tests can significantly increase the cost of testing \citeG{G8,G36}, and makes it hard to debug and reproduce the CUT \citeG{G11,G52,G95}. In general, flaky tests can be very expensive to repair and often require time and resources to debug \citeG{G95,G114}. They can also make end-of-end testing useless \citeG{G74}, which can reduce test reliability \citeG{G27,G103}. One notable area that flaky tests compromise is coverage -  if the test is flaky enough that it can fail even when retried, then coverage is already considered lost \citeG{G129}.
Flaky tests can also spread and accumulate, with some unfixed flaky tests can lead to more flaky tests in the test suite \citeG{G7,G8}. Fowler describe them as \textit{``virulent infection that can completely ruin your entire test suite"} \citeG{G4}.\\
% \begin{quote}
%   \textit{``\dots flaky tests are an incredible cost to businesses. They are very expensive to repair often requiring hours or even days to debug and they jam the continuous deployment pipeline making shipping features slower.''} \citeG{G95} \\
% \end{quote}

Flaky tests can have serious implications in terms of time and resources required to identify and fix potential bugs in the CUT, and can directly impact production reliability \citeG{G36}. However, detecting and fixing flaky tests can help in finding underlying flaws and issues in the tested application and CUT that is otherwise much harder to detect \citeG{G95}.\\

\noindent \textbf{Impact on developers:} 
We observed that several of the blog posts we analysed here are written by developers and discuss the impact of flaky tests on their productivity and confidence.  
Developers noted that flaky tests can cause them to lose confidence in the `usefulness' of the test suite in general \citeG{G2}, and to lose trust in their builds \citeG{G74}. Flaky tests may also lead to a ``collateral damage'' for developers: if they are left uncontrolled or unresolved, they can have a bigger impact and may ruin the value of an entire test suite \citeG{G8}.
They are also reported to be disruptive and counter-productive, as they can waste developers' time as they try to debug and fix those flaky tests \citeG{G95,G107,G26,G30}.\\

\begin{quote}
 \textit{``The real cost of test flakiness is a lack of confidence in your tests..... If you don’t have confidence in your tests, then you are in no better position than a team that has zero tests. Flaky tests will significantly impact your ability to confidently continuously deliver.''} (Spotify Engineering, \citeG{G2}).\\
\end{quote}

Another experience report from Microsoft explained the practices followed and tools used to manage flaky tests at Microsoft in order to boost developers' productivity:
\begin{quote}
\textit{``Flaky tests.... negatively impact developers’ productivity by providing misleading signals about their recent changes ... developers may end up spending time investigating those failures, only to discover that the failures have nothing to do with their changes and may simply go away by rerunning the tests.'' (Engineering@Microsoft, \citeG{G134})
}\end{quote}

\noindent \textbf{Impact on delivery:} 
Developers and managers also presented evidence of how flaky tests can delay developments and have a wider impact on the delivery (e.g., \citeG{G36}) - mostly by slowing down the development \citeG{G45,G95,G98} and delaying products' release \citeG{G107,G108}. 
They can also reduce the value of an automated regression suite \citeG{G4} and lead organization and testing teams to lose faith that their tests will actually find bugs \citeG{G30,G89}. 
Some developers also noted that if flaky tests left unchecked, or untreated, they can lead to a completely useless test suites, as this is the case with some organisations:

\begin{quote}
\emph{`` We've talked to some organizations that reached 50\%+ flaky tests in their codebase, and now developers hardly ever write any tests and don’t bother looking at the results. Testing is no longer a useful tool to improve code quality within that organization.'' (Product Manager at Datadog, \citeG{G8})}\end{quote}

 Another impact of flaky tests is that it could slow down deployment pipeline which can decrease confidence in the correctness of changes in the software \citeG{G22,G114}. They could even block deployment until spotted and resolved \citeG{G5}.\\

\begin{qoutebox}{white}{}
\textbf{RQ3 summary.}
The impact of flaky tests has been the subject of discussion in both academic and grey literature. Flaky tests are reported to have an impact on the products under development, the quality of CUT and the tests themselves and on the delivery pipelines. %This also impacts the developer who write those tests.  
Techniques  that  rely  on tests such as test-based program repair, crash reproduction and fault detection and localization can be negatively  impacted by the presence of flaky tests.  
\end{qoutebox}

\subsection{Responses to Flaky Tests (RQ4)}
The way that developers and teams respond to flaky tests has been discussed in detail in both academic and grey literature. However, the type of applied/recommended response is slightly different from one study to another as this also depends on the context of the causes of flaky tests, and also the methods used to detect them. Below we discuss the responses as noted in academic and grey literature, separately:

\subsubsection{Response Noted in Academic Literature}

We classify responses to flaky tests as follows:

\begin{itemize}
    \item Modifying the test.
    \item Modifying the program/code under test.
    \item Process response.
\end{itemize}

\begin{figure*}[h]
\centering
\includegraphics[width=0.7\linewidth]{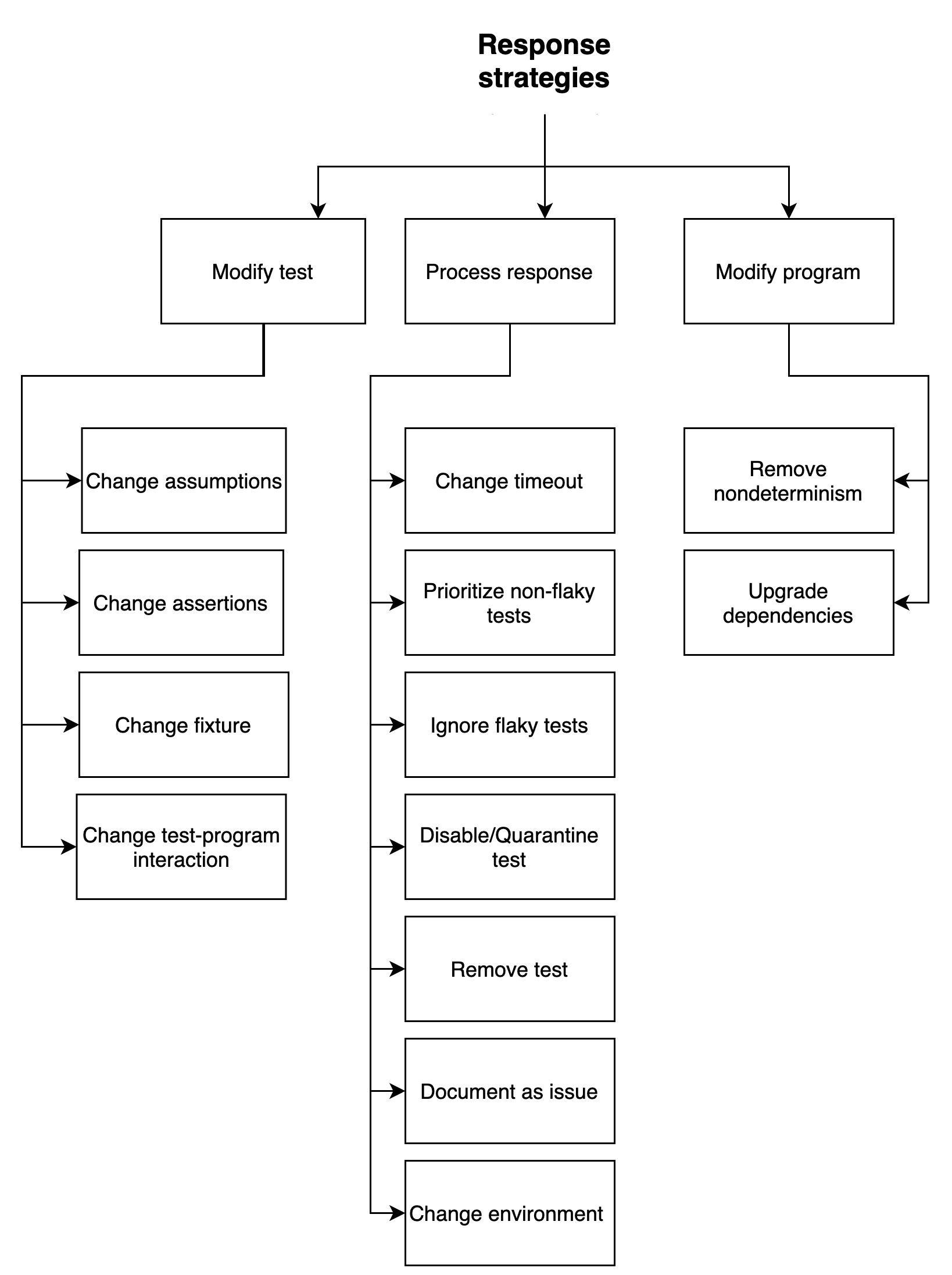}
\caption{Taxonomy of response strategies}
 \label{fig:taxonomyresponse}
\end{figure*}

We provide a general taxonomy of the responses to flaky tests as noted in the reviewed studies in  Figure~\ref{fig:taxonomyresponse}.

\begin{table*}[h]
\centering
\caption{Summary of the response strategies to flaky tests in academic literature.}
\resizebox{\linewidth}{!}{
\begin{tabular}{@{}lll@{}}
\toprule
\textbf{Strategy}                    & \textbf{Description}                                                                                                                                       & \textbf{Articles}                                                                                           \\ \toprule
\textbf{Modify test} & \textbf{Change assumptions} & \\
& Fix assumptions about library API's & \citeS{S152} \\
& automatically repair implementation-dependent tests & \citeS{S1013} \\
& Replace test & \citeS{S8} \\
  & Merge dependent tests & \citeS{S1} \\
 & \textbf{Change assertions} & \\
 & Modify assertion bounds (e.g., to accommodate wider ranges of outputs) & \citeS{S1,S7,S8,S14,S1009,S1044} \\
 
 & \textbf{Change fixture} & \\
 & Removing shared dependency between tests & \citeS{S1} \\
 & Global time as system variable & \citeS{S950} \\
 & Setup/clean shared state between tests & \citeS{S1}\\
 & Modify test parameters & \citeS{S14} \\
 & Modify test fixture & \citeS{S9} \\
 & Fix defective tests & \citeS{S446,S963,S1} \\
 & Make behaviour deterministic & \citeS{S39} \\
 & Change delay for async-wait & \citeS{S72,S1045} \\
 & Concurrency-related fixes & \citeS{S8} \\
  & \textbf{Change test-program interaction} & \\
 & Mock use of environment/concurrency & \citeS{S179} \\ \midrule
\textbf{Modify program} 
 & Concurrency-related fixes & \citeS{S1} \\
& Replace dependencies & \citeS{S7,S14} \\ 
& Remove nondeterminism & \citeS{S8} \\ \midrule
\textbf{Process response}  & Rerun tests & \citeS{S7,S269,S234} \\ % rerun before accepting code patch
 & Ignore/Disable & \citeS{S8,S359,S143} \\
 & Quarantine & \citeS{S106} \\
 & Add annotation to mark test as flaky & \citeS{S14} \\
 & Increase test time-outs & \citeS{S8} \\
 & Reconfigure test environment (e.g., containerize or virtualize unit tests) & \citeS{SB29} \\
 & Remove & \citeS{S164,S145,S39,S165,S71} \\
 & Improve bots to detect flakiness & \citeS{S100} \\
 & Responsibility of CI to deal with it & \citeS{S67} \\
 %& Separate test suites by goal & \citeS{S35} \\
 & Prioritize tests & \citeS{S137,S1021} \\
 
\bottomrule
\end{tabular}}
\label{tab:responses_academic}
\end{table*}

A summary of the responses found in academic articles is presented in Table \ref{tab:responses_academic}. The three major strategies are to fix tests, to modify the CUT or putting in a mechanism to deal with flaky tests (e.g., retry or quarantine tests). Berglund and Vateman \citeS{S39} listed some strategies for avoiding non-deterministic behaviour in tests: minimising variations in testing environment, avoiding asynchronous implementations, testing in isolation, aiming for deterministic assertions and limiting the use of third party dependencies. Other measures include mocking to reduce flakiness, for instance EvoSuite \cite{fraser2011evosuite} uses mocking for this. Zhu et al. \citeS{S179} proposed a tool for identifying and proposing mocks for unit tests. A wider list of  specific fixes to the different types of flaky tests is provided in \cite{S1}. Shi et al. \citeS{S9} presented a tool, iFixFlakies, to fix order dependent tests.

Fixes in the CUT are not discussed as much in academic articles. The closest mention in relation to this is in \citeS{S7}, which finds instances in flaky test fix commits where the CUT is improved and dependencies are changed to fix flakiness. 

Another strategy, removing flaky tests, was also identified in \citeS{S7}. The study found that developers commented out flaky tests in 10/77 of examined commits. Removing flaky tests is also a strategy cited in papers that discuss testing related techniques \citeS{S164,S165}. Quarantining, ignoring or disabling flaky tests is also discussed as responses. Memon et al. \citeS{S137} detailed the approach at Google for dealing with flaky tests. They use multiple factors (e.g., more frequently modified files are more likely to cause faults) to prioritize the tests to rerun rather than a simple test selection heuristic such as rerun tests that failed recently, which is sensitive to flakiness. 

A number of tools have been proposed recently for automatically repairing flaky tests. They can fix flakiness due to these causes: randomness in ML projects, order dependence and implementation dependence.  
Dutta et al. \citeS{S1009,S1044} conducted an empirical analysis of seeds in machine learning projects and propose approaches to repair flaky tests due to randomness by tuning hyperparameters, fixing seeds and modifying assertions bounds. Zhang et al. \citeS{S1013} proposed a tool for fixing flaky tests that are caused by implementation dependencies of the type explored by NonDex \citeS{S152}. Wang et al.  \citeS{S1031} proposed iPFlakies for Python that fixes order dependent tests that fail due to state that is polluted by other tests. This is related to their work, iFixFlakies for repairing order dependent tests in Java programs. The Python tool discovers existing tests or helper methods that clean the state before successfully rerunning the order dependent test. ODRepair from Li et al.  \citeS{S1018} is an approach that uses automatic test generation to clean the state rather than existing code. Mondal et al. \citeS{S1058} proposed an approach to fixing flakiness due to parallelizing dependent tests by adding a test from the same class to correct dependency failure.

\subsubsection{Response Noted in Grey Literature}
Here we look at the methods and strategies followed to deal with flaky tests as noted in the grey literature. We classified those strategies into the following categories: 
\begin{enumerate}
    \item \textbf{Quarantine:} keep flaky tests in a different test suite to other `healthy' tests in a quarantined area in order to diagnose and then fix those tests.
    \item \textbf{Fix immediately:} fix any flaky test that has been found immediately, but first developers will need to reproduce the flaky behaviour .
    \item \textbf{Skip and ignore:} provide an option to developers to ignore flaky tests from the build and suppress the test failures. This is usually in the form of annotation.
    In some cases, especially when developers are fully aware of the flaky behaviour of the tests and the implications of those tests have been considered, they may decide to ignore those flaky tests and continue with the test run as planned.
    \item \textbf{Remove:} remove any test that is flaky from the test suite once detected.  
\end{enumerate}

\begin{table*}[!h]
\centering
\caption{Summary of the response strategies followed by some organisations to deal with flaky tests, as discussed in grey literature.}

\resizebox{\linewidth}{!}{
\begin{tabular}{@{}lll@{}}
\toprule
\textbf{Strategy}                    & \textbf{Description}                                                                                                                                       & \textbf{Example}                                                                                           \\ \toprule
Quarantine    & \begin{tabular}[c]{@{}l@{}}Keep flaky tests in a different test suite to \\ other healthy tests in a quarantined area.\end{tabular}              & \begin{tabular}[c]{@{}l@{}}\citeG{G1,G4,G5,G36,G104,G106,G108}\\ \citeG{G35,G38,G67,G70,G79,G89,G111,G114} \\ \citeG{G149,G154,G164,G213,G134,G202} \end{tabular} \\ \midrule
\begin{tabular}[c]{@{}l@{}}Fix and replace immediately, \\ or remove if not fixed \end{tabular}&                               \begin{tabular}[c]{@{}l@{}}         Test with flaky behaviour are given priority \\ and fixed/removed once detected.      \end{tabular}                                                                                                   & \citeG{G20,G101,G102,G37,G147,G189,G111}                                                                                          \\ \midrule

Label flaky tests & Leave it to developers to decide & \citeG{G22,G67,G129,G134,G202} \\ \midrule
Ignore/Skip                     & \begin{tabular}[c]{@{}l@{}}Provide an option to developers to ignore  \\ flaky tests from the build (e.g., though the \\ use of annotations)  and suppress the test failures.\end{tabular} & \citeG{G6,G8,G70,G129} \\
\bottomrule
\end{tabular} }
\label{tab:responses_grey}
\end{table*}

A summary of the response found in grey literature is shown in Table \ref{tab:responses_grey}. The most common strategy that has been discussed is to quarantine and then fix flaky tests. As explained by Fowler \citeG{G4}, this strategy indicates that developers should follow a number of steps once a flaky test has been identified:  \textit{Quarantine} $\rightarrow$ \textit{Determine the cause} $\rightarrow$ \textit{Report/Document}
$\rightarrow$ \textit{Isolate and run locally} $\rightarrow$ \textit{Reproduce} $\rightarrow$ \textit{Decide (fix/ ignore)}.
% \begin{quote}
%   \textit{``Place any non-deterministic test in a quarantined area. (But fix quarantined tests quickly.) .... 
%     That way you'll you can continue to pay attention to what's going on with your healthy tests and get good feedback from them.'' }
% \end{quote}
This is the same strategy that Google (and many other organisations) has been employing to deal with any flaky tests detected in the pipelines \citeG{G1}. A report from Google reported that they use a tool that monitors all potential flaky tests, and then automatically quarantines the test in case flakiness is found to be high. The quarantining works by removing ``\emph{the test from the critical path and files a bug for developers to reduce the flakiness. This prevents it from becoming a problem for developers, but could easily mask a real race condition or some other bug in the code being tested}'' \citeG{G1}. Other organizations also follow the same strategy e.g., Flexport \citeG{G36} and Dropbox \citeG{G108}.
Flexport \citeG{G36} have even included a mechanism to automate the process of quarantining and skipping flaky tests. The Ruby gem, Quarantine\footnote{\url{https://github.com/flexport/quarantine}}, used to maintain a list of flaky tests by automatically ``detects flaky tests and disables them until they are proven reliable''. 

It has been suggested by some developers and managers that all identified flaky tests should be labelled  by their severity. This can be determined by which specific component they impact, the frequency of a flaky test, or the flakiness rate of a given test.
One approach that has been suggested is not only to quarantine and treat all flaky tests equally, but to quantify the level of flakiness of each flaky test so that those tests can be priorities for fixing. A report from Facebook engineers proposed a statistical metric called the Probabilistic Flakiness Score (PFS), with the aim to quantify flakiness by measuring test reliability based on how flaky they are \citeG{G127}. Using this metric, developers can \textit{``test the tests to measure and monitor their reliability, and thus be able to react quickly to any regressions in the quality of our test suite. PFS ... quantify the degree of flakiness for each individual test at Facebook and to monitor changes in its reliability over time. If we detect specific tests that became unreliable soon after they were created, we can direct engineers’ attention to repairing them.''} \citeG{G127}. GitHub reported a similar metrics-based approach to determine the level of flakiness for each flaky test. An impact score is given to each flaky test based on how many times it changed its outcomes, as well as how many branches, developers, and deployments were affected by it. The higher the impact score, the more important the flaky test and thus the highest priority for fix is given to this test \citeG{G147}.

At Spotify \citeG{G2}, engineers use Odeneye, a system that visualises an entire test suite running in the CI, and can point out developers to tests with flaky outcomes as the results of different runs. Another tool used at Spotify is Flakybot\footnote{\url{https://www.flakybot.com}}, which is designed to help developers determine if their tests are flaky before merging their code to the master/main branch. The tool can be self-invoked by a developer in a pull request, which will exercise all tests and provide a report of their success/failure and possible flakiness.

There are a number of issues to consider when quarantining flaky tests though, such as how many tests should be quarantined (having too many tests in the quarantine can be considered as counterproductive) and how long a test should stay in quarantine. Fowler \citeG{G4} suggested that not more than 8 tests in the quarantine at one time, and not to keep those tests for a long period of time. It was suggested to have a dashboard to track progress of all flaky tests so that they are not forgotten \citeG{G8}, and have an automated approach not to only quarantine flaky tests, but also to de-quarantine them once fixed or decided to be ignored \citeG{G202}.

Regarding the different causes of flaky tests, there are different strategies that are recommended to deal with the specific sources of test flakiness. For example, to deal with flakiness due to state-dependent scenarios such as if there is an ``Inconsistent assertion timing'' (i.e., state is not consistent between test runs that can cause tests to fail randomly), one solution is to ``construct tests so that you wait for the application to be in a consistent state before asserting'' \citeG{G36}. If the test depends on specific test order (i.e., global state shared between tests as one test may depend on compilation of another one), an obvious solution is to ``reset the state between each test run and reduce the need for global state'' \citeG{G36}. Table \ref{tab:fixing-grey} provides a brief summary of flaky tests' fixing strategies due to the most common causes as noted in grey literature articles.\\

\begin{table*}[]
\caption{Some fixing strategies for some common flaky tests noted in grey literature }
\label{tab:fixing-grey}
\resizebox{\linewidth}{!}{
\begin{tabular}{lll}
\toprule
\textbf{Cause of flakiness} & \textbf{Suggested fix} & \textbf{Example} \\ \toprule
Asynchronous wait & \begin{tabular}[c]{@{}l@{}}Wait for a specified period of time before it checks if the action has been \\ successful (with callbacks and polling) .\end{tabular} & \citeG{G7,G74} \\ \midrule
Inconsistent assertion timing & \begin{tabular}[c]{@{}l@{}}Construct tests so that you wait for the application to be in a consistent \\ state before asserting.\end{tabular} & \citeG{G7} \\ \midrule
Concurrency & \begin{tabular}[c]{@{}l@{}}Make tests more robust, so that it accepts all valid results . \\ Avoid running tests in parallel .\end{tabular} & \begin{tabular}[c]{@{}l@{}}\citeG{G7}\\ \citeG{G74}\end{tabular} \\ \midrule
Order dependency & \begin{tabular}[c]{@{}l@{}}Run a test in a database transaction that’s rolled back once the test has \\ finished executing .\\ Clean up  the environment (i.e., reset state)  and prepare it before every \\ test (and reduce the need for global state in general).   \\ Run test in isolation. \\ Run test in random order to find out if they are still flaky .\end{tabular} & \begin{tabular}[c]{@{}l@{}}\citeG{G7}\\ \\ \citeG{G7,G2,G75}\\ \\ \citeG{G5,G101}\\ \citeG{G96}\end{tabular} \\ \midrule
Time-dependent tests & \begin{tabular}[c]{@{}l@{}}Wrapping the system clock with routines that can be replaced with a \\ seeded value for testing. \\ Use a tool  to control for time variables such as  freeze time helper in\\ Ruby and Sinon.JS in JavaScript.\end{tabular} & \begin{tabular}[c]{@{}l@{}}\citeG{G5}\\ \\ \citeG{G95}\end{tabular} \\ \midrule
Randomization & Avoid the use of random seeds.  & \citeG{G38} \\ \midrule
Environmental & \begin{tabular}[c]{@{}l@{}}Limit dependency on environments in the test.\\ limit calls to external resources and build a mocking server  for \\ tests .\end{tabular} & \citeG{G27, G81,G98} \\ \midrule
Leak global state & Run test in random order.  & \citeG{G95} \\ \bottomrule
\end{tabular}}
\end{table*}

% \todo[inline]{@Amjed add new results on specific fixes to table 12 }

\begin{qoutebox}{white}{}
\textbf{RQ4 summary.}
Quarantining flaky tests (for a later investigation and fix) is a common strategy that is widely used in practice. This is now supported by many tools that can integrate with modern CI tooling (able to automatically detect changes in test outcomes to identify flaky tests). Understanding the main cause of the flaky behaviour is a key to reproducing flakiness and  identifying an appropriate fix, which remains a challenge.\end{qoutebox}

\section{Discussion}
\label{sec:discussion}
In this section, we discuss the results of the review and present possible challenges in detecting and managing flaky tests.
We also provide our own perspective on current limitations of existing approaches, and discuss potential future research directions.
\subsection{Flaky Tests in Research and Practice}
The problem with flaky tests has been a subject of a wide discussion among researchers and practitioners. Dealing with flaky tests is a real issue that is impacting developers and test engineers on a daily basis. It can undermine the validity of test suites and make them almost useless \citeG{G8,G74}. The topic of flaky tests has been a research focus, with a noticeable increase in the number of publications over the last four years (between 2017 and 2021). We observed that the way the issue of test flakiness is being discussed is slightly different in academic and grey literature. The majority of research articles discuss mainly the impact of flaky test on different software engineering techniques and applications. Research that focuses solely on flaky tests mainly tackles new methods that are employed to detect flaky tests, aiming to increase speed (i.e., how fast flakiness can be manifested) and accuracy of flakiness detection.
Many of those studies have focused on specific causes of flakiness (either in terms of detection or fixes) - namely those related to order-dependency in test execution or to concurrency. There is generally a lack of studies that investigate the impact of other causes of test flakiness, such as those related to variation in the environment or in the network. This is an area that can be addressed by future tools designed specifically to detect test flakiness due to those factors. Our recent work targets this by designing a tool, \textit{saflate}, that is aimed at reducing test flakiness by sanitising failures induced by network connectivity problems \cite{dietrich2022flaky}.

On the other hand, the discussion in grey literature focused more on the general strategies that are being followed in practice to deal with any flaky tests once detected in the CI pipeline. Those are usually detected by checking if the test outcomes have changed between different runs (e.g., between \texttt{PASS} to \texttt{FAIL}). Several strategies that have been followed by software development teams are discussed in grey literature, especially around what to do with flaky tests once they have been identified. A notable approach is quarantining flaky tests in an isolated `staging' area before they are fixed \citeG{G4,S106}.  

The gap between academic research and practice when it comes to the way flaky tests are viewed has also been discussed in some of the most recent articles. An experience report published by managers and engineers at Facebook \citeG{G127} explained how real world applications can \textit{always} be flaky (e.g., due to the non-determinism of algorithms), and what we should be focusing on is not when or if tests are flaky, but rather how flaky those tests can be. This supports Harman and O'Hearn's view \cite{harman2018start} that all test should, by default, considered to be flaky, which provides a defensive mechanism that can help in managing flaky tests in general.  

\subsection{Identifying and Detecting Flaky Tests}
In RQ1, we surveyed the common causes of flaky tests, whether in the CUT or in the tests themselves. We observe that there are a variety of causes for flakiness, from the use of specific programming language features to the reliance on external resources. It is clear that there are common factors that are responsible for flaky test behaviours, regardless of the programming language used or the application domains. Factors like test order dependency, concurrency, randomness, network and reliance on external resources are common across almost all domains, and are responsible for a high proportion of flaky tests.

Beyond the list of causes noted in the first empirical study in this topic of Luo et al. \citeS{luo2014empirical}, we found evidence of a number of additional causes, namely flakiness due to algorithmic nondeterminism (related to randomness), variations of hardware, environment and those related to the use of ML applications (which are nondeterministic in nature).
Some causes identified in academic literature overlap, and causes can also be interconnected. For example, UI flakiness can in turn be due to a platform dependency (e.g. dependency on a specific browser) or because of event races.

With this large number of causes of flaky tests, further in-depth investigation into the different causes is needed to understand how flaky tests are introduced in the code base, and better understand the root causes of flaky tests in general. This also includes studies of test flakiness in the context of variety of programming languages (as opposite to Java or Python, which most flakiness studies have covered).

Another point worth mentioning here is how \emph{flakiness} is being defined in different studies -- in general, a test is considered flaky if it has a different outcome on different runs with the same input data. Academic literature refers to tests having binary outcomes, i.e., \emph{PASS} or \emph{FAIL}. In practice, however, tests can have multiple outcomes on execution (pass, fail, error or skip). For instance, tests may be skipped/ignored (potentially non-deterministically) or may not terminate (or timeout, depending on the configuration of tests or test runners). A more concise and consistent definition of the different variants of flakiness is needed.

%\subsection{Limitations of the Current Detection Approaches}

%repetitive text below? 

\subsection{The Impact of and Response to Flaky Tests}
It is clear that flaky tests are known to have a negative impact on the validity of the tests, or the quality of the software as a whole. 
A few impact points have been discussed in both academic and grey literature. Notable areas that are impacted by flaky tests are test-dependent techniques, such as fault localization, program repair and test selection. 
An important impact area that has not been widely acknowledged is how flaky tests affect developers. Although the impact on developers was mentioned in developers'
surveys \cite{S8,S1019}, and in many grey literature articles (e.g., \citeG{G2,G8,G134}), such an impact has not been explicitly studied in more detail -- an area that should be explored further in the future.

In terms of responses to flaky tests, it seems that the most common approach is to quarantine flaky tests once they are detected. The recommendation is to keep tests with flaky outcomes in a separate quarantine area from other ``healthy'' tests. This way, those flaky tests are not forgotten and the cause of flakiness can be investigated later to apply a suitable fix. On the other hand, other non-flaky tests can still run so that it does not cause any delay in development pipelines. 
However, there remains some open questions about how to deal with quarantined tests, how long those tests should stay in the designated quarantine area, and how many tests can be quarantined at once. A strategy (that can be implemented into tools) on how to process quarantined flaky tests and remove them from the designated quarantine area (i.e., de-quarantining) also needs further investigation.

One interesting area for future research is to study the long-term impact of flaky tests. For example, what is the impact of flaky tests on the validity of test suites if left unfixed or unchanged. Do a few flaky tests that are left in the test suite untreated have a wider impact on the presence of bugs as the development progresses?
It is also interesting to see, when flaky tests are flagged and quarantined, how long it will take developers to fix those tests. This can be viewed at as a technical debt that will need to be paid back. Therefore, a study on whether this is actually been paid back, and how long it takes, will be valuable.\\

 \subsection{Implications on Research and Practice}
This study yields some actionable insights and opportunities for future research. We discuss those implications in the following:
\begin{enumerate}
 \item The review clearly demonstrates that academic research on test flakiness focuses mostly on Java, with limited studies done in other popular languages\footnote{Based on Stack Overflow language popularity statistics \url{https://insights.stackoverflow.com/survey/2021\#technology-most-popular-technologies}} i.e., JavaScript and Python. The likely reasons are a combination of existing expertise, the availability of open-source datasets, and the availability of high-quality and low cost (often free) program analysis tools. However, our grey literature review shows that the focus among practitioners is more on the ``big picture'', and flakiness has been discussed in the context of a variety of programming languages. 
\item  Different programming languages have different features, and it is not obvious how results observed in Java programs carry over to other languages. For instance: flakiness caused by test order dependencies and shared (memory) state are not possible in a pure functional language (like Haskell), and at least less likely in a language that manages memory more actively to restrict aliasing (like Rust using ownership\footnote{\url{https://doc.rust-lang.org/book/ch04-00-understanding-ownership.html}}).  In languages with different concurrency models such single-threaded languages (e.g., JavaScript), some flakiness caused by concurrency is less likely to occur. For instance, deadlocks are more common in multithreaded applications \cite{wang2017comprehensive}. Still, this does not mean that flakiness cannot occur due to concurrency, but it is likely to happen to a lesser extent compared to multithreaded languages such as Java. Similarly, languages (like Java) that use a virtual machine decoupling the runtime from operating systems and hardware are less likely to produce flakiness due to variability in those platforms than low-level languages lacking such a feature, like C.  Languages with strong integrated dynamic/meta programming features to facilitate testing like mock testing, which when used may help avoid certain kinds of flakiness, for instance, flakiness caused by network dependencies.

\item There seems to be an imbalance in the way to respond to flaky tests between what have been discussed in academic and industry articles. Industry responses have focused on processes to deal with flaky tests (such as quarantining strategies), and academic research have focused more on causes detection (note that there are some recent studies on automatically repairing flakiness in ML projects and order dependent tests). This is not unexpected; however, and may also indicate opportunities for future academic research to provide tools that can help automate quarantining (and de-quarantining). Furthermore, it appears that some industrial practices such as rerunning failed tests until they pass may require a deeper theoretical foundation. For instance, does a test that only passes after several reruns provide the same level of assurance as a test that always passes provides? The same question can be asked for  entire test suites: what is the quality of a test suite that never passes in its entirety, but each individual test is observed to pass in some configuration.

\item Another question arises from this, what is the  number of reruns required to assure (with a high level of confidence level) that a test is not flaky? From what we observed in the studies that used a rerun approach to manifest flakiness, the number of reruns used differ from one study to another (with some studies noting 2 \citeS{S16}, 10 \citeS{S99} or even 100 \citeS{S6} reruns as baselines). A recent study on Python projects reported that $\sim$170 reruns are required to ensure a test is not flaky due to non-order-dependent reasons \cite{gruber2021empirical}. 
We believe that the number of reruns required will depend largely on the cause of flakiness. Some rerun based tools, such as rspec-retry\footnote{\url{https://github.com/NoRedInk/rspec-retry}} for Ruby or the \texttt{\@RepeatedTest}\footnote{\url{https://junit.org/junit5/docs/5.0.1/api/org/junit/jupiter/api/RepeatedTest.html}} annotation in JUnit, provides an option to rerun tests a specified \textit{n} number of times (set by the developer). However, it is unknown what is a suitable threshold for the number of reruns required for different types of flakiness. 
Further empirical investigation to quantify the minimum number of reruns required to manifest flakiness (for the different causes and in different contexts) is required. Alshammari et al. \citeS{S1008} is a step in this direction where they rerun tests in 24 Java projects 10,000 times to find out how many flaky tests can be found with different numbers of reruns.

\end{enumerate}

\section{Validity Threats}
\label{sec:threats}

We discuss a number of potential threats to the validity of the study below, and explain the steps taken to mitigate them. \\

\noindent\textbf{Incomplete or inappropriate selection of articles:} As with any systematic review study, due to the use of an automatic search it is possible that we may have missed some articles that were not either covered by our search string, or were not captured by our search tool. 
We mitigated this threat by first running and refining our search string multiple times. We piloted the search string on Google Scholar to check what the string will return. We cross-validated this by checking if the search string would return well-known, highly cited articles of test flakiness (e.g., \cite{luo2014empirical,memon2013automated,eck2019understanding}). We believe this iterative approach has improved our search string and reduced the risk of missing key articles.

There is also a chance that some related articles have used terms other than those we used in our search string. If terms other than ``flaky'', ``flakiness'' or  ``non-deterministic'' were used, then the possibility of missing those studies increases. To avoid such a limitation we repeatedly refined our search string and performed sequential testing in order to recognize and include as many relevant studies as possible.\\ %In addition, we conducted snowballing on the references of the selected ``focused'' articles in order to locate any missing influential articles. \\

\noindent\textbf{Manual analysis of articles:}
We read through each of the academic and grey literature articles in order to  answer our research questions. 
This was done manually with at least one of the authors reading through articles and then the overall results are verified by another co-author. This manual analysis could introduce bias due to multiple interpretations and/or oversight. We are aware that human interpretation introduces bias, and thus we attempted to account for it via cross-validation involving multiple evaluators and by cross-checking the results from the classification stage, by involving at least two coders.\\

\noindent\textbf{Classification and reliability:} We have performed a number of classifications based on findings from different academic and grey literature articles to answer our four research questions. %Some of these classification were based on our manual analysis of the articles. 
We extracted information such as causes of flakiness (RQ1), detection methods and tools (RQ2), impact of flakiness (RQ3) and responses (RQ4). This information was obtained by reading through the articles, extracting the relevant information, and then classifying the articles by one of the author. Another author then cross-validated the overall classification of articles. We made sure that at least two of the co-authors will check each results and discuss any difference until a 100\% agreement between the two is reached.\\

%\textbf{Reliability of grey literature sources:}
\section{Conclusion}
\label{sec:conclusion}
In this paper, we systematically studied how test flakiness has been addressed in academic and grey literature. We provide a comprehensive view of flaky tests, their common causes, their impact on other techniques/artefacts, and discuss response strategies in research and practice. We also studied methods and tools that have been used to detect and locate flaky tests and strategies followed to respond to flaky tests.\\
This review covers 560 academic literature and 91 grey literature articles. The results show that most academic studies that covered test flakiness has focused more on Java compared to other programming languages. In terms of common causes, we observed that flakiness due to test order dependency and concurrency have been studied more widely compared to other noted source of flakiness. However, this depends mainly on the focus of the studies that reported those causes. For example, studies that used Android as their subject systems has focused mostly on flakiness in UI (which concurrency issues are attributed as the root cause).
Correspondingly, methods to detect flaky tests have focused more on specific types of flaky tests, with the dynamic rerun-based approach noted as the main proposed method for flaky tests detection. The intention is to provide approaches (either static or dynamic) that are less expensive to run by accelerating ways to manifest flakiness with running fewer tests.\\
This paper outlines some limitations in test flakiness research that should be addressed by researchers in the future.

\section{Acknowledgements}
This work is funded by  Science for Technological Innovation National Science Challenge of New Zealand, grant number MAUX2004.

\bibliographystyleS{IEEEtran}
\bibliographyS{literature}
\Urlmuskip=0mu plus 1mu\relax
\bibliographystyleG{IEEEtran}
\bibliographyG{greyliterature}
\bibliographystyle{IEEEtran}
\bibliography{main.bib}

\begin{thebibliography}{10}
\providecommand{\url}[1]{#1}
\csname url@samestyle\endcsname
\providecommand{\newblock}{\relax}
\providecommand{\bibinfo}[2]{#2}
\providecommand{\BIBentrySTDinterwordspacing}{\spaceskip=0pt\relax}
\providecommand{\BIBentryALTinterwordstretchfactor}{4}
\providecommand{\BIBentryALTinterwordspacing}{\spaceskip=\fontdimen2\font plus
\BIBentryALTinterwordstretchfactor\fontdimen3\font minus
  \fontdimen4\font\relax}
\providecommand{\BIBforeignlanguage}[2]{{%
\expandafter\ifx\csname l@#1\endcsname\relax
\typeout{** WARNING: IEEEtran.bst: No hyphenation pattern has been}%
\typeout{** loaded for the language `#1'. Using the pattern for}%
\typeout{** the default language instead.}%
\else
\language=\csname l@#1\endcsname
\fi
#2}}
\providecommand{\BIBdecl}{\relax}
\BIBdecl

\bibitem{G2}
\BIBentryALTinterwordspacing
J.~Palmer, ``Test {Flakiness} - {Methods} for identifying and dealing with
  flaky tests,'' 2019. [Online]. Available:
  \url{https://engineering.atspotify.com/2019/11/18/test-flakiness-methods-for-identifying-and-dealing-with-flaky-tests/}
\BIBentrySTDinterwordspacing

\bibitem{G6}
\BIBentryALTinterwordspacing
``\BIBforeignlanguage{en-us}{Manage flaky tests - {Azure} {Pipelines}},'' 2020.
  [Online]. Available:
  \url{https://docs.microsoft.com/en-us/azure/devops/pipelines/test/flaky-test-management}
\BIBentrySTDinterwordspacing

\bibitem{G8}
\BIBentryALTinterwordspacing
B.~Lee, ``\BIBforeignlanguage{en}{We have a flaky test problem},'' 2020.
  [Online]. Available:
  \url{https://medium.com/scopedev/how-can-we-peacefully-co-exist-with-flaky-tests-3c8f94fba166}
\BIBentrySTDinterwordspacing

\bibitem{G9}
\BIBentryALTinterwordspacing
A.~Rustamzadeh, ``Introducing {Flaky} {Test} {Detection} \& {Alerts},'' 2020.
  [Online]. Available:
  \url{https://www.cypress.io/blog/2020/10/20/introducing-flaky-test-detection-alerts/}
\BIBentrySTDinterwordspacing

\bibitem{G22}
\BIBentryALTinterwordspacing
E.~Wendelin, ``\BIBforeignlanguage{en-US}{Identifying and analyzing flaky tests
  in {Maven} and {Gradle} builds},'' 2019. [Online]. Available:
  \url{https://gradle.com/blog/flaky-tests/}
\BIBentrySTDinterwordspacing

\bibitem{G31}
\BIBentryALTinterwordspacing
``Flaky tests — pytest documentation.'' [Online]. Available:
  \url{https://docs.pytest.org/en/stable/flaky.html}
\BIBentrySTDinterwordspacing

\bibitem{G92}
\BIBentryALTinterwordspacing
``\BIBforeignlanguage{en}{Detect, track and eliminate flaky tests},'' 2020.
  [Online]. Available: \url{https://buildpulse.io/}
\BIBentrySTDinterwordspacing

\bibitem{G35}
\BIBentryALTinterwordspacing
D.~Mayer, ``\BIBforeignlanguage{en}{Flaky {Ruby} {Tests}},'' 2019. [Online].
  Available: \url{https://www.mayerdan.com/ruby/2019/09/07/flaky-ruby-tests}
\BIBentrySTDinterwordspacing

\bibitem{G36}
\BIBentryALTinterwordspacing
E.~Zhu, ``Solving flaky tests in rspec.'' [Online]. Available:
  \url{https://flexport.engineering/solving-flaky-tests-in-rspec-9ceadedeaf0e}
\BIBentrySTDinterwordspacing

\bibitem{G50}
\BIBentryALTinterwordspacing
``automated testing - {How} to deal with flaky tests that have intermittent
  failures?'' 2017. [Online]. Available:
  \url{https://sqa.stackexchange.com/questions/28204/how-to-deal-with-flaky-tests-that-have-intermittent-failures}
\BIBentrySTDinterwordspacing

\bibitem{G57}
\BIBentryALTinterwordspacing
D.~Andrawis, ``\BIBforeignlanguage{en-US}{Tips on how to reduce {Selenium}
  flaky tests},'' 2020. [Online]. Available:
  \url{https://www.shield34.com/blog/specific-tips-on-how-to-reduce-selenium-flaky-tests-that-every-coder-should-know/}
\BIBentrySTDinterwordspacing

\bibitem{G58}
\BIBentryALTinterwordspacing
J.~McCrary, ``\BIBforeignlanguage{en}{Using {Bazel} to help fix flaky tests},''
  2020. [Online]. Available:
  \url{https://jakemccrary.com/blog/2020/06/28/using-bazel-to-help-fix-flaky-tests/}
\BIBentrySTDinterwordspacing

\bibitem{G71}
\BIBentryALTinterwordspacing
K.~Chodorow, ``\BIBforeignlanguage{en}{Debugging flaky tests with {Bazel}},''
  2015. [Online]. Available:
  \url{https://kchodorow.com/2015/09/17/debugging-flaky-tests-with-bazel/}
\BIBentrySTDinterwordspacing

\bibitem{G59}
\BIBentryALTinterwordspacing
J.~Meadows, ``\BIBforeignlanguage{en}{Introducing flaky - a nose test plugin
  for automatically rerunning flaky tests},'' 2014. [Online]. Available:
  \url{https://blog.box.com/introducing-flaky-a-nose-test-plugin-for-automatically-rerunning-flaky-tests}
\BIBentrySTDinterwordspacing

\bibitem{G67}
\BIBentryALTinterwordspacing
``\BIBforeignlanguage{en-us}{Strategies for handling flaky test suites -
  {Redshiftzero}},'' 2019. [Online]. Available:
  \url{https://www.redshiftzero.com/test-flakes/}
\BIBentrySTDinterwordspacing

\bibitem{G68}
\BIBentryALTinterwordspacing
K.~Hoa, ``\BIBforeignlanguage{en-gb}{Flaky tests \& {Capybara} best
  practices},'' 2015. [Online]. Available:
  \url{https://www.simplybusiness.co.uk/about-us/tech/2015/02/flaky-tests-and-capybara-best-practices/}
\BIBentrySTDinterwordspacing

\bibitem{G70}
\BIBentryALTinterwordspacing
L.~Richter, ``\BIBforeignlanguage{en}{Handle flaky tests with quarantine and
  {Xunit}.{SkippableFact}},'' 2020. [Online]. Available:
  \url{https://dev.to/n_develop/handle-flaky-tests-with-quarantine-and-xunit-skippablefact-3a14}
\BIBentrySTDinterwordspacing

\bibitem{G81}
\BIBentryALTinterwordspacing
I.~Tse, ``\BIBforeignlanguage{en-US}{How {I} tracked down a flaky test},''
  2016. [Online]. Available:
  \url{https://www.paperlesspost.com/blog/teams/how-i-tracked-down-a-flaky-test/}
\BIBentrySTDinterwordspacing

\bibitem{G96}
\BIBentryALTinterwordspacing
K.~Shatrov, ``\BIBforeignlanguage{en-us}{Five ways to write a flaky test},''
  2016. [Online]. Available:
  \url{http://kirshatrov.com/2016/10/21/flaky-tests/}
\BIBentrySTDinterwordspacing

\bibitem{G108}
\BIBentryALTinterwordspacing
U.~Shah, ``\BIBforeignlanguage{en}{Athena: {Our} automated build health
  management system},'' 2019. [Online]. Available:
  \url{https://dropbox.tech/infrastructure/athena-our-automated-build-health-management-system}
\BIBentrySTDinterwordspacing

\bibitem{G116}
\BIBentryALTinterwordspacing
{Datadog}, ``\BIBforeignlanguage{en}{Flaky {Test} {Management}}.'' [Online].
  Available:
  \url{https://docs.datadoghq.com/continuous_integration/guides/flaky_test_management/}
\BIBentrySTDinterwordspacing

\bibitem{G122}
\BIBentryALTinterwordspacing
``\BIBforeignlanguage{en}{Introducing {Test} {Insights} with flaky test
  detection},'' Oct. 2021. [Online]. Available:
  \url{https://circleci.com/blog/introducing-test-insights-with-flaky-test-detection/}
\BIBentrySTDinterwordspacing

\bibitem{G140}
\BIBentryALTinterwordspacing
``Flaky test extractor maven plugin.'' [Online]. Available:
  \url{https://github.com/zeebe-io/flaky-test-extractor-maven-plugin}
\BIBentrySTDinterwordspacing

\bibitem{G213}
\BIBentryALTinterwordspacing
``\BIBforeignlanguage{en-US}{{eBay} {Launches} {Targeted} {Auto} {Retry}},''
  2021. [Online]. Available:
  \url{https://tech.ebayinc.com/engineering/ebay-launches-targeted-auto-retry/}
\BIBentrySTDinterwordspacing

\bibitem{G192}
\BIBentryALTinterwordspacing
``Maven {Surefire} {Plugin} – {Rerun} failing tests.'' [Online]. Available:
  \url{https://maven.apache.org/surefire/maven-surefire-plugin/examples/rerun-failing-tests.html}
\BIBentrySTDinterwordspacing

\bibitem{G142}
\BIBentryALTinterwordspacing
``Gradle {Enterprise} {Flaky} {Test} {Detection} {Guide} {\textbar} {Gradle}
  {Enterprise} {Docs}.'' [Online]. Available:
  \url{https://docs.gradle.com/enterprise/flaky-test-detection/}
\BIBentrySTDinterwordspacing

\bibitem{G149}
\BIBentryALTinterwordspacing
R.~Agarwal, ``\BIBforeignlanguage{en-US}{Handling {Flaky} {Unit} {Tests} in
  {Java}},'' Jun. 2021. [Online]. Available:
  \url{https://eng.uber.com/handling-flaky-tests-java/}
\BIBentrySTDinterwordspacing

\bibitem{G202}
A.~Klotz.

\bibitem{G203}
\BIBentryALTinterwordspacing
``\BIBforeignlanguage{en}{Katalon {Studio} 8.1 – {Ways} to {Handle} {Flaky}
  {Tests} {Smarter}}.'' [Online]. Available:
  \url{https://katalon.com/resources-center/blog/studio-8-1-handle-flaky-tests}
\BIBentrySTDinterwordspacing

\bibitem{G11}
\BIBentryALTinterwordspacing
L.~Eloussi, ``\BIBforeignlanguage{en}{Flaky {Tests} ({And} {How} {To} {Avoid}
  {Them})},'' 2016. [Online]. Available:
  \url{https://engineering.salesforce.com/flaky-tests-and-how-to-avoid-them-25b84b756f60}
\BIBentrySTDinterwordspacing

\bibitem{G52}
\BIBentryALTinterwordspacing
M.~Otonelli, ``\BIBforeignlanguage{en}{'{Flaky}' {Tests}: {A} {Short} {Story} -
  {The} {Lean} {Software} {Boutique}},'' 2017. [Online]. Available:
  \url{https://www.ombulabs.com/blog/rspec/continuous-integration/how-to-track-down-a-flaky-test.html}
\BIBentrySTDinterwordspacing

\bibitem{G95}
\BIBentryALTinterwordspacing
S.~Saffron, ``Tests that sometimes fail,'' 2019. [Online]. Available:
  \url{https://samsaffron.com/archive/2019/05/15/tests-that-sometimes-fail}
\BIBentrySTDinterwordspacing

\bibitem{G27}
\BIBentryALTinterwordspacing
``Managing {Test} {Flakiness}.'' [Online]. Available:
  \url{https://smartbear.com/resources/ebooks/managing-ui-test-flakiness/}
\BIBentrySTDinterwordspacing

\bibitem{G103}
\BIBentryALTinterwordspacing
J.~Listfield, ``\BIBforeignlanguage{en}{Where do our flaky tests come from?}''
  2017. [Online]. Available:
  \url{https://testing.googleblog.com/2017/04/where-do-our-flaky-tests-come-from.html}
\BIBentrySTDinterwordspacing

\bibitem{G114}
\BIBentryALTinterwordspacing
``\BIBforeignlanguage{en}{Flaky {Tests}: {Getting} {Rid} {Of} {A} {Living}
  {Nightmare} {In} {Testing}},'' Apr. 2021. [Online]. Available:
  \url{https://www.smashingmagazine.com/2021/04/flaky-tests-living-nightmare/}
\BIBentrySTDinterwordspacing

\bibitem{G210}
\BIBentryALTinterwordspacing
``\BIBforeignlanguage{en-US}{Tips on {Treating} {Flakiness} in your {Rails}
  {Test} {Suite}},'' Aug. 2017. [Online]. Available:
  \url{https://semaphoreci.com/blog/2017/08/03/tips-on-treating-flakiness-in-your-test-suite.html}
\BIBentrySTDinterwordspacing

\bibitem{G74}
\BIBentryALTinterwordspacing
J.~Grant, ``\BIBforeignlanguage{en}{What {Flaky} {Tests} {Can} {Tell} {You}},''
  2016. [Online]. Available:
  \url{https://www.stickyminds.com/article/what-flaky-tests-can-tell-you}
\BIBentrySTDinterwordspacing

\bibitem{G127}
\BIBentryALTinterwordspacing
``\BIBforeignlanguage{en-US}{How do you test your tests?}'' Dec. 2020.
  [Online]. Available:
  \url{https://engineering.fb.com/2020/12/10/developer-tools/probabilistic-flakiness/}
\BIBentrySTDinterwordspacing

\bibitem{G152}
\BIBentryALTinterwordspacing
``\BIBforeignlanguage{en-US}{10 {Reasons} for {Flaky} {Tests} {Automation} +
  {Tips} {\textbar} {Test} {Guild}},'' Dec. 2015. [Online]. Available:
  \url{https://testguild.com/top-10-reasons-for-flaky-automated-tests/}
\BIBentrySTDinterwordspacing

\bibitem{G165}
\BIBentryALTinterwordspacing
``\BIBforeignlanguage{en}{Fixing a flaky test - how hard can it be?}''
  [Online]. Available:
  \url{https://www.qt.io/blog/2017/05/12/fixing-a-flaky-test-how-hard-can-it-be}
\BIBentrySTDinterwordspacing

\bibitem{G107}
\BIBentryALTinterwordspacing
J.~Micco, ``The state of continuous integration testing @google,'' 2017.
  [Online]. Available: \url{https://research.google/pubs/pub45880/}
\BIBentrySTDinterwordspacing

\bibitem{G134}
\BIBentryALTinterwordspacing
``\BIBforeignlanguage{en-US}{Improving developer productivity via flaky test
  management},'' Feb. 2022. [Online]. Available:
  \url{https://devblogs.microsoft.com/engineering-at-microsoft/improving-developer-productivity-via-flaky-test-management/}
\BIBentrySTDinterwordspacing

\bibitem{G144}
\BIBentryALTinterwordspacing
{TestProject}, ``\BIBforeignlanguage{en-US}{Top {Reasons} \& {Solutions} {For}
  {Test} {Flakiness}},'' Aug. 2021. [Online]. Available:
  \url{https://blog.testproject.io/2021/08/25/top-reasons-and-solutions-for-test-flakiness/}
\BIBentrySTDinterwordspacing

\bibitem{G147}
\BIBentryALTinterwordspacing
``\BIBforeignlanguage{en-US}{Reducing flaky builds by 18x},'' Dec. 2020.
  [Online]. Available:
  \url{https://github.blog/2020-12-16-reducing-flaky-builds-by-18x/}
\BIBentrySTDinterwordspacing

\bibitem{G26}
\BIBentryALTinterwordspacing
P.~Sudarshan, ``\BIBforeignlanguage{en}{No more flaky tests on the {Go}
  team},'' 2012. [Online]. Available:
  \url{https://qa.webteam.thoughtworks.com/insights/blog/no-more-flaky-tests-go-team}
\BIBentrySTDinterwordspacing

\bibitem{G30}
\BIBentryALTinterwordspacing
J.~Yarn, ``\BIBforeignlanguage{en-US}{Flaky {Tests}: {The} {Tester}'s {F}
  {Word}},'' 2016. [Online]. Available:
  \url{https://www.lucidchart.com/techblog/2016/12/28/flaky-the-testers-f-word/}
\BIBentrySTDinterwordspacing

\bibitem{G129}
\BIBentryALTinterwordspacing
``\BIBforeignlanguage{en-US}{A {Pragmatist}'s {Guide} to {Flaky} {Test}
  {Management}},'' Aug. 2021. [Online]. Available:
  \url{https://gradle.com/blog/a-pragmatists-guide-to-flaky-test-management/}
\BIBentrySTDinterwordspacing

\bibitem{G154}
\BIBentryALTinterwordspacing
``\BIBforeignlanguage{en-US}{How to {Deal} with {Flaky} {Tests}},'' Mar. 2021.
  [Online]. Available:
  \url{https://thenewstack.io/how-to-deal-with-flaky-tests/}
\BIBentrySTDinterwordspacing

\bibitem{G45}
\BIBentryALTinterwordspacing
S.~Peterson, ``\BIBforeignlanguage{en}{Fixing {Flaky} {Tests} {Like} a
  {Detective} ({RailsConf} 2019)},'' 2019. [Online]. Available:
  \url{https://sonja.codes/fixing-flaky-tests-like-a-detective}
\BIBentrySTDinterwordspacing

\bibitem{G98}
\BIBentryALTinterwordspacing
{Testinium}, ``\BIBforeignlanguage{en-GB}{Flaky {Tests} and {How} to {Reduce}
  them},'' Sep. 2018. [Online]. Available:
  \url{https://testinium.com/blog/flaky-tests-and-how-to-reduce-them/}
\BIBentrySTDinterwordspacing

\bibitem{G110}
\BIBentryALTinterwordspacing
``\BIBforeignlanguage{en}{Test {Flakiness} - {One} of the main challenges of
  automated testing}.'' [Online]. Available:
  \url{https://testing.googleblog.com/2020/12/test-flakiness-one-of-main-challenges.html}
\BIBentrySTDinterwordspacing

\bibitem{G89}
\BIBentryALTinterwordspacing
A.~Sandhu, ``How to fix flaky tests,'' 2015. [Online]. Available:
  \url{https://tech.justeattakeaway.com/2015/03/30/how-to-fix-flaky-tests/}
\BIBentrySTDinterwordspacing

\bibitem{G7}
\BIBentryALTinterwordspacing
N.~Stri\^cevi\^c, ``\BIBforeignlanguage{en-US}{How to {Deal} {With} and
  {Eliminate} {Flaky} {Tests}},'' 2015. [Online]. Available:
  \url{https://semaphoreci.com/community/tutorials/how-to-deal-with-and-eliminate-flaky-tests}
\BIBentrySTDinterwordspacing

\bibitem{G4}
\BIBentryALTinterwordspacing
M.~Fowler, ``Eradicating {Non}-{Determinism} in {Tests},'' 2011. [Online].
  Available: \url{https://martinfowler.com/articles/nonDeterminism.html}
\BIBentrySTDinterwordspacing

\bibitem{G5}
\BIBentryALTinterwordspacing
``\BIBforeignlanguage{en}{Flaky {Tests} - {A} {War} that {Never} {Ends}
  {\textbar} {Hacker} {Noon}},'' 2017. [Online]. Available:
  \url{https://hackernoon.com/flaky-tests-a-war-that-never-ends-9aa32fdef359}
\BIBentrySTDinterwordspacing

\bibitem{G1}
\BIBentryALTinterwordspacing
J.~Micco, ``Flaky {Tests} at {Google} and {How} {We} {Mitigate} {Them},'' 2016.
  [Online]. Available:
  \url{https://testing.googleblog.com/2016/05/flaky-tests-at-google-and-how-we.html}
\BIBentrySTDinterwordspacing

\bibitem{G104}
\BIBentryALTinterwordspacing
{SamGu}, ``\BIBforeignlanguage{en-us}{Eliminating {Flaky} {Tests} - {Azure}
  {DevOps}},'' 2020. [Online]. Available:
  \url{https://docs.microsoft.com/en-us/azure/devops/learn/devops-at-microsoft/eliminating-flaky-tests}
\BIBentrySTDinterwordspacing

\bibitem{G106}
\BIBentryALTinterwordspacing
M.~Lapierre, ``\BIBforeignlanguage{en}{Pros and {Cons} of {Quarantined}
  {Tests}},'' 2017. [Online]. Available:
  \url{https://dev.to/mlapierre/pros-and-cons-of-quarantined-tests-2emj}
\BIBentrySTDinterwordspacing

\bibitem{G38}
\BIBentryALTinterwordspacing
J.~Hallam, ``\BIBforeignlanguage{en-us}{Avoiding {Flaky} {Tests}},'' 2019.
  [Online]. Available:
  \url{https://developers.mattermost.com/blog/avoiding-flaky-tests/}
\BIBentrySTDinterwordspacing

\bibitem{G79}
\BIBentryALTinterwordspacing
A.~McPeak, ``How to {Fix} a {Flaky} {Selenium} {Suite},'' 2018. [Online].
  Available:
  \url{https://smartbear.com/en/blog/eliminating-flaky-selenium-tests-forever/}
\BIBentrySTDinterwordspacing

\bibitem{G111}
\BIBentryALTinterwordspacing
``\BIBforeignlanguage{en-us}{Flaky tests {\textbar} {GitLab}}.'' [Online].
  Available:
  \url{https://docs.gitlab.com/ee/development/testing_guide/flaky_tests.html}
\BIBentrySTDinterwordspacing

\bibitem{G164}
\BIBentryALTinterwordspacing
``Testing in {Chromium} - {Fixing} web test flakiness.'' [Online]. Available:
  \url{https://chromium.googlesource.com/chromium/src/+/HEAD/docs/testing/identifying_tests_that_depend_on_order.md}
\BIBentrySTDinterwordspacing

\bibitem{G20}
\BIBentryALTinterwordspacing
M.~Rushakoff, ``\BIBforeignlanguage{en-US}{Reproducing a {Flaky} {Test} in {Go}
  {\textbar} {Blog}},'' 2019. [Online]. Available:
  \url{https://www.influxdata.com/blog/reproducing-a-flaky-test-in-go/}
\BIBentrySTDinterwordspacing

\bibitem{G101}
\BIBentryALTinterwordspacing
D.~Stosik, ``\BIBforeignlanguage{en}{A {Methodological} {Approach} to {Fixing}
  flaky tests},'' Mar. 2020. [Online]. Available:
  \url{https://sourcediving.com/a-methodological-approach-to-fixing-flaky-tests-92a39162b769}
\BIBentrySTDinterwordspacing

\bibitem{G102}
\BIBentryALTinterwordspacing
P.~Raki\^c, ``\BIBforeignlanguage{en-US}{Flaky {Tests}: {Are} {You} {Sure}
  {You} {Want} to {Rerun} {Them}?}'' Apr. 2017. [Online]. Available:
  \url{https://semaphoreci.com/blog/2017/04/20/flaky-tests.html}
\BIBentrySTDinterwordspacing

\bibitem{G37}
\BIBentryALTinterwordspacing
R.~Raposa, ``Flaky {Test} {Process} - {Test} {Engineering} - {Confluence},''
  2020. [Online]. Available:
  \url{https://openedx.atlassian.net/wiki/spaces/TE/pages/161427235/Flaky+Test+Process}
\BIBentrySTDinterwordspacing

\bibitem{G189}
\BIBentryALTinterwordspacing
``\BIBforeignlanguage{en}{How to {Debug} {Non}-{Deterministic} {Test}
  {Failures} {With} {RSpec} - {FastRuby}.io {\textbar} {Rails} {Upgrade}
  {Service}},'' Sep. 2019. [Online]. Available:
  \url{https://fastruby.io/blog/rspec/debug/how-to-debug-non-deterministic-specs.html}
\BIBentrySTDinterwordspacing

\bibitem{G75}
\BIBentryALTinterwordspacing
S.~Berczuk, ``\BIBforeignlanguage{en}{Defensive {Design} {Strategies} to
  {Prevent} {Flaky} {Tests}},'' 2020. [Online]. Available:
  \url{https://www.techwell.com/techwell-insights/2020/04/defensive-design-strategies-prevent-flaky-tests}
\BIBentrySTDinterwordspacing

\end{thebibliography}


\begin{thebibliography}{100}
\providecommand{\url}[1]{#1}
\csname url@samestyle\endcsname
\providecommand{\newblock}{\relax}
\providecommand{\bibinfo}[2]{#2}
\providecommand{\BIBentrySTDinterwordspacing}{\spaceskip=0pt\relax}
\providecommand{\BIBentryALTinterwordstretchfactor}{4}
\providecommand{\BIBentryALTinterwordspacing}{\spaceskip=\fontdimen2\font plus
\BIBentryALTinterwordstretchfactor\fontdimen3\font minus
  \fontdimen4\font\relax}
\providecommand{\BIBforeignlanguage}[2]{{%
\expandafter\ifx\csname l@#1\endcsname\relax
\typeout{** WARNING: IEEEtran.bst: No hyphenation pattern has been}%
\typeout{** loaded for the language `#1'. Using the pattern for}%
\typeout{** the default language instead.}%
\else
\language=\csname l@#1\endcsname
\fi
#2}}
\providecommand{\BIBdecl}{\relax}
\BIBdecl

\bibitem{S1}
Q.~Luo, F.~Hariri, L.~Eloussi, and D.~Marinov, ``An empirical analysis of flaky
  tests,'' in \emph{Proceedings of the 22nd ACM SIGSOFT International Symposium
  on Foundations of Software Engineering}, 2014, pp. 643--653.

\bibitem{S7}
S.~Thorve, C.~Sreshtha, and N.~Meng, ``An empirical study of flaky tests in
  android apps,'' in \emph{2018 IEEE International Conference on Software
  Maintenance and Evolution (ICSME)}, 2018, pp. 534--538.

\bibitem{S14}
S.~Dutta, A.~Shi, R.~Choudhary, Z.~Zhang, A.~Jain, and S.~Misailovic,
  ``Detecting flaky tests in probabilistic and machine learning applications,''
  in \emph{Proceedings of the 29th ACM SIGSOFT International Symposium on
  Software Testing and Analysis}.\hskip 1em plus 0.5em minus 0.4em\relax
  Association for Computing Machinery, 2020, p. 211–224.

\bibitem{S10}
\BIBentryALTinterwordspacing
W.~Lam, K.~Mu\c{s}lu, H.~Sajnani, and S.~Thummalapenta, ``A study on the
  lifecycle of flaky tests,'' in \emph{Proceedings of the ACM/IEEE 42nd
  International Conference on Software Engineering}, ser. ICSE '20.\hskip 1em
  plus 0.5em minus 0.4em\relax New York, NY, USA: Association for Computing
  Machinery, 2020, p. 1471–1482. [Online]. Available:
  \url{https://doi.org/10.1145/3377811.3381749}
\BIBentrySTDinterwordspacing

\bibitem{S6}
\BIBentryALTinterwordspacing
W.~Lam, P.~Godefroid, S.~Nath, A.~Santhiar, and S.~Thummalapenta, ``Root
  causing flaky tests in a large-scale industrial setting,'' in
  \emph{Proceedings of the 28th ACM SIGSOFT International Symposium on Software
  Testing and Analysis}, ser. ISSTA 2019.\hskip 1em plus 0.5em minus
  0.4em\relax New York, NY, USA: Association for Computing Machinery, 2019, p.
  101–111. [Online]. Available: \url{https://doi.org/10.1145/3293882.3330570}
\BIBentrySTDinterwordspacing

\bibitem{S15}
\BIBentryALTinterwordspacing
W.~Lam, S.~Winter, A.~Wei, T.~Xie, D.~Marinov, and J.~Bell, ``A large-scale
  longitudinal study of flaky tests,'' \emph{Proc. ACM Program. Lang.}, vol.~4,
  2020. [Online]. Available: \url{https://doi.org/10.1145/3428270}
\BIBentrySTDinterwordspacing

\bibitem{S72}
\BIBentryALTinterwordspacing
J.~Malm, A.~Causevic, B.~Lisper, and S.~Eldh, ``Automated analysis of
  flakiness-mitigating delays,'' in \emph{Proceedings of the IEEE/ACM 1st
  International Conference on Automation of Software Test}, ser. AST '20.\hskip
  1em plus 0.5em minus 0.4em\relax New York, NY, USA: Association for Computing
  Machinery, 2020, p. 81–84. [Online]. Available:
  \url{https://doi.org/10.1145/3387903.3389320}
\BIBentrySTDinterwordspacing

\bibitem{S1008}
\BIBentryALTinterwordspacing
Z.~Dong, A.~Tiwari, X.~L. Yu, and A.~Roychoudhury, ``Flaky test detection in
  android via event order exploration,'' in \emph{Proceedings of the 29th ACM
  Joint Meeting on European Software Engineering Conference and Symposium on
  the Foundations of Software Engineering}, ser. ESEC/FSE 2021.\hskip 1em plus
  0.5em minus 0.4em\relax New York, NY, USA: Association for Computing
  Machinery, 2021, p. 367–378. [Online]. Available:
  \url{https://doi.org/10.1145/3468264.3468584}
\BIBentrySTDinterwordspacing

\bibitem{S210}
\BIBentryALTinterwordspacing
P.~E. Strandberg, T.~J. Ostrand, E.~J. Weyuker, W.~Afzal, and D.~Sundmark,
  ``Intermittently failing tests in the embedded systems domain,'' in
  \emph{Proceedings of the 29th ACM SIGSOFT International Symposium on Software
  Testing and Analysis}, ser. ISSTA 2020.\hskip 1em plus 0.5em minus
  0.4em\relax New York, NY, USA: Association for Computing Machinery, 2020, p.
  337–348. [Online]. Available: \url{https://doi.org/10.1145/3395363.3397359}
\BIBentrySTDinterwordspacing

\bibitem{S147}
A.~Gambi, J.~Bell, and A.~Zeller, ``Practical test dependency detection,'' in
  \emph{2018 IEEE 11th International Conference on Software Testing,
  Verification and Validation (ICST)}, 2018, pp. 1--11.

\bibitem{S4}
W.~Lam, R.~Oei, A.~Shi, D.~Marinov, and T.~Xie, ``idflakies: A framework for
  detecting and partially classifying flaky tests,'' in \emph{2019 12th IEEE
  Conference on Software Testing, Validation and Verification (ICST)}, 2019,
  pp. 312--322.

\bibitem{S110}
S.~Paydar and A.~Azamnouri, ``An experimental study on flakiness and fragility
  of randoop regression test suites,'' in \emph{International Conference on
  Fundamentals of Software Engineering}.\hskip 1em plus 0.5em minus 0.4em\relax
  Springer, 2019, pp. 111--126.

\bibitem{S99}
J.~Mor{\'a}n~Barb{\'o}n, C.~Augusto~Alonso, A.~Bertolino, C.~A.
  Riva~{\'A}lvarez, P.~J. Tuya~Gonz{\'a}lez \emph{et~al.}, ``Flakyloc:
  flakiness localization for reliable test suites in web applications,''
  \emph{Journal of Web Engineering, 2}, 2020.

\bibitem{S8}
\BIBentryALTinterwordspacing
M.~Eck, F.~Palomba, M.~Castelluccio, and A.~Bacchelli, ``Understanding flaky
  tests: The developer’s perspective,'' in \emph{Proceedings of the 2019 27th
  ACM Joint Meeting on European Software Engineering Conference and Symposium
  on the Foundations of Software Engineering}, ser. ESEC/FSE 2019.\hskip 1em
  plus 0.5em minus 0.4em\relax New York, NY, USA: Association for Computing
  Machinery, 2019, p. 830–840. [Online]. Available:
  \url{https://doi.org/10.1145/3338906.3338945}
\BIBentrySTDinterwordspacing

\bibitem{S16}
Z.~Dong, A.~Tiwari, X.~L. Yu, and A.~Roychoudhury, ``Concurrency-related flaky
  test detection in android apps,'' 2021.

\bibitem{S93}
A.~T. Endo and A.~Møller, ``Noderacer: Event race detection for node.js
  applications,'' in \emph{2020 IEEE 13th International Conference on Software
  Testing, Validation and Verification (ICST)}, 2020, pp. 120--130.

\bibitem{S357}
S.~Person and S.~Elbaum, ``Test analysis: Searching for faults in tests (n),''
  in \emph{2015 30th IEEE/ACM International Conference on Automated Software
  Engineering (ASE)}.\hskip 1em plus 0.5em minus 0.4em\relax IEEE, 2015, pp.
  149--154.

\bibitem{S545}
S.~L. Eddins, ``Automated software testing for matlab,'' \emph{Computing in
  Science Engineering}, vol.~11, no.~6, pp. 48--55, 2009.

\bibitem{S217}
\BIBentryALTinterwordspacing
E.~O. Scott and S.~Luke, ``Ecj at 20: Toward a general metaheuristics
  toolkit,'' in \emph{Proceedings of the Genetic and Evolutionary Computation
  Conference Companion}, ser. GECCO '19.\hskip 1em plus 0.5em minus 0.4em\relax
  New York, NY, USA: Association for Computing Machinery, 2019, p. 1391–1398.
  [Online]. Available: \url{https://doi.org/10.1145/3319619.3326865}
\BIBentrySTDinterwordspacing

\bibitem{S548}
T.~M{\aa}rtensson, D.~St{\aa}hl, and J.~Bosch, ``Continuous integration applied
  to software-intensive embedded systems -- problems and experiences,'' in
  \emph{Product-Focused Software Process Improvement}, P.~Abrahamsson,
  A.~Jedlitschka, A.~Nguyen~Duc, M.~Felderer, S.~Amasaki, and T.~Mikkonen,
  Eds.\hskip 1em plus 0.5em minus 0.4em\relax Cham: Springer International
  Publishing, 2016, pp. 448--457.

\bibitem{S42}
J.~Mor{\'a}n~Barb{\'o}n, C.~Augusto~Alonso, A.~Bertolino, C.~A.
  Riva~{\'A}lvarez, J.~F. Garc{\'\i}a~Tuya \emph{et~al.}, ``Debugging flaky
  tests on web applications,'' in \emph{Proceedings of the 15th International
  Conference on Web Information Systems and Technologies-Volume 1: APMDWE},
  2019.

\bibitem{S1019}
M.~Gruber and G.~Fraser, ``A survey on how test flakiness affects developers
  and what support they need to address it,'' \emph{arXiv preprint
  arXiv:2203.00483}, 2022.

\bibitem{S1007}
M.~Gruber, S.~Lukasczyk, F.~Kroiß, and G.~Fraser, ``An empirical study of
  flaky tests in python,'' in \emph{2021 14th IEEE Conference on Software
  Testing, Verification and Validation (ICST)}, 2021, pp. 148--158.

\bibitem{S23}
M.~A. Mascheroni and E.~Irraz{\'a}bal, ``Identifying key success factors in
  stopping flaky tests in automated rest service testing,'' \emph{Journal of
  Computer Science and Technology}, vol.~18, no.~02, pp. e16--e16, 2018.

\bibitem{S39}
A.~Berglund and O.~Vateman, ``Mitigation and handling of non-deterministic
  tests in automatic regression testing,'' \emph{LU-CS-EX}, 2020.

\bibitem{S713}
D.~J.~G. Mendes, ``Automated testing for provisioning systems of complex cloud
  products,'' Ph.D. dissertation, 2019.

\bibitem{S1025}
S.~Habchi, G.~Haben, M.~Papadakis, M.~Cordy, and Y.~L. Traon, ``A qualitative
  study on the sources, impacts, and mitigation strategies of flaky tests,''
  \emph{arXiv preprint arXiv:2112.04919}, 2021.

\bibitem{S258}
\BIBentryALTinterwordspacing
M.~Ivankovi\'{c}, G.~Petrovi\'{c}, R.~Just, and G.~Fraser, ``Code coverage at
  google,'' in \emph{Proceedings of the 2019 27th ACM Joint Meeting on European
  Software Engineering Conference and Symposium on the Foundations of Software
  Engineering}, ser. ESEC/FSE 2019.\hskip 1em plus 0.5em minus 0.4em\relax New
  York, NY, USA: Association for Computing Machinery, 2019, p. 955–963.
  [Online]. Available: \url{https://doi.org/10.1145/3338906.3340459}
\BIBentrySTDinterwordspacing

\bibitem{S28}
\BIBentryALTinterwordspacing
O.~Parry, G.~M. Kapfhammer, M.~Hilton, and P.~McMinn, ``Flake it 'till you make
  it: Using automated repair to induce and fix latent test flakiness,'' in
  \emph{Proceedings of the IEEE/ACM 42nd International Conference on Software
  Engineering Workshops}, ser. ICSEW'20.\hskip 1em plus 0.5em minus 0.4em\relax
  New York, NY, USA: Association for Computing Machinery, 2020, p. 11–12.
  [Online]. Available: \url{https://doi.org/10.1145/3387940.3392177}
\BIBentrySTDinterwordspacing

\bibitem{S2}
J.~Bell, O.~Legunsen, M.~Hilton, L.~Eloussi, T.~Yung, and D.~Marinov,
  ``Deflaker: Automatically detecting flaky tests,'' in \emph{2018 IEEE/ACM
  40th International Conference on Software Engineering (ICSE)}.\hskip 1em plus
  0.5em minus 0.4em\relax IEEE, 2018, pp. 433--444.

\bibitem{S607}
A.~Ahmad, O.~Leifler, and K.~Sandahl, ``An evaluation of machine learning
  methods for predicting flaky tests,'' in \emph{QuASoQ@APSEC}, 2020.

\bibitem{S118}
R.~Verdecchia, E.~Cruciani, B.~Miranda, and A.~Bertolino, ``Know you neighbor:
  Fast static prediction of test flakiness,'' \emph{IEEE Access}, 2021.

\bibitem{S31}
T.~M. King, D.~Santiago, J.~Phillips, and P.~J. Clarke, ``Towards a bayesian
  network model for predicting flaky automated tests,'' in \emph{2018 IEEE
  International Conference on Software Quality, Reliability and Security
  Companion (QRS-C)}, 2018, pp. 100--107.

\bibitem{SB1}
K.~Herzig and N.~Nagappan, ``Empirically detecting false test alarms using
  association rules,'' in \emph{2015 IEEE/ACM 37th IEEE International
  Conference on Software Engineering}, vol.~2, 2015, pp. 39--48.

\bibitem{SB28}
\BIBentryALTinterwordspacing
A.~Gyori, B.~Lambeth, S.~Khurshid, and D.~Marinov, ``Exploring underdetermined
  specifications using java pathfinder,'' \emph{SIGSOFT Softw. Eng. Notes},
  vol.~41, no.~6, p. 1–5, jan 2017. [Online]. Available:
  \url{https://doi.org/10.1145/3011286.3011295}
\BIBentrySTDinterwordspacing

\bibitem{S11}
G.~Pinto, B.~Miranda, S.~Dissanayake, M.~d'Amorim, C.~Treude, and A.~Bertolino,
  ``What is the vocabulary of flaky tests?'' in \emph{Proceedings of the 17th
  International Conference on Mining Software Repositories}, 2020, pp.
  492--502.

\bibitem{S591}
S.~Zhang, D.~Jalali, J.~Wuttke, K.~Mu{\c{s}}lu, W.~Lam, M.~D. Ernst, and
  D.~Notkin, ``Empirically revisiting the test independence assumption,'' in
  \emph{Proceedings of the 2014 International Symposium on Software Testing and
  Analysis}, 2014, pp. 385--396.

\bibitem{S1014}
A.~Wei, P.~Yi, T.~Xie, D.~Marinov, and W.~Lam, ``Probabilistic and systematic
  coverage of consecutive test-method pairs for detecting order-dependent flaky
  tests,'' in \emph{International Conference on Tools and Algorithms for the
  Construction and Analysis of Systems}.\hskip 1em plus 0.5em minus 0.4em\relax
  Springer, 2021, pp. 270--287.

\bibitem{S1011}
A.~Wei, P.~Yi, Z.~Li, T.~Xie, D.~Marinov, and W.~Lam, ``Preempting flaky tests
  via non-idempotent-outcome tests,'' in \emph{International Conference on
  Software Engineering (ICSE’22)}, 2022.

\bibitem{S1027}
A.~Alshammari, C.~Morris, M.~Hilton, and J.~Bell, ``Flakeflagger: Predicting
  flakiness without rerunning tests,'' in \emph{2021 IEEE/ACM 43rd
  International Conference on Software Engineering (ICSE)}, 2021, pp.
  1572--1584.

\bibitem{S27}
L.~Eloussi, ``Determining flaky tests from test failures,'' 2015.

\bibitem{S24}
D.~Silva, L.~Teixeira, and M.~d’Amorim, ``Shake it! detecting flaky tests
  caused by concurrency with shaker,'' in \emph{2020 IEEE International
  Conference on Software Maintenance and Evolution (ICSME)}, 2020, pp.
  301--311.

\bibitem{S78}
S.~V. Vaidhyam~Subramanian, S.~McIntosh, and B.~Adams, ``Quantifying,
  characterizing, and mitigating flakily covered program elements,'' \emph{IEEE
  Transactions on Software Engineering}, pp. 1--1, 2020.

\bibitem{S152}
\BIBentryALTinterwordspacing
A.~Gyori, B.~Lambeth, A.~Shi, O.~Legunsen, and D.~Marinov, ``Nondex: A tool for
  detecting and debugging wrong assumptions on java api specifications,'' in
  \emph{Proceedings of the 2016 24th ACM SIGSOFT International Symposium on
  Foundations of Software Engineering}, ser. FSE 2016.\hskip 1em plus 0.5em
  minus 0.4em\relax New York, NY, USA: Association for Computing Machinery,
  2016, p. 993–997. [Online]. Available:
  \url{https://doi.org/10.1145/2950290.2983932}
\BIBentrySTDinterwordspacing

\bibitem{S135}
A.~W. Shi, ``Improving regression testing efficiency and reliability via
  test-suite transformations,'' Ph.D. dissertation, University of Illinois at
  Urbana-Champaign, 2020.

\bibitem{S25}
C.~Ziftci and D.~Cavalcanti, ``De-flake your tests : Automatically locating
  root causes of flaky tests in code at google,'' in \emph{2020 IEEE
  International Conference on Software Maintenance and Evolution (ICSME)},
  2020, pp. 736--745.

\bibitem{S460}
A.~Gyori, A.~Shi, F.~Hariri, and D.~Marinov, ``Reliable testing: Detecting
  state-polluting tests to prevent test dependency,'' in \emph{Proceedings of
  the 2015 International Symposium on Software Testing and Analysis}, 2015, pp.
  223--233.

\bibitem{S1022}
\BIBentryALTinterwordspacing
S.~Fatima, T.~A. Ghaleb, and L.~Briand, ``Flakify: A black-box, language
  model-based predictor for flaky tests,'' 2021. [Online]. Available:
  \url{https://arxiv.org/abs/2112.12331}
\BIBentrySTDinterwordspacing

\bibitem{S19}
V.~Terragni, P.~Salza, and F.~Ferrucci, ``A container-based infrastructure for
  fuzzy-driven root causing of flaky tests,'' in \emph{Proceedings of the
  ACM/IEEE 42nd International Conference on Software Engineering: New Ideas and
  Emerging Results}, 2020, pp. 69--72.

\bibitem{S274}
\BIBentryALTinterwordspacing
M.~Biagiola, A.~Stocco, A.~Mesbah, F.~Ricca, and P.~Tonella, ``Web test
  dependency detection,'' in \emph{Proceedings of the 2019 27th ACM Joint
  Meeting on European Software Engineering Conference and Symposium on the
  Foundations of Software Engineering}, ser. ESEC/FSE 2019.\hskip 1em plus
  0.5em minus 0.4em\relax New York, NY, USA: Association for Computing
  Machinery, 2019, p. 154–164. [Online]. Available:
  \url{https://doi.org/10.1145/3338906.3338948}
\BIBentrySTDinterwordspacing

\bibitem{S780}
\BIBentryALTinterwordspacing
O.~Schwahn, N.~Coppik, S.~Winter, and N.~Suri, ``Assessing the state and
  improving the art of parallel testing for c,'' in \emph{Proceedings of the
  28th ACM SIGSOFT International Symposium on Software Testing and Analysis},
  ser. ISSTA 2019.\hskip 1em plus 0.5em minus 0.4em\relax New York, NY, USA:
  Association for Computing Machinery, 2019, p. 123–133. [Online]. Available:
  \url{https://doi.org/10.1145/3293882.3330573}
\BIBentrySTDinterwordspacing

\bibitem{S391}
J.~Bell, G.~Kaiser, E.~Melski, and M.~Dattatreya, ``Efficient dependency
  detection for safe java test acceleration,'' in \emph{Proceedings of the 2015
  10th Joint Meeting on Foundations of Software Engineering}, 2015, pp.
  770--781.

\bibitem{S1031}
R.~Wang, Y.~Chen, and W.~Lam, ``ipflakies: A framework for detecting and fixing
  python order-dependent flaky tests,'' 2022.

\bibitem{S1050}
G.~Haben, S.~Habchi, M.~Papadakis, M.~Cordy, and Y.~L. Traon, ``Discerning
  legitimate failures from false alerts: A study of chromium's continuous
  integration,'' 2021.

\bibitem{S1051}
\BIBentryALTinterwordspacing
P.~Yi, A.~Wei, W.~Lam, T.~Xie, and D.~Marinov, ``Finding polluter tests using
  java pathfinder,'' \emph{SIGSOFT Softw. Eng. Notes}, vol.~46, no.~3, p.
  37–41, jul 2021. [Online]. Available:
  \url{https://doi.org/10.1145/3468744.3468756}
\BIBentrySTDinterwordspacing

\bibitem{S1054}
R.~Mudduluru, J.~Waataja, S.~Millstein, and M.~Ernst, ``Verifying determinism
  in sequential programs,'' in \emph{2021 IEEE/ACM 43rd International
  Conference on Software Engineering (ICSE)}, 2021, pp. 37--49.

\bibitem{S1004}
\BIBentryALTinterwordspacing
V.~Pontillo, F.~Palomba, and F.~Ferrucci, ``Toward static test flakiness
  prediction: A feasibility study,'' in \emph{Proceedings of the 5th
  International Workshop on Machine Learning Techniques for Software Quality
  Evolution}, ser. MaLTESQuE 2021.\hskip 1em plus 0.5em minus 0.4em\relax New
  York, NY, USA: Association for Computing Machinery, 2021, p. 19–24.
  [Online]. Available: \url{https://doi.org/10.1145/3472674.3473981}
\BIBentrySTDinterwordspacing

\bibitem{S1005}
G.~Haben, S.~Habchi, M.~Papadakis, M.~Cordy, and Y.~Le~Traon, ``A replication
  study on the usability of code vocabulary in predicting flaky tests,'' in
  \emph{2021 IEEE/ACM 18th International Conference on Mining Software
  Repositories (MSR)}, 2021, pp. 219--229.

\bibitem{S1012}
B.~H.~P. Camara, M.~A.~G. Silva, A.~T. Endo, and S.~R. Vergilio, ``What is the
  vocabulary of flaky tests? an extended replication,'' in \emph{2021 IEEE/ACM
  29th International Conference on Program Comprehension (ICPC)}, 2021, pp.
  444--454.

\bibitem{S1034}
O.~Parry, G.~M. Kapfhammer, M.~Hilton, and P.~McMinn, ``Evaluating features for
  machine learning detection of order-and non-order-dependent flaky tests.''

\bibitem{S1017}
\BIBentryALTinterwordspacing
B.~Camara, M.~Silva, A.~Endo, and S.~Vergilio, \emph{On the Use of Test Smells
  for Prediction of Flaky Tests}.\hskip 1em plus 0.5em minus 0.4em\relax New
  York, NY, USA: Association for Computing Machinery, 2021, p. 46–54.
  [Online]. Available: \url{https://doi.org/10.1145/3482909.3482916}
\BIBentrySTDinterwordspacing

\bibitem{S12}
F.~Palomba and A.~Zaidman, ``Notice of retraction: Does refactoring of test
  smells induce fixing flaky tests?'' in \emph{2017 IEEE international
  conference on software maintenance and evolution (ICSME)}.\hskip 1em plus
  0.5em minus 0.4em\relax IEEE, 2017, pp. 1--12.

\bibitem{S136}
\BIBentryALTinterwordspacing
T.~Durieux, C.~Le~Goues, M.~Hilton, and R.~Abreu, ``Empirical study of
  restarted and flaky builds on travis ci,'' in \emph{Proceedings of the 17th
  International Conference on Mining Software Repositories}, ser. MSR
  '20.\hskip 1em plus 0.5em minus 0.4em\relax New York, NY, USA: Association
  for Computing Machinery, 2020, p. 254–264. [Online]. Available:
  \url{https://doi.org/10.1145/3379597.3387460}
\BIBentrySTDinterwordspacing

\bibitem{S330}
M.~Linares-V{\'a}squez, K.~Moran, and D.~Poshyvanyk, ``Continuous, evolutionary
  and large-scale: A new perspective for automated mobile app testing,'' in
  \emph{2017 IEEE International Conference on Software Maintenance and
  Evolution (ICSME)}.\hskip 1em plus 0.5em minus 0.4em\relax IEEE, 2017, pp.
  399--410.

\bibitem{S57}
M.~B{\"o}hme, ``Assurances in software testing: A roadmap,'' in \emph{2019
  IEEE/ACM 41st International Conference on Software Engineering: New Ideas and
  Emerging Results (ICSE-NIER)}.\hskip 1em plus 0.5em minus 0.4em\relax IEEE,
  2019, pp. 5--8.

\bibitem{S430}
A.~Lassila \emph{et~al.}, ``Opportunities and challenges in adopting continuous
  end-to-end testing: A case study,'' 2019.

\bibitem{S40}
\BIBentryALTinterwordspacing
W.~Lam, A.~Shi, R.~Oei, S.~Zhang, M.~D. Ernst, and T.~Xie,
  ``Dependent-test-aware regression testing techniques,'' in \emph{Proceedings
  of the 29th ACM SIGSOFT International Symposium on Software Testing and
  Analysis}, ser. ISSTA 2020.\hskip 1em plus 0.5em minus 0.4em\relax New York,
  NY, USA: Association for Computing Machinery, 2020, p. 298–311. [Online].
  Available: \url{https://doi.org/10.1145/3395363.3397364}
\BIBentrySTDinterwordspacing

\bibitem{S1156}
\BIBentryALTinterwordspacing
M.~Abdi, H.~Rocha, S.~Demeyer, and A.~Bergel, ``Small-amp: Test amplification
  in a dynamically typed language,'' 2021. [Online]. Available:
  \url{https://arxiv.org/abs/2108.05663}
\BIBentrySTDinterwordspacing

\bibitem{S1041}
J.~Ahlgren, M.~Berezin, K.~Bojarczuk, E.~Dulskyte, I.~Dvortsova, J.~George,
  N.~Gucevska, M.~Harman, M.~Lomeli, E.~Meijer, S.~Sapora, and
  J.~Spahr-Summers, ``Testing web enabled simulation at scale using metamorphic
  testing,'' in \emph{2021 IEEE/ACM 43rd International Conference on Software
  Engineering: Software Engineering in Practice (ICSE-SEIP)}, 2021, pp.
  140--149.

\bibitem{S1081}
\BIBentryALTinterwordspacing
R.~Haas, D.~Elsner, E.~Juergens, A.~Pretschner, and S.~Apel, \emph{How Can
  Manual Testing Processes Be Optimized? Developer Survey, Optimization
  Guidelines, and Case Studies}.\hskip 1em plus 0.5em minus 0.4em\relax New
  York, NY, USA: Association for Computing Machinery, 2021, p. 1281–1291.
  [Online]. Available: \url{https://doi.org/10.1145/3468264.3473922}
\BIBentrySTDinterwordspacing

\bibitem{S336}
A.~Vehabovic, ``The process of changing out expandable elements in a
  large-scale web application,'' 2020.

\bibitem{S256}
\BIBentryALTinterwordspacing
D.~G. Widder, M.~Hilton, C.~K\"{a}stner, and B.~Vasilescu, ``A conceptual
  replication of continuous integration pain points in the context of travis
  ci,'' in \emph{Proceedings of the 2019 27th ACM Joint Meeting on European
  Software Engineering Conference and Symposium on the Foundations of Software
  Engineering}, ser. ESEC/FSE 2019.\hskip 1em plus 0.5em minus 0.4em\relax New
  York, NY, USA: Association for Computing Machinery, 2019, p. 647–658.
  [Online]. Available: \url{https://doi.org/10.1145/3338906.3338922}
\BIBentrySTDinterwordspacing

\bibitem{S429}
U.~Zdun, E.~Wittern, and P.~Leitner, ``Emerging trends, challenges, and
  experiences in devops and microservice apis,'' \emph{IEEE Software}, vol.~37,
  no.~1, pp. 87--91, 2019.

\bibitem{S197}
T.~Hirsch, C.~Schindler, M.~M{\"u}ller, T.~Schranz, and W.~Slany, ``An approach
  to test classification in big android applications,'' in \emph{2019 IEEE 19th
  International Conference on Software Quality, Reliability and Security
  Companion (QRS-C)}.\hskip 1em plus 0.5em minus 0.4em\relax IEEE, 2019, pp.
  300--308.

\bibitem{S556}
\BIBentryALTinterwordspacing
C.~Vassallo, S.~Proksch, A.~Jancso, H.~C. Gall, and M.~Di~Penta,
  ``Configuration smells in continuous delivery pipelines: A linter and a
  six-month study on gitlab,'' in \emph{Proceedings of the 28th ACM Joint
  Meeting on European Software Engineering Conference and Symposium on the
  Foundations of Software Engineering}, ser. ESEC/FSE 2020.\hskip 1em plus
  0.5em minus 0.4em\relax New York, NY, USA: Association for Computing
  Machinery, 2020, p. 327–337. [Online]. Available:
  \url{https://doi.org/10.1145/3368089.3409709}
\BIBentrySTDinterwordspacing

\bibitem{S1043}
M.~A. Mascheroni, E.~Irrazábal, and G.~Rossi, ``Continuous testing improvement
  model,'' in \emph{2021 IEEE/ACM International Conference on Automation of
  Software Test (AST)}, 2021, pp. 109--112.

\bibitem{S750}
J.~Koivuniemi, ``Shortening feedback time in continuous integration environment
  in large-scale embedded software development with test selection,''
  \emph{University of Oulu repository}, pp. 16--18, 2017.

\bibitem{S252}
M.~Martinez, T.~Durieux, R.~Sommerard, J.~Xuan, and M.~Monperrus, ``Automatic
  repair of real bugs in java: A large-scale experiment on the defects4j
  dataset,'' \emph{Empirical Software Engineering}, vol.~22, no.~4, pp.
  1936--1964, 2017.

\bibitem{S962}
J.~Chen, W.~Shang, and E.~Shihab, ``Perfjit: Test-level just-in-time prediction
  for performance regression introducing commits,'' \emph{IEEE Transactions on
  Software Engineering}, 2020.

\bibitem{S1049}
B.~Dorward, C.~Johnston, E.~Nickell, and T.~Henderson, ``Flake-aware culprit
  finding,'' 2021.

\bibitem{S143}
W.~Chan, M.~Cheng, S.~Cheung, and T.~Tse, ``Automated goal-oriented
  classification of failure behaviors for testing xml-based multimedia software
  applications: An experimental case study,'' \emph{Quality control and applied
  statistics}, vol.~52, no.~1, pp. 113--114, 2007.

\bibitem{S395}
S.~Cox and N.~Chen, ``Improving client side web testing automation in
  continuous integration-a case study,'' in \emph{Proceedings of the
  International Conference on Software Engineering Research and Practice
  (SERP)}.\hskip 1em plus 0.5em minus 0.4em\relax The Steering Committee of The
  World Congress in Computer Science, Computer~…, 2019, pp. 41--47.

\bibitem{S159}
J.~Candido, L.~Melo, and M.~d'Amorim, ``Test suite parallelization in
  open-source projects: a study on its usage and impact,'' in \emph{2017 32nd
  IEEE/ACM International Conference on Automated Software Engineering
  (ASE)}.\hskip 1em plus 0.5em minus 0.4em\relax IEEE, 2017, pp. 838--848.

\bibitem{S662}
\BIBentryALTinterwordspacing
A.~Shi, A.~Gyori, S.~Mahmood, P.~Zhao, and D.~Marinov, ``Evaluating test-suite
  reduction in real software evolution,'' in \emph{Proceedings of the 27th ACM
  SIGSOFT International Symposium on Software Testing and Analysis}, ser. ISSTA
  2018.\hskip 1em plus 0.5em minus 0.4em\relax New York, NY, USA: Association
  for Computing Machinery, 2018, p. 84–94. [Online]. Available:
  \url{https://doi.org/10.1145/3213846.3213875}
\BIBentrySTDinterwordspacing

\bibitem{S444}
\BIBentryALTinterwordspacing
D.~R. MacIver and A.~F. Donaldson, ``{Test-Case Reduction via Test-Case
  Generation: Insights from the Hypothesis Reducer (Tool Insights Paper)},'' in
  \emph{34th European Conference on Object-Oriented Programming (ECOOP 2020)},
  ser. Leibniz International Proceedings in Informatics (LIPIcs), R.~Hirschfeld
  and T.~Pape, Eds., vol. 166.\hskip 1em plus 0.5em minus 0.4em\relax Dagstuhl,
  Germany: Schloss Dagstuhl--Leibniz-Zentrum f{\"u}r Informatik, 2020, pp.
  13:1--13:27. [Online]. Available:
  \url{https://drops.dagstuhl.de/opus/volltexte/2020/13170}
\BIBentrySTDinterwordspacing

\bibitem{S174}
S.~Demeyer, A.~Parsai, S.~Vercammen, B.~van Bladel, and M.~Abdi, ``Formal
  verification of developer tests: a research agenda inspired by mutation
  testing,'' T.~Margaria and B.~Steffen, Eds.

\bibitem{S234}
T.~Laurent, F.~Wall, and A.~Ventresque, ``On the impact of timeouts and jvm
  crashes in pitest,'' in \emph{2020 IEEE International Conference on Software
  Testing, Verification and Validation Workshops (ICSTW)}.\hskip 1em plus 0.5em
  minus 0.4em\relax IEEE, 2020, pp. 247--253.

\bibitem{S526}
A.~Vahabzadeh, A.~Stocco, and A.~Mesbah, ``Fine-grained test minimization,'' in
  \emph{2018 IEEE/ACM 40th International Conference on Software Engineering
  (ICSE)}, 2018, pp. 210--221.

\bibitem{S211}
\BIBentryALTinterwordspacing
E.~Gabrielova, ``End-to-end regression testing for distributed systems,'' in
  \emph{Proceedings of the 18th Doctoral Symposium of the 18th International
  Middleware Conference}, ser. Middleware '17.\hskip 1em plus 0.5em minus
  0.4em\relax New York, NY, USA: Association for Computing Machinery, 2017, p.
  9–12. [Online]. Available: \url{https://doi.org/10.1145/3152688.3152692}
\BIBentrySTDinterwordspacing

\bibitem{S207}
\BIBentryALTinterwordspacing
S.~Elbaum, G.~Rothermel, and J.~Penix, ``Techniques for improving regression
  testing in continuous integration development environments,'' in
  \emph{Proceedings of the 22nd ACM SIGSOFT International Symposium on
  Foundations of Software Engineering}, ser. FSE 2014.\hskip 1em plus 0.5em
  minus 0.4em\relax New York, NY, USA: Association for Computing Machinery,
  2014, p. 235–245. [Online]. Available:
  \url{https://doi.org/10.1145/2635868.2635910}
\BIBentrySTDinterwordspacing

\bibitem{S584}
S.~Mansky and E.~L. Gunter, ``Safety of a smart classes-used regression test
  selection algorithm,'' \emph{Electronic Notes in Theoretical Computer
  Science}, vol. 351, pp. 51--73, 2020.

\bibitem{S269}
D.~Ginelli, M.~Martinez, L.~Mariani, and M.~Monperrus, ``A comprehensive study
  of code-removal patches in automated program repair,'' \emph{arXiv preprint
  arXiv:2012.06264}, 2020.

\bibitem{S704}
\BIBentryALTinterwordspacing
R.~Ramler, C.~Klammer, and T.~Wetzlmaier, ``Lessons learned from making the
  transition to model-based gui testing,'' in \emph{Proceedings of the 10th ACM
  SIGSOFT International Workshop on Automating TEST Case Design, Selection, and
  Evaluation}, ser. A-TEST 2019.\hskip 1em plus 0.5em minus 0.4em\relax New
  York, NY, USA: Association for Computing Machinery, 2019, p. 22–27.
  [Online]. Available: \url{https://doi.org/10.1145/3340433.3342823}
\BIBentrySTDinterwordspacing

\bibitem{S565}
\BIBentryALTinterwordspacing
P.~Alvaro, A.~Hutchinson, N.~Conway, W.~R. Marczak, and J.~M. Hellerstein,
  ``Bloomunit: Declarative testing for distributed programs,'' in
  \emph{Proceedings of the Fifth International Workshop on Testing Database
  Systems}, ser. DBTest '12.\hskip 1em plus 0.5em minus 0.4em\relax New York,
  NY, USA: Association for Computing Machinery, 2012. [Online]. Available:
  \url{https://doi.org/10.1145/2304510.2304512}
\BIBentrySTDinterwordspacing

\bibitem{S73}
F.~G. de~Oliveira~Neto, F.~Dobslaw, and R.~Feldt, ``Using mutation testing to
  measure behavioural test diversity,'' in \emph{2020 IEEE International
  Conference on Software Testing, Verification and Validation Workshops
  (ICSTW)}.\hskip 1em plus 0.5em minus 0.4em\relax IEEE, 2020, pp. 254--263.

\bibitem{S348}
\BIBentryALTinterwordspacing
J.~Malm, A.~Causevic, B.~Lisper, and S.~Eldh, ``Automated analysis of
  flakiness-mitigating delays,'' in \emph{Proceedings of the IEEE/ACM 1st
  International Conference on Automation of Software Test}, ser. AST '20.\hskip
  1em plus 0.5em minus 0.4em\relax New York, NY, USA: Association for Computing
  Machinery, 2020, p. 81–84. [Online]. Available:
  \url{https://doi.org/10.1145/3387903.3389320}
\BIBentrySTDinterwordspacing

\bibitem{S520}
M.~White, M.~Tufano, M.~Martinez, M.~Monperrus, and D.~Poshyvanyk, ``Sorting
  and transforming program repair ingredients via deep learning code
  similarities,'' in \emph{2019 IEEE 26th International Conference on Software
  Analysis, Evolution and Reengineering (SANER)}.\hskip 1em plus 0.5em minus
  0.4em\relax IEEE, 2019, pp. 479--490.

\bibitem{S1070}
H.~Ye, M.~Martinez, and M.~Monperrus, ``Automated patch assessment for program
  repair at scale,'' \emph{Empirical Software Engineering}, vol.~26, no.~2, pp.
  1--38, 2021.

\bibitem{S229}
S.~Urli, Z.~Yu, L.~Seinturier, and M.~Monperrus, ``How to design a program
  repair bot? insights from the repairnator project,'' in \emph{2018 IEEE/ACM
  40th International Conference on Software Engineering: Software Engineering
  in Practice Track (ICSE-SEIP)}.\hskip 1em plus 0.5em minus 0.4em\relax IEEE,
  2018, pp. 95--104.

\bibitem{S762}
M.~Soltani, P.~Derakhshanfar, X.~Devroey, and A.~Van~Deursen, ``A
  benchmark-based evaluation of search-based crash reproduction,''
  \emph{Empirical Software Engineering}, vol.~25, no.~1, pp. 96--138, 2020.

\bibitem{S17}
B.~Vancsics, T.~Gergely, and {\'A}.~Besz{\'e}des, ``Simulating the effect of
  test flakiness on fault localization effectiveness,'' in \emph{2020 IEEE
  Workshop on Validation, Analysis and Evolution of Software Tests
  (VST)}.\hskip 1em plus 0.5em minus 0.4em\relax IEEE, 2020, pp. 28--35.

\bibitem{S692}
\BIBentryALTinterwordspacing
J.~Jeon and S.~Hong, ``Threats to validity in experimenting mutation-based
  fault localization,'' in \emph{Proceedings of the ACM/IEEE 42nd International
  Conference on Software Engineering: New Ideas and Emerging Results}, ser.
  ICSE-NIER '20.\hskip 1em plus 0.5em minus 0.4em\relax New York, NY, USA:
  Association for Computing Machinery, 2020, p. 1–4. [Online]. Available:
  \url{https://doi.org/10.1145/3377816.3381746}
\BIBentrySTDinterwordspacing

\bibitem{S1013}
P.~Zhang, Y.~Jiang, A.~Wei, V.~Stodden, D.~Marinov, and A.~Shi,
  ``Domain-specific fixes for flaky tests with wrong assumptions on
  underdetermined specifications,'' in \emph{2021 IEEE/ACM 43rd International
  Conference on Software Engineering (ICSE)}, 2021, pp. 50--61.

\bibitem{S1009}
\BIBentryALTinterwordspacing
R.~Haas, D.~Elsner, E.~Juergens, A.~Pretschner, and S.~Apel, \emph{How Can
  Manual Testing Processes Be Optimized? Developer Survey, Optimization
  Guidelines, and Case Studies}.\hskip 1em plus 0.5em minus 0.4em\relax New
  York, NY, USA: Association for Computing Machinery, 2021, p. 1281–1291.
  [Online]. Available: \url{https://doi.org/10.1145/3468264.3473922}
\BIBentrySTDinterwordspacing

\bibitem{S1044}
S.~Dutta, A.~Arunachalam, and S.~Misailovic, ``To seed or not to seed? an
  empirical analysis of usage of seeds for testing in machine learning
  projects,'' in \emph{15th IEEE International Conference on Software Testing,
  Verification and Validation}, 2022.

\bibitem{S950}
B.~C.~F. Silva, G.~Carvalho, and A.~Sampaio, ``Cpn simulation-based test case
  generation from controlled natural-language requirements,'' \emph{Science of
  Computer Programming}, vol. 181, pp. 111--139, 2019.

\bibitem{S9}
A.~Shi, W.~Lam, R.~Oei, T.~Xie, and D.~Marinov, ``ifixflakies: A framework for
  automatically fixing order-dependent flaky tests,'' in \emph{Proceedings of
  the 2019 27th ACM Joint Meeting on European Software Engineering Conference
  and Symposium on the Foundations of Software Engineering}, 2019, pp.
  545--555.

\bibitem{S446}
N.~Wild, H.~Lichter, and P.~Kehren, ``Test automation challenges for
  application landscape frameworks,'' in \emph{2020 IEEE International
  Conference on Software Testing, Verification and Validation Workshops
  (ICSTW)}.\hskip 1em plus 0.5em minus 0.4em\relax IEEE, 2020, pp. 330--333.

\bibitem{S963}
G.~Grano, F.~Palomba, D.~Di~Nucci, A.~De~Lucia, and H.~C. Gall, ``Scented since
  the beginning: On the diffuseness of test smells in automatically generated
  test code,'' \emph{Journal of Systems and Software}, vol. 156, pp. 312--327,
  2019.

\bibitem{S1045}
D.~Olianas, M.~Leotta, F.~Ricca, and L.~Villa, ``Reducing flakiness in
  end-to-end test suites: An experience report,'' in \emph{International
  Conference on the Quality of Information and Communications
  Technology}.\hskip 1em plus 0.5em minus 0.4em\relax Springer, 2021, pp.
  3--17.

\bibitem{S179}
H.~Zhu, L.~Wei, M.~Wen, Y.~Liu, S.-C. Cheung, Q.~Sheng, and C.~Zhou,
  ``Mocksniffer: Characterizing and recommending mocking decisions for unit
  tests,'' in \emph{2020 35th IEEE/ACM International Conference on Automated
  Software Engineering (ASE)}.\hskip 1em plus 0.5em minus 0.4em\relax IEEE,
  2020, pp. 436--447.

\bibitem{S359}
M.~Beller, C.-P. Wong, J.~Bader, A.~Scott, M.~Machalica, S.~Chandra, and
  E.~Meijer, ``What it would take to use mutation testing in industry—a study
  at facebook,'' in \emph{2021 IEEE/ACM 43rd International Conference on
  Software Engineering: Software Engineering in Practice (ICSE-SEIP)}.\hskip
  1em plus 0.5em minus 0.4em\relax IEEE, 2021, pp. 268--277.

\bibitem{S106}
\BIBentryALTinterwordspacing
M.-A. Storey and A.~Zagalsky, ``Disrupting developer productivity one bot at a
  time,'' in \emph{Proceedings of the 2016 24th ACM SIGSOFT International
  Symposium on Foundations of Software Engineering}, ser. FSE 2016.\hskip 1em
  plus 0.5em minus 0.4em\relax New York, NY, USA: Association for Computing
  Machinery, 2016, p. 928–931. [Online]. Available:
  \url{https://doi.org/10.1145/2950290.2983989}
\BIBentrySTDinterwordspacing

\bibitem{SB29}
J.~Bell and G.~Kaiser, ``Unit test virtualization with vmvm,'' in
  \emph{Proceedings of the 36th International Conference on Software
  Engineering}, 2014, pp. 550--561.

\bibitem{S164}
G.~Gay, ``Generating effective test suites by combining coverage criteria,'' in
  \emph{International Symposium on Search Based Software Engineering}.\hskip
  1em plus 0.5em minus 0.4em\relax Springer, 2017, pp. 65--82.

\bibitem{S145}
D.~A. Tomassi and C.~Rubio-Gonz{\'a}lez, ``A note about: Critical review of
  bugswarm for fault localization and program repair,'' \emph{arXiv preprint
  arXiv:1910.13058}, 2019.

\bibitem{S165}
B.~Danglot, O.~L. Vera-P{\'e}rez, B.~Baudry, and M.~Monperrus, ``Automatic test
  improvement with dspot: a study with ten mature open-source projects,''
  \emph{Empirical Software Engineering}, vol.~24, no.~4, pp. 2603--2635, 2019.

\bibitem{S71}
M.~A. Mascheroni, E.~Irraz{\'a}bal, J.~A. Carruthers, and J.~A. Pinto, ``Rapid
  releases and testing problems at the industry: A survey,'' in \emph{XXV
  Congreso Argentino de Ciencias de la Computaci{\'o}n (CACIC)(Universidad
  Nacional de R{\'\i}o Cuarto, C{\'o}rdoba, 14 al 18 de octubre de 2019)},
  2019.

\bibitem{S100}
\BIBentryALTinterwordspacing
L.~Erlenhov, F.~G. de~Oliveira~Neto, M.~Chukaleski, and S.~Daknache,
  ``Challenges and guidelines on designing test cases for test bots,'' in
  \emph{Proceedings of the IEEE/ACM 42nd International Conference on Software
  Engineering Workshops}, ser. ICSEW'20.\hskip 1em plus 0.5em minus 0.4em\relax
  New York, NY, USA: Association for Computing Machinery, 2020, p. 41–45.
  [Online]. Available: \url{https://doi.org/10.1145/3387940.3391535}
\BIBentrySTDinterwordspacing

\bibitem{S67}
M.~Machalica, A.~Samylkin, M.~Porth, and S.~Chandra, ``Predictive test
  selection,'' in \emph{2019 IEEE/ACM 41st International Conference on Software
  Engineering: Software Engineering in Practice (ICSE-SEIP)}.\hskip 1em plus
  0.5em minus 0.4em\relax IEEE, 2019, pp. 91--100.

\bibitem{S137}
A.~Memon, Z.~Gao, B.~Nguyen, S.~Dhanda, E.~Nickell, R.~Siemborski, and
  J.~Micco, ``Taming google-scale continuous testing,'' in \emph{2017 IEEE/ACM
  39th International Conference on Software Engineering: Software Engineering
  in Practice Track (ICSE-SEIP)}.\hskip 1em plus 0.5em minus 0.4em\relax IEEE,
  2017, pp. 233--242.

\bibitem{S1021}
\BIBentryALTinterwordspacing
M.~H.~U. Rehman and P.~C. Rigby, ``Quantifying no-fault-found test failures to
  prioritize inspection of flaky tests at ericsson,'' in \emph{Proceedings of
  the 29th ACM Joint Meeting on European Software Engineering Conference and
  Symposium on the Foundations of Software Engineering}, ser. ESEC/FSE
  2021.\hskip 1em plus 0.5em minus 0.4em\relax New York, NY, USA: Association
  for Computing Machinery, 2021, p. 1371–1380. [Online]. Available:
  \url{https://doi.org/10.1145/3468264.3473930}
\BIBentrySTDinterwordspacing

\bibitem{S1018}
C.~Li, C.~Zhu, W.~Wang, and A.~Shi, ``Repairing order-dependent flaky tests via
  test generation.''\hskip 1em plus 0.5em minus 0.4em\relax ICSE, 2022.

\bibitem{S1058}
S.~Mondal, D.~Silva, and M.~d'Amorim, ``Soundy automated parallelization of
  test execution,'' in \emph{2021 IEEE International Conference on Software
  Maintenance and Evolution (ICSME)}, 2021, pp. 309--319.

\bibitem{luo2014empirical}


\end{thebibliography}


\begin{thebibliography}{10}
\providecommand{\url}[1]{#1}
\csname url@samestyle\endcsname
\providecommand{\newblock}{\relax}
\providecommand{\bibinfo}[2]{#2}
\providecommand{\BIBentrySTDinterwordspacing}{\spaceskip=0pt\relax}
\providecommand{\BIBentryALTinterwordstretchfactor}{4}
\providecommand{\BIBentryALTinterwordspacing}{\spaceskip=\fontdimen2\font plus
\BIBentryALTinterwordstretchfactor\fontdimen3\font minus
  \fontdimen4\font\relax}
\providecommand{\BIBforeignlanguage}[2]{{%
\expandafter\ifx\csname l@#1\endcsname\relax
\typeout{** WARNING: IEEEtran.bst: No hyphenation pattern has been}%
\typeout{** loaded for the language `#1'. Using the pattern for}%
\typeout{** the default language instead.}%
\else
\language=\csname l@#1\endcsname
\fi
#2}}
\providecommand{\BIBdecl}{\relax}
\BIBdecl

\bibitem{harman2018start}
M.~Harman and P.~O'Hearn, ``From start-ups to scale-ups: Opportunities and open
  problems for static and dynamic program analysis,'' in \emph{2018 IEEE 18th
  International Working Conference on Source Code Analysis and Manipulation
  (SCAM)}.\hskip 1em plus 0.5em minus 0.4em\relax IEEE, 2018, pp. 1--23.

\bibitem{Labuschagne2017}
\BIBentryALTinterwordspacing
A.~Labuschagne, L.~Inozemtseva, and R.~Holmes, ``Measuring the cost of
  regression testing in practice: A study of java projects using continuous
  integration,'' in \emph{Proceedings of the 2017 11th Joint Meeting on
  Foundations of Software Engineering}, ser. ESEC/FSE 2017.\hskip 1em plus
  0.5em minus 0.4em\relax New York, NY, USA: Association for Computing
  Machinery, 2017, p. 821–830. [Online]. Available:
  \url{https://doi.org/10.1145/3106237.3106288}
\BIBentrySTDinterwordspacing

\bibitem{googleFlaky2016}
\BIBentryALTinterwordspacing
J.~Micco. (2016) Flaky tests at google and how we mitigate them. [Online].
  Available:
  \url{https://testing.googleblog.com/2016/05/flaky-tests-at-google-and-how-we.html}
\BIBentrySTDinterwordspacing

\bibitem{github2020reducing}
\BIBentryALTinterwordspacing
J.~Raine, ``\BIBforeignlanguage{en-US}{Reducing flaky builds by 18x},'' Dec.
  2020. [Online]. Available:
  \url{https://github.blog/2020-12-16-reducing-flaky-builds-by-18x/}
\BIBentrySTDinterwordspacing

\bibitem{fowler2011eradicating}
M.~Fowler, ``Eradicating non-determinism in tests,'' 2011.

\bibitem{SandhuTesting2015}
\BIBentryALTinterwordspacing
A.~Sandhu, ``How to fix flaky tests,'' 2015. [Online]. Available:
  \url{https://tech.justeattakeaway.com/2015/03/30/how-to-fix-flaky-tests/}
\BIBentrySTDinterwordspacing

\bibitem{palmer2019}
\BIBentryALTinterwordspacing
J.~Palmer, ``Test flakiness – methods for identifying and dealing with flaky
  tests,'' 2019. [Online]. Available:
  \url{https://engineering.atspotify.com/2019/11/18/test-flakiness-methods-for-identifying-and-dealing-with-flaky-tests/}
\BIBentrySTDinterwordspacing

\bibitem{luo2014empirical}
Q.~Luo, F.~Hariri, L.~Eloussi, and D.~Marinov, ``An empirical analysis of flaky
  tests,'' in \emph{Proceedings of the 22nd ACM SIGSOFT International Symposium
  on Foundations of Software Engineering}, 2014, pp. 643--653.

\bibitem{gruber2021empirical}
M.~Gruber, S.~Lukasczyk, F.~Kroi{\ss}, and G.~Fraser, ``An empirical study of
  flaky tests in python,'' in \emph{2021 14th IEEE Conference on Software
  Testing, Verification and Validation (ICST)}.\hskip 1em plus 0.5em minus
  0.4em\relax IEEE, 2021, pp. 148--158.

\bibitem{Hashemi2022flakyJS}
N.~Hashemi, A.~Tahir, and S.~Rasheed, ``An empirical study of flaky tests in
  javascript,'' in \emph{2022 38th IEEE International Conference on Software
  Maintenance and Evolution (ICSME)}, 2022.

\bibitem{gambi2018practical}
A.~Gambi, J.~Bell, and A.~Zeller, ``Practical test dependency detection,'' in
  \emph{2018 IEEE 11th International Conference on Software Testing,
  Verification and Validation (ICST)}.\hskip 1em plus 0.5em minus 0.4em\relax
  IEEE, 2018, pp. 1--11.

\bibitem{dong2021flaky}
Z.~Dong, A.~Tiwari, X.~L. Yu, and A.~Roychoudhury, ``Flaky test detection in
  android via event order exploration,'' in \emph{Proceedings of the 29th ACM
  Joint Meeting on European Software Engineering Conference and Symposium on
  the Foundations of Software Engineering}, 2021, pp. 367--378.

\bibitem{memon2013automated}
A.~M. Memon and M.~B. Cohen, ``Automated testing of gui applications: models,
  tools, and controlling flakiness,'' in \emph{2013 35th International
  Conference on Software Engineering (ICSE)}.\hskip 1em plus 0.5em minus
  0.4em\relax IEEE, 2013, pp. 1479--1480.

\bibitem{romano2021empirical}
A.~Romano, Z.~Song, S.~Grandhi, W.~Yang, and W.~Wang, ``An empirical analysis
  of ui-based flaky tests,'' in \emph{2021 IEEE/ACM 43rd International
  Conference on Software Engineering (ICSE)}.\hskip 1em plus 0.5em minus
  0.4em\relax IEEE, 2021, pp. 1585--1597.

\bibitem{kitchenham2007guidelines}
B.~Kitchenham and S.~Charters, ``Guidelines for performing systematic
  literature reviews in software engineering,'' EBSE Technical Report,
  EBSE-2007-01, 2007.

\bibitem{Garousi2019Guidelines}
V.~Garousi, M.~Felderer, and M.~V. M{\"a}ntyl{\"a}, ``Guidelines for including
  grey literature and conducting multivocal literature reviews in software
  engineering,'' \emph{Information and Software Technology}, vol. 106, pp.
  101--121, Feb. 2019.

\bibitem{Tom2013TD}
E.~Tom, A.~Aurum, and R.~Vidgen, ``An exploration of technical debt,'' \emph{J.
  Syst. Softw.}, vol.~86, no.~6, pp. 1498--1516, Jun. 2013.

\bibitem{garousi2018smell}
V.~Garousi and B.~K{\"u}{\c c}{\"u}k, ``Smells in software test code: A survey
  of knowledge in industry and academia,'' \emph{J. Syst. Softw.}, vol. 138,
  pp. 52--81, 2018.

\bibitem{Islam2019Security}
C.~Islam, M.~A. Babar, and S.~Nepal, ``A {Multi-Vocal} review of security
  orchestration,'' \emph{ACM Comput. Surv.}, vol.~52, no.~2, pp. 1--45, Apr.
  2019.

\bibitem{Butijn2020Blockchains}
B.-J. Butijn, D.~A. Tamburri, and W.-J. van~den Heuvel, ``Blockchains: A
  systematic multivocal literature review,'' \emph{ACM Comput. Surv.}, vol.~53,
  no.~3, pp. 1--37, Jun. 2020.

\bibitem{Garousi2016Multivocalreviews}
V.~Garousi, M.~Felderer, and M.~V. M{\"a}ntyl{\"a}, ``The need for multivocal
  literature reviews in software engineering: complementing systematic
  literature reviews with grey literature,'' in \emph{Proceedings of the 20th
  International Conference on Evaluation and Assessment in Software
  Engineering}, ser. EASE '16, no. Article 26.\hskip 1em plus 0.5em minus
  0.4em\relax New York, NY, USA: Association for Computing Machinery, Jun.
  2016, pp. 1--6.

\bibitem{Glass2006Creativity}
R.~L. Glass and T.~DeMarco, \emph{Software Creativity 2.0}.\hskip 1em plus
  0.5em minus 0.4em\relax developer.* Books, 2006.

\bibitem{Williams2019Grey}
A.~Williams, ``\BIBforeignlanguage{en}{Finding high-quality grey literature for
  use as evidence in software engineering research},'' Ph.D. dissertation,
  University of Canterbury, 2019.

\bibitem{zolfaghari2020root}
B.~Zolfaghari, R.~M. Parizi, G.~Srivastava, and Y.~Hailemariam, ``Root causing,
  detecting, and fixing flaky tests: State of the art and future roadmap,''
  \emph{Software: Practice and Experience}, 2020.

\bibitem{zheng2021research}
W.~Zheng, G.~Liu, M.~Zhang, X.~Chen, and W.~Zhao, ``Research progress of flaky
  tests,'' in \emph{3rd International Workshop on Intelligent Bug
  Fixing}.\hskip 1em plus 0.5em minus 0.4em\relax IEEE, 2021, pp. 639--646.

\bibitem{parry2021survey}
O.~Parry, G.~M. Kapfhammer, M.~Hilton, and P.~McMinn, ``A survey of flaky
  tests,'' \emph{ACM Transactions on Software Engineering and Methodology},
  vol.~31, no.~1, pp. 1--74, 2021.

\bibitem{eck2019understanding}
M.~Eck, F.~Palomba, M.~Castelluccio, and A.~Bacchelli, ``Understanding flaky
  tests: The developer’s perspective,'' in \emph{Proceedings of the 2019 27th
  ACM Joint Meeting on European Software Engineering Conference and Symposium
  on the Foundations of Software Engineering}, 2019, pp. 830--840.

\bibitem{habchi2022qualitative}
S.~Habchi, G.~Haben, M.~Papadakis, M.~Cordy, and Y.~Le~Traon, ``A qualitative
  study on the sources, impacts, and mitigation strategies of flaky tests,'' in
  \emph{2022 IEEE Conference on Software Testing, Verification and Validation
  (ICST)}.\hskip 1em plus 0.5em minus 0.4em\relax IEEE, 2022, pp. 244--255.

\bibitem{ahmad2021empirical}
A.~Ahmad, O.~Leifler, and K.~Sandahl, ``Empirical analysis of practitioners'
  perceptions of test flakiness factors,'' \emph{Software Testing, Verification
  and Reliability}, vol.~31, no.~8, p. e1791, 2021.

\bibitem{barboni2021we}
M.~Barboni, A.~Bertolino, and G.~D. Angelis, ``What we talk about when we talk
  about software test flakiness,'' in \emph{International Conference on the
  Quality of Information and Communications Technology}.\hskip 1em plus 0.5em
  minus 0.4em\relax Springer, 2021, pp. 29--39.

\bibitem{myrbakken2017devsecops}
H.~Myrbakken and R.~Colomo-Palacios, ``Devsecops: a multivocal literature
  review,'' in \emph{International Conference on Software Process Improvement
  and Capability Determination}.\hskip 1em plus 0.5em minus 0.4em\relax
  Springer, 2017, pp. 17--29.

\bibitem{garousi2016and}
V.~Garousi and M.~V. M{\"a}ntyl{\"a}, ``When and what to automate in software
  testing? a multi-vocal literature review,'' \emph{Information and Software
  Technology}, vol.~76, pp. 92--117, 2016.

\bibitem{pereira2021security}
A.~Pereira-Vale, E.~B. Fernandez, R.~Monge, H.~Astudillo, and G.~M{\'a}rquez,
  ``Security in microservice-based systems: A multivocal literature review,''
  \emph{Computers \& Security}, p. 102200, 2021.

\bibitem{neuhaus2006depth}
C.~Neuhaus, E.~Neuhaus, A.~Asher, and C.~Wrede, ``The depth and breadth of
  google scholar: An empirical study,'' \emph{portal: Libraries and the
  Academy}, vol.~6, no.~2, pp. 127--141, 2006.

\bibitem{yasin2020using}
A.~Yasin, R.~Fatima, L.~Wen, W.~Afzal, M.~Azhar, and R.~Torkar, ``On using grey
  literature and google scholar in systematic literature reviews in software
  engineering,'' \emph{IEEE Access}, vol.~8, pp. 36\,226--36\,243, 2020.

\bibitem{mahood2014searching}
Q.~Mahood, D.~Van~Eerd, and E.~Irvin, ``Searching for grey literature for
  systematic reviews: challenges and benefits,'' \emph{Research synthesis
  methods}, vol.~5, no.~3, pp. 221--234, 2014.

\bibitem{adams2016searching}
J.~Adams, F.~C. Hillier-Brown, H.~J. Moore, A.~A. Lake, V.~Araujo-Soares,
  M.~White, and C.~Summerbell, ``Searching and synthesising ‘grey
  literature’and ‘grey information’in public health: critical reflections
  on three case studies,'' \emph{Systematic reviews}, vol.~5, no.~1, pp. 1--11,
  2016.

\bibitem{mcguinness2004owl}
D.~L. McGuinness, F.~Van~Harmelen \emph{et~al.}, ``Owl web ontology language
  overview,'' \emph{W3C recommendation}, vol.~10, no.~10, p. 2004, 2004.

\bibitem{visser2003model}
W.~Visser, K.~Havelund, G.~Brat, S.~Park, and F.~Lerda, ``Model checking
  programs,'' \emph{Automated software engineering}, vol.~10, no.~2, pp.
  203--232, 2003.

\bibitem{ernst2003static}
M.~D. Ernst, ``Static and dynamic analysis: Synergy and duality,'' in \emph{IN
  WODA 2003: ICSE WORKSHOP ON DYNAMIC ANALYSIS}, 2003.

\bibitem{wong2016survey}
W.~E. Wong, R.~Gao, Y.~Li, R.~Abreu, and F.~Wotawa, ``A survey on software
  fault localization,'' \emph{IEEE Transactions on Software Engineering},
  vol.~42, no.~8, pp. 707--740, 2016.

\bibitem{fraser2011evosuite}
G.~Fraser and A.~Arcuri, ``Evosuite: automatic test suite generation for
  object-oriented software,'' in \emph{Proceedings of the 19th ACM SIGSOFT
  symposium and the 13th European conference on Foundations of software
  engineering}, 2011, pp. 416--419.

\bibitem{dietrich2022flaky}
J.~Dietrich, S.~Rasheed, and A.~Tahir, ``Flaky test sanitisation via on-the-fly
  assumption inference for tests with network dependencies,'' in \emph{22nd
  IEEE International Working Conference on Source Code Analysis and
  Manipulation (SCAM)}.\hskip 1em plus 0.5em minus 0.4em\relax IEEE, 2022.

\bibitem{wang2017comprehensive}
J.~Wang, W.~Dou, Y.~Gao, C.~Gao, F.~Qin, K.~Yin, and J.~Wei, ``A comprehensive
  study on real world concurrency bugs in node. js,'' in \emph{2017 32nd
  IEEE/ACM International Conference on Automated Software Engineering
  (ASE)}.\hskip 1em plus 0.5em minus 0.4em\relax IEEE, 2017, pp. 520--531.

\end{thebibliography}

\end{document}